\documentclass[
 reprint,
 amsmath,amssymb,
 aps,
 nofootinbib,
 superscriptaddress,
 a4paper
]{article}

\usepackage{amsthm}
\usepackage{graphicx}
\usepackage{xcolor}
\usepackage{braket}
\usepackage{amsmath}
\usepackage{amssymb}
\usepackage{amsthm}
\usepackage{bbm}
\usepackage{enumerate}
\usepackage{booktabs}
\usepackage[subrefformat=parens]{subcaption}
\usepackage{caption}
\usepackage{authblk}
\usepackage{cite}
\usepackage{mathtools}
\usepackage[a4paper]{geometry}
\usepackage{algorithm}
\usepackage{algorithmicx}
\usepackage{algpseudocode}
\usepackage{acronym}
\usepackage{array}
\usepackage{nicematrix}
\usepackage{makecell}
\usepackage{multirow}
\usepackage{mathrsfs}
\usepackage{tikz}
\usetikzlibrary{quantikz2}
\usetikzlibrary{decorations.pathreplacing}
\usepackage[colorlinks=true
,linktocpage=true
,pdfproducer=medialab
,pdfa=true
,hyperfootnotes=false]
{hyperref}

\newtheorem{theorem}{Theorem}
\newtheorem{lemma}{Lemma}

\newtheorem{assum}{Assumption}
\theoremstyle{definition}
\newtheorem{definition}{Definition}
\newtheorem{remark}{Remark}

\newcommand{\N}{\mathbb{N}}
\newcommand{\R}{\mathbb{R}}
\newcommand{\Prob}{\mathbb{P}}
\newcommand{\E}{\mathbb{E}}
\newcommand{\drm}{\mathrm{d}}
\newcommand{\vecb}{\mathbf{b}}
\newcommand{\vecc}{\mathbf{c}}
\newcommand{\vecm}{\mathbf{m}}
\newcommand{\vecx}{\mathbf{x}}
\newcommand{\vecy}{\mathbf{y}}
\newcommand{\vecJ}{\mathbf{J}}
\newcommand{\vecX}{\mathbf{X}}
\newcommand{\vecY}{\mathbf{Y}}

\newcommand{\vecW}{\mathbf{W}}
\newcommand{\vecmu}{\boldsymbol{\mu}}
\newcommand{\veceta}{\boldsymbol{\eta}}
\newcommand{\vecDelta}{\boldsymbol{\Delta}}
\newcommand{\NC}{J}
\newcommand{\NumIntItoIntegUB}{L}
\newcommand{\calA}{\mathcal{A}}
\newcommand{\calB}{\mathcal{B}}
\newcommand{\calC}{\mathcal{C}}
\newcommand{\calD}{\mathcal{D}}
\newcommand{\calE}{\mathcal{E}}
\newcommand{\veccalE}{\boldsymbol{\mathcal{E}}}
\newcommand{\calL}{\mathcal{L}}
\newcommand{\calM}{\mathcal{M}}
\newcommand{\calN}{\mathcal{N}}
\newcommand{\calS}{\mathcal{S}}
\newcommand{\calU}{\mathcal{U}}

\newcommand{\calY}{\mathcal{Y}}

\newcommand{\veccalX}{\boldsymbol{\mathcal{X}}}
\newcommand{\veccalY}{\boldsymbol{\mathcal{Y}}}
\newcommand{\vecXtrunc}{\bar{\vecX}}
\newcommand{\SampCarErrProof}{S}

\begin{document}

\title{Quantum algorithm for high-dimensional nonlinear stochastic differential equations based on Carleman linearization}
\renewcommand{\thefootnote}{\fnsymbol{footnote}}

\author{
Koichi Miyamoto$^{1}$\footnote{miyamoto.kouichi.qiqb@osaka-u.ac.jp}

\small $^{1}$ Center for Quantum Information and Quantum Biology, The University of Osaka, Toyonaka, Osaka, Japan
}

\date{\today}
\maketitle

\acrodef{EM}{Euler-Maruyama}
\acrodef{LCU}{linear combination of unitaries}
\acrodef{PRN}{pseudorandom number}
\acrodef{PRNG}{pseudorandom number generator}
\acrodef{PCG}{permutated congruential generator}
\acrodef{ODE}{ordinary differential equation}
\acrodef{SDE}{stochastic differential equation}
\acrodef{OU}{Ornstein–Uhlenbeck}
\acrodef{CL}{Carleman linearization}

\renewcommand{\thefootnote}{\arabic{footnote}}

\begin{abstract}
    Addressing nonlinear numerical problems in quantum computing remains challenging, but several approaches, including Carleman linearization (CL), have been proposed to construct quantum algorithms for such problems.
    In this work, we develop a CL-based method for a high-dimensional nonlinear stochastic differential equation (SDE) $\drm \vecX_t = \left(A(t) \vecX_t + F(t) \vecX_t^{\otimes s}\right) \drm t + B(t) \drm \vecW_t$ for an $\R^N$-valued random process $\vecX_t$, aiming at quantum speed-up with respect to the dimension $N$.
    We introduce auxiliary processes $\vecY^j_t$ that approximate $\vecX_t^{\otimes j}$, derive a linearized system of SDEs for them, and obtain the linear ordinary differential equation (ODE) governing their expectations.
    We then apply a quantum ODE solver to it and, from the resulting quantum state, extract an estimate for the expectation of a linear function of $\vecX_t$ by overlap estimation.
    We show that, provided the nonlinear and noise terms in the SDE are sufficiently small relative to the linear term, the CL error decreases exponentially with the truncation order $\NC$ in CL, and we can estimate the target expectation with a ${\rm polylog}(N)$ number of queries to unitaries encoding coefficient matrices $A$, $F$, and $B$.
    Finally, we also run a numerical demonstration for an SDE related to Langevin dynamics, observing the exponential decay of the CL error as $\NC$ increases.
\end{abstract}

\section{Introduction}

Among the active research on quantum algorithms for high-dimensional numerical problems, including linear equation systems \cite{HHL,Childs2017QLSS,Subasi2019,Lin2020optimalpolynomial,Martyn2021,An2022QLSS,costa2022optimal} (see Ref. \cite{morales2024quantum} as a survey) and \acp{ODE} \cite{Berry_2014,berry2017quantum,childs2020quantum,Liu2021,Fang2023timemarchingbased,Krovi2023improvedquantum,liu2023efficient,An2023LCHS,Berry2024quantumalgorithm,Jennings2024costofsolvinglinear,low2025optimal,an2025quantum,costa2025further,an2026fast,an2026quantum}, those for \acp{SDE} have recently attracted attention.
There are several proposals~\cite{bravyi2025quantum,bravyi2026quantum,wu2026universal,miyamoto2026quantum} on quantum algorithms for \acp{SDE} that provide quantum speed-up with respect to the dimension of the \ac{SDE}.
In this paper, we consider a quantum algorithm for high-dimensional nonlinear \acp{SDE}, based on \ac{CL}. 

Concretely, we consider the following problem.
Let $T\in(0,\infty)$, $N,m\in\N$, and $(\Omega,\mathcal{F},\Prob,(\mathcal{F}_t)_{t\in[0,T]})$ be a filtered probability space equipped with a standard $m$-dimensional Brownian motion $\vecW:[0,T] \times \Omega \rightarrow \R^m$.
We consider a stochastic process $\vecX=(X^1,\ldots,X^N)^T:[0,T] \times \Omega \rightarrow \R^N$ that follows the \ac{SDE}
\begin{align}
    \drm \vecX_t = \vecmu(\vecX_t) \drm t + \Sigma(\vecX_t) \drm \vecW_t
    \label{eq:SDEGen}
\end{align}
with the drift function $\vecmu=(\mu_1,\ldots,\mu_N)^T:\R^N \rightarrow \R^N$ and the diffusion function $\Sigma=(\Sigma_{i,j})_{1 \le i \le N, 1 \le j \le m}:\R^N \rightarrow \R^{N \times m}$.
In particular, this study focuses on the \ac{SDE}
\begin{align}
    \drm \vecX_t = \left(A(t) \vecX_t + F(t) \vecX_t^{\otimes s}\right) \drm t + B(t) \drm \vecW_t
    \label{eq:SDE}
\end{align}
with an integer $s \ge 3$~\footnote{We assume this because the condition in Theorem~\ref{th:SDESolExist}, under which the \ac{SDE} \eqref{eq:SDE} has a solution, does not hold if $s=2$.}, $A:[0,T]\rightarrow\R^{N \times N}$, $F:[0,T]\rightarrow\R^{N \times N^s}$, $B:[0,T]\rightarrow\R^{N \times m}$, and the initial value $\vecX_0=\vecX_{\rm ini}=(X^1_{\rm ini},\ldots,X^N_{\rm ini})^T$, which is assumed to be deterministic for simplicity.
Our aim is to analyze $\vecX$, namely, to obtain some information that characterizes $\vecX$ as classical values.

Because of the unitarity of quantum mechanics, addressing nonlinear systems by quantum computing inevitably accompanies challenges.
\ac{CL}, which was originally proposed nearly a century ago \cite{Carleman}, is a widely used method to map a nonlinear \ac{ODE} to a linear one with higher dimension so that quantum algorithms can be applied \cite{Liu2021,Krovi2023improvedquantum,liu2023efficient,costa2025further}.
In this method, to linearize an \ac{ODE} $\frac{\drm}{\drm t}\vecX_t = \mathbf{f}(\vecX_t)$ for a $\R^N$-valued process $\vecX_t$ with a polynomial $\mathbf{f}$, we introduce an auxiliary variables $\vecY^1_t,\vecY^2_t,\ldots$ that corresponds to $\vecX_t^{\otimes 1}, \vecX_t^{\otimes 2}, \ldots$, tensor products of $\vecX_t$.
By regarding $\vecX_t^{\otimes 1}, \vecX_t^{\otimes 2}, \ldots$ as separate variables, the system consisting of them apparently becomes linear, but any finite collection of $\vecX_t^{\otimes 1},\ldots,\vecX_t^{\otimes J}$ fails to form a closed system, since $\frac{\drm}{\drm t}\vecX_t^{\otimes j}$ depends on higher-order tensor products $\vecX_t^{\otimes k},k>j$.
We thus make a truncation at a finite order, namely, neglect $\vecX_t^{\otimes j},j>\NC$ with some $\NC$.
If the nonlinear term in $\mathbf{f}$ is {\it smaller} than the linear one in some sense, the truncated system is a good approximation of the original one, making the difference between $\vecY^1_t$ and $\vecX_t$ decrease exponentially in $J$.

In this work, we take the same approach for the \ac{SDE} \eqref{eq:SDE}.
We can derive the \acp{SDE} for $\vecX_t^{\otimes 2}, \vecX_t^{\otimes 3}, \ldots$ from Eq.~\eqref{eq:SDE} and make a truncation at order $\NC$, obtaining a closed \ac{SDE} system for the approximating processes $\vecY^1_t,\ldots,\vecY^\NC_t$.
Then, we need to address the following issues:
\begin{itemize}
    \item As $J$ increases, does $\vecY^1_t$ converge to $\vecX_t$ in some sense under some condition?
    \item Can we obtain some information about $\vecX_t$ as a classical value using a quantum algorithm with logarithmic complexity in $N$?
\end{itemize}
For the first point, we show that $\E[\vecY^1_t]$, the expectation of $\vecY^1_t$, gets close to that of (rescaled) $\vecX_t$ exponentially in $J$, under a condition about the smallness of the nonlinear and noise terms.
We then construct a quantum algorithm in the following way to achieve the second point.
We derive an \ac{ODE} that $\E[\vecY^1_t],\ldots,\E[\vecY^\NC_t]$ follows from the \ac{SDE} of $\vecY^1_t,\ldots,\vecY^\NC_t$ and use a quantum \ac{ODE} solver in Ref.~\cite{Berry2024quantumalgorithm} to obtain the history state, which encodes the values of $\E[\vecY^j_t]$ at time grid points $t=t_1,\ldots,t_r$.
Then, from this quantum state, we extract by overlap estimation \cite{Knill2007} the expectation $\E[f]$ of a linear function of $\{\vecX_{t_n}\}$, that is, $f=\sum_{n=1}^r \sum_{i=1}^N C_{n,i} X^i_{t_n}$ with $C=(C_{n,i})\in\R^{r \times N}$.
Quantities of practical interest, such as weighted averages of $\vecX_t$ over time and across its entries, take this form, and even expectations of linear functions do not have a closed formula as in the linear case, requiring numerical methods for them.
In the proposed method, both preparing the history state and estimating the expectation take logarithmic query complexities in $N$.

Let us compare this work with existing studies on quantum algorithms for high-dimensional \acp{SDE}.
The author's previous work \cite{miyamoto2026quantum} has focused on linear \acp{SDE} with additive noise.
It has proposed methods to prepare the history state that encodes sample paths of $\vecX_t$ and estimate expectations of polynomials of $\vecX_t$.
Apparently, this approach can be applied to the \ac{SDE} of $\vecY^1_t,\ldots,\vecY^\NC_t$, enabling us to estimate expectations of polynomials of $\vecX_t$ in the nonlinear case as well.
However, as seen below, the \ac{SDE} of $\vecY^1_t,\ldots,\vecY^\NC_t$ has multiplicative noise, which prevents us from using the methods in Ref.~\cite{miyamoto2026quantum} straightforwardly.
Therefore, as the first step, we now focus on linear functions, which is still motivating in the nonlinear case, and leave preparing the state encoding sample paths and estimating expectations of polynomials as future work.

To the best of the author's knowledge, a series of papers \cite{bravyi2025quantum,bravyi2026quantum} is the only work that considers quantum algorithms for high-dimensional nonlinear \acp{SDE}.
The quantum algorithm proposed in the papers estimates expectations of functions of $\vecX_t$ via switching from \acp{SDE} to the Kolmogorov equation, which describes the time evolution of the expectations.
This algorithm outputs expectations of functions of $\vecX_t$ at a single terminal time, while our method can address functions depending on the values of $\vecX_t$ at multiple times\footnote{Although this paper focuses on multi-time functions for simplicity, we expect that extending our algorithm to terminal-time functions is possible similarly to Ref.~\cite{miyamoto2026quantum}.}.
Besides, it requires $\mathbf{X}_t$ to be random at the initial time, unlike our methods, which can address deterministic initial values\footnote{Although this paper concentrates on deterministic initial values, we can extend our algorithm to the case of the random initial value, if we have access to the oracle to generate the superposition of the amplitude-encoding states of possible initial values.}.
It also requires the drift function to be quadratic and satisfy the divergence-free condition, whereas our method can deal with drift functions containing the higher-order term and does not impose such a condition (but the smallness of nonlinearity and noise instead).
A comparison of the existing and currently proposed quantum methods, along with the classical method, is summarized in Table~\ref{tab:compare}. 

\begin{table}[tp]
  \footnotesize
  \centering
  \makebox[\textwidth][c]{
  \begin{tabular}{
  |c|c|c|c|c|c|
  }
    \hline
    \multicolumn{2}{|c|}{Method} & \ac{SDE} & \makecell{Test func. $f$} & Query complexity & Assumed oracles\\
    \hline \hline
    \multirow{4}{*}{Quantum} &
    \makecell{\\ Carleman \\ linearization- \\ based \\ (this paper) \\ ~} & \makecell{Eq.~\eqref{eq:SDE}\\with sufficiently small\\nonlinearity \& noise\\(Eqs.~\eqref{eq:CarlConvCond1}-\eqref{eq:XiniBd})}
    & \makecell{Linear\\(multi-time)} & \makecell{$\tilde{O}\left(\frac{(\alpha_A T)^2\|C\|_{\rm F}\sqrt{\sigma^2T}}{c(1-c)^{9/2}\epsilon}\right)$ \\ (Theorem~\ref{th:MeanEstim})} & \makecell[l]{
      - Block-encodings of \\
      $A$, $BB^T$, $F$ \\
      - State preparation \\ for $\vecX_{\rm ini}$ \\
      - State preparation \\
      for the coefficients of $f$
    }
  \\
    \cline{2-6}
    & \makecell{Kolmogorov\\eq.-based\\ \cite{bravyi2026quantum}} &
    \makecell[l]{$dX^i_t=\left(-\lambda_iX^i_t\right.$ \\
    $ + \sum\nolimits_j b_{ij}X^j_t$ \\ $\left.+\sum\nolimits_{j,k} c_{ijk}X^j_tX^k_t\right)\drm t$\\$  +\sqrt{q}\drm W^i_t$\\with div-free condition}
    & \makecell{Polynomial\\(terminal-time)} & \makecell{linear with $T$,\\${\rm poly}$ with\\$\log N,\epsilon,J,\lambda_1^{-1}$,\\$\exp$ with\\ $\sum_i 2\lambda_i (X^i_{\rm ini})^2/q$} & \makecell{Block-encodings of\\the matrices  in\\Kolmogorov eq.} \\
    \hline
    Classical& \makecell{Euler-Maruyama \\ + \\ Monte Carlo} & Nonlinear\footnotemark~(Eq.~\eqref{eq:SDEGen}) & Nonlinear{\textsuperscript{4}} &
    $\tilde{O}\left(K_\vecX N/\epsilon^3\right)$ \cite{KloedenPlaten}
    & --- \\
    \hline
  \end{tabular}
  }
  \caption{Summary of the methods for estimating $\mathbb{E}[f(\{\mathbf{X}_t\})]$, the expectation of a test function $f$ of the solution of the nonlinear \ac{SDE}. The query complexity means the number of queries to the assumed oracles for the quantum methods and the number of arithmetic operations in the classical method\protect\footnotemark. $\alpha_A$ and $\sigma$ upper bound $A$'s spectral norm and $B$'s Schatten 1-norm, respectively. $\epsilon$ is the required accuracy, and $c\in(0,1)$ is a parameter in the rescaling factor of $\vecX_t$. The detailed definitions are given later.
  For the method of Ref.~\cite{bravyi2026quantum}, $J$ denotes the drift strength (for the detailed definitions, see Ref.~\cite{bravyi2026quantum}). For the classical method, $K_\vecX$ is a positive number depending on the drift and noise function, and it is assumed that evaluating $\vecmu$ and $\Sigma$ takes $O(N)$ arithmetic operations, which applies to Eq.~\eqref{eq:SDE} with $O(1)$-sparse $A$, $B$, and $F$.}
  \label{tab:compare}
\end{table}
\normalsize
\footnotetext[4]{For general coefficient functions in the \ac{SDE} and/or general test functions, the standard \ac{EM} method may not work, or the \ac{SDE} may not even have a solution (e.g., non-globally Lipschitz continuous coefficients). Nonetheless, various modified numerical integration schemes have been proposed. See Ref.~\cite{hutzenthaler2015numerical} and the reference therein.}
\footnotetext[5]{Note that we are comparing the number of queries to the block-encoding quantum circuits and the number of classical arithmetic operations, which are not equivalent. Nonetheless, the block-encoding can be efficiently constructed in some cases, which justifies the above comparison. For example, the block-encoding can be constructed with $O(1)$ queries to the sparse-access oracles and a logarithmic number of additional elementary gates~\cite{Gilyen2019}. If the sparse-access oracles for $A$, $BB^T$, and $F$ are constructed with $\tilde{O}(1)$ arithmetic circuits, the current comparison amounts to comparing the number of quantum arithmetic operations with that of classical ones.}

Although there are some other papers related to our work, their targets somewhat differ from ours.
Ref.~\cite{li2026efficient} focuses on nonlinear \acp{ODE} with the inhomogeneous term being an \ac{OU} process, which differ from nonlinear \acp{SDE} considered in this paper.
Ref.~\cite{wu2026nonlinear} proposes a quantum algorithm for nonlinear \acp{SDE} based on Kolmogorov–Lindblad mapping but does not explicitly establish the scaling of its complexity with the dimension, although logarithmic scaling is conjectured to be possible.

The organization of this paper is as follows.
Sec.~\ref{sec:Pre} is a preliminary one, where some mathematical facts and quantum building blocks are introduced.
Sec.~\ref{sec:main} is the main part, where we develop our quantum algorithm in the aforementioned approach.
In Sec.~\ref{sec:exp}, we conduct a numerical experiment on a nonlinear \ac{SDE} related to overdamped Langevin dynamics to demonstrate that \ac{CL} actually works in this test case.

\section{Preliminaries \label{sec:Pre}}

\subsection{Notation}

We denote the set of all the nonnegative integers by $\N_0$ and the set of all positive real numbers by $\R_+$.

For $n\in\N$, we define $[n] \coloneqq \{1,\ldots,n\}$ and $[n]_0 \coloneqq \{0\} \cup [n]$, and denote the $n \times n$ identity matrix and the $n$-dimensional zero vector by $I_n$ and $\mathbf{0}_n$, respectively.
For $n,n^\prime\in\N$, $O_{n,n^\prime}$ denotes the $n \times n^\prime$ zero matrix.
The subscripts of these may be omitted when clear from context.

For $n \in \N$, $p\in(0,\infty)$, and $\vecx = (x_1,\ldots,x_n)^T \in \mathbb{C}^n$, we define the $p$-norm of $\vecx$ by $\|\vecx\|_p \coloneqq \left(\sum_{i=1}^n |x_i|^p\right)^{1/p}$.
When $p=2$, this is the Euclidean norm, and we omit the subscript: $\|\vecx\|$ denotes $\|\vecx\|_2$.
For $p=\infty$, we define $\|\vecx\|_\infty \coloneqq \max_{i\in[n]} |x_i|$.
We denote by $\calS^n_p$ the $n$-dimensional ``sphere" with respect to $p$-norm: $\calS^n_p \coloneqq \{\vecx\in\R^{n+1}~|~\|\vecx\|_p=1\}$.

For $n,n^\prime\in\N$ and $A\in\R^{n \times n^\prime}$, $A_{:,j},j\in[n^\prime]$ is the $j$-th column of $A$ as a $n$-dimensional vector, and ${\rm vec}(A)$ denotes vectorized $A$, an $nn^\prime$-dimensional vector such that for $i\in[n]$ and $i^\prime\in[n^\prime]$, the $((i-1)n^\prime+i^\prime)$-th entry of ${\rm vec}(A)$ is $A_{i,i^\prime}$.

$\lg$ denotes the base-2 logarithm.

\subsection{Matrix norms}

For $m,n \in \N$, $p\in \R \cup \{\infty\}$, and $A =(A_{i,j})_{i,j} \in \R^{m \times n}$, the $A$'s norm induced by the vector $p$-norms is given by
\begin{align}
     \|A\|_p \coloneqq \sup_{\vecx \in \R^n \setminus \{\mathbf{0}\}} \frac{\|A\vecx\|_p}{\|\vecx\|_p} = \sup_{\vecx \in \calS^n_p} \|A\vecx\|_p.
    \label{eq:pqNormDef}
\end{align}
We hereafter omit the subscript 2 of $\| \cdot \|_2$ for simplicity.

Another type of matrix norm is the Schatten norm: for $p$ and $A$ as above,
\begin{align}
    \|A\|_{S_p} \coloneqq
    \begin{dcases}
        \left(\sum_{i=1}^{\min\{m,n\}} (\sigma_i(A))^p \right)^{1/p}, & ~ p < \infty \\
        \sigma_1(A), & ~ p=\infty
    \end{dcases}
    \label{eq:Schatten}
\end{align}
where $\sigma_i$ is the $i$-th largest singular value of $A$.
Note that $\|A\|_{S_\infty}$ and $\|A\|_{S_2}$ are equal to $\|A\|$ and the Frobenius norm $\|A\|_{\rm F} \coloneqq \left(\sum_{i=1}^m \sum_{j=1}^n |A_{i,j}|^2\right)^{1/2}$, respectively.
By the definition \eqref{eq:Schatten}, we have
\begin{align}
    \|A\|_{S_q} \le \|A\|_{S_p}
    \label{eq:SchNormOrd}
\end{align}
for $p,q\in\R\cup\{\infty\}$ such that $p \le q$.
Besides, by combining von Neumann's trace inequality and H\"{o}lder's inequality, we see that if $p$ and $q$ satisfy $1/p+1/q=1$,
\begin{align}
    |{\rm Tr} (A^\dagger B)| \le \|A\|_{S_p} \|B\|_{S_q}
\end{align}
holds for matrices $A$ and $B$ of the same size.
In particular, for $p=\infty$ and $q=1$, this becomes
\begin{align}
    |{\rm Tr} (A^\dagger B)| \le \|A\| \|B\|_{S_1}.
    \label{eq:TrBound}
\end{align}

We also introduce the logarithmic norm of matrices~\cite{soderlind2024}, which is, in fact, not a norm despite its name: for a square matrix $A \in \R^{n \times n}$, we define
\begin{align}
    \mu(A) \coloneqq \lim_{h \rightarrow 0^+} \frac{\|I+hA\| - 1}{h}.
    \label{eq:logNormDef}
\end{align}
This definition implies that
\begin{align}
    \|I+hA\|  = 1 + h\mu(A) + o(h)
    \label{eq:|I+hA|smallh}
\end{align}
holds for sufficiently small $h>0$.
$\mu(A)$ can be written as~\cite{soderlind2024}
\begin{align}
    \mu(A) = \sup_{\vecx \in \mathbb{C}^n \setminus \mathbf{0}} \frac{\Re (\vecx^\dagger A \vecx)}{\|\vecx\|^2}.
    \label{eq:mu2Formula}
\end{align}
Other properties that we will use later are as follows~\cite{soderlind2024}: for $A,B\in\R^{n \times n}$ and $a\in[0,\infty)$, we have
\begin{align}
    -\|A\| \le \mu(A) \le \|A\|,
    \label{eq:muAle|A|}
\end{align}
\begin{align}
    \mu(A + B) \le \mu(A) + \mu(B),
    \label{eq:muAdd}
\end{align}
\begin{align}
    \mu(aA) =a\mu(A),
    \label{eq:muRNProd}
\end{align}
\begin{align}
    \|e^{aA}\| \le e^{a\mu(A)}.
    \label{eq:e^AtNorm}
\end{align}

\subsection{Solution of stochastic differential equations}

The \ac{SDE} in Eq.~\eqref{eq:SDEGen} does not necessarily have a solution in general.
Nevertheless, it is known that under the local and global one-sided Lipschitz conditions on the coefficient functions, the \ac{SDE} is guaranteed to have a solution.

\begin{theorem}[\cite{mao2007stochastic}, Theorem 2.3.5]
    Consider the \ac{SDE} in Eq.~\eqref{eq:SDEGen}.
    Assume that $\vecmu$ and $\Sigma$ are locally Lipschitz continuous and satisfy the global one-sided Lipschitz condition: there exists $C>0$ independent of $\vecx$ such that 
    \begin{align}
        \vecx^T \vecmu(\vecx) + \frac{1}{2}\|\Sigma(\vecx)\| \le C (1+\|\vecx\|^2)
    \end{align}
    holds for any $\vecx \in \R^N$.
    Then, there exists a unique adapted $\R^d$-valued stochastic process $\vecX$ that solves Eq.~\eqref{eq:SDEGen}.
    \label{th:SDESolExist}
\end{theorem}

For the \ac{SDE} in the form of Eq.~\eqref{eq:SDE}, the solution has an implicit expression in terms of the time-evolution operator $\Phi_A$, which is defined as follows: for $0 \le u \le t \le T$,
\begin{align}
    \Phi_A(t,u) \coloneqq \sum_{n=0}^\infty \int_u^t \drm \tau_1 \int_u^{\tau_1} \drm \tau_2 \cdots \int_s^{\tau_{n-1}} \drm \tau_n A(\tau_1) \cdots A(\tau_n).
\end{align}
We see that the solution of Eq.~\eqref{eq:SDE} satisfies
\begin{align}
    \vecX_t = \Phi_A(t,0)\vecX_{\rm ini} + \int_0^t \Phi_A(t,u)F(u)\vecX_u^{\otimes s} \drm u + \int_0^t \Phi_A(t,u)B(u) \drm \vecW_u,
    \label{eq:SDESol}
\end{align}
by using the property \cite[Theorem 1 on p. 20]{brockett2015finite}
\begin{align}
    \frac{\drm}{\drm t}\Phi_A(t,u)=A\Phi_A(t,u).
\end{align}
As another property used later, we note that the norm of $\Phi_A(t,u)$ is bounded as \cite[Eq. (6.6.42)]{soderlind2024}
\begin{align}
    \|\Phi_A(t,u)\| \le \exp\left(\int_u^t \mu(A(\tau)) \drm\tau\right).
    \label{eq:PhiNormBnd}
\end{align}

\subsection{Quantum building blocks}

\subsubsection{Representing numbers on qubits and arithmetic circuits}

In this paper, we consider computation on systems consisting of quantum registers, each of which is a set of qubits.
We can represent an integer $i$ on a quantum register as a binary: we denote by $\ket{i}$ the computational basis state on a quantum register that holds the bit string equal to the binary representation of $i$.
Then, we can implement quantum circuits for arithmetic operations on numbers represented on registers, such as addition $O_{\mathrm{Add}}:\ket{a}\ket{b}\mapsto\ket{a}\ket{a+b}$ and modular addition $O_{\mathrm{MAdd}}:\ket{a}\ket{b}\mapsto\ket{a}\ket{a+b~{\rm mod}~c}$, where $a,b\in\mathbb{Z}$ and $c\in\mathbb{N}$.
These circuits can be constructed with a polynomial number of elementary gates with respect to the number $n$ of bits in the binary representation of real numbers; for example, modular addition costs $O(n)$ elementary gates.
For concrete implementations, see Ref.~\cite{wang2025comprehensive} and the references therein.

Note that modular addition can be used to shift rows and columns.
Suppose that we have access to an $n$-qubit unitary $U=(u_{i,j})_{i,j=1,\ldots,2^n}$.
By operating $U$ followed by the controlled modular addition that maps $\ket{i},i\in[2^n]$ to $\ket{f_{\rm MAdd}^{a,c,n}(i)}$, where $a\in[2^n],c\in[2^n-a]$, and
\begin{align}
    f_{\rm MAdd}^{a,c,n}(i) \coloneqq
    \begin{dcases}
        \left(i + c - a ~ {\rm mod} ~ 2^n-a+1\right) + a & ; ~ i \ge a\\
        i & ; ~ {\rm otherwise}
    \end{dcases},
\end{align}
we can implement a unitary
\begin{align}
    \begin{pmatrix}
        u_{1,1} & \cdots & u_{1,2^n} \\
        & \vdots & \\
        u_{a-1,1} & \cdots & u_{a-1,2^n} \\
        u_{2^n-c+1,1} & \cdots & u_{2^n-c+1,2^n} \\
        & \vdots & \\
        u_{2^n,1} & \cdots & u_{2^n,2^n} \\
        u_{a,1} & \cdots & u_{a,2^n} \\
        & \vdots & \\
        u_{2^n-c,1} & \cdots & u_{2^n-c,2^n}
    \end{pmatrix},
\end{align}
which is obtained by cyclically shifting the $a$-th through $2^n$-th rows of $U$, while leaving the first through $(a-1)$-th rows unchanged.
On the other hand, $U$ preceded by the same modular addition yields the shift of the columns, which results in the unitary
\begin{align}
    \begin{pmatrix}
        u_{1,1} &  & u_{1,a-1} & u_{1,2^n-c+1} &  & u_{1,2^n} & u_{1,a} &  & u_{1,2^n-c}\\
        \vdots & \cdots & \vdots & \vdots & \cdots & \vdots & \vdots & \cdots & \vdots \\
        u_{2^n,1} & & u_{2^n,a-1} & u_{2^n,2^n-c+1} & & u_{2^n,2^n} & u_{2^n,a} & & u_{2^n,2^n-c}
    \end{pmatrix}.
\end{align}

Also note that, by using the fixed-point binary representation, we can extend the above representation of integers on a quantum register to real numbers: for $x\in\R$, we denote by $\ket{x}$ the computational basis state on a quantum register that holds the bit string equal to the binary representation of $x$.
We assume that every register has a sufficiently large number of qubits and thus neglect errors caused by finite-precision representation hereafter.

\subsubsection{Block-encoding}

Block-encoding~\cite{chakraborty2019power,Gilyen2019}, which is mathematically defined as follows, has now been a common tool for developing quantum algorithms.

\begin{definition}
    Let $L_1,L_2\in\mathbb{N}$ and $A\in\mathbb{C}^{L_1 \times L_2}$.
    Let $a\in\mathbb{N}$, $\alpha \ge \|A\|$, and $\epsilon \ge 0$.
    If an $(l+a)$-qubit unitary $U$ has the upper-left $L_1 \times L_2$ block $U_{\rm ul}$ satisfying $\|A-\alpha U_{\rm ul}\|\le\epsilon$, where $l=\left\lceil \lg \left(\max\{L_1,L_2\}\right) \right\rceil$, we say that $U$ is an $(\alpha,a,\epsilon)$-block-encoding of $A$.
    We call $\alpha$, $a$, and $\epsilon$ the normalization factor, the ancilla qubit number, and the accuracy of the block-encoding, respectively.
\end{definition}

Let us also define the time-dependent version of block-encoding, which gives the block-encoding of the value of a time-dependent matrix at the specified time.

\begin{definition}
    Let $L_1,L_2\in\mathbb{N}$ and $A:[0,T]\rightarrow\mathbb{C}^{L_1 \times L_2}$.
    Let $a\in\mathbb{N}$, $\alpha \ge \max_{t\in[0,T]}\|A(t)\|$, and $\epsilon \ge 0$.
    Let $l=\left\lceil \lg \left(\max\{L_1,L_2\}\right) \right\rceil$.
    If a quantum circuit $U$ acts on a system consisting of two registers, the latter of which has $(l+a)$ qubits, as
    \begin{align}
        U \ket{t}\ket{\psi} = \ket{t}U_{A(t)}\ket{\psi}
    \end{align}
    for any $t\in[0,T]$ and any $(l+a)$-qubit state $\ket{\psi}$, where $U_{A(t)}$ is an $(\alpha,a,\epsilon)$-block-encoding of $A(t)$, we say that $U$ is an $(\alpha,a,\epsilon)$-time-dependent-block-encoding of $A(t)$.
\end{definition}

We now present some techniques related to block-encoding for later use.
First, if we have block-encodings of matrices, we can construct the block-encoding of their linear combination.

\begin{theorem}[Lemma 29 in Ref.~\cite{Gilyen2019}, modified]
    Let $n\in\N$, $A_1,\ldots,A_n$ be complex matrices of the same size, $\vecy=(y_1,\ldots,y_n)^T \in \R_+^n$, and $A=\sum_{i=1}^n y_i A_i$.
    Let $U_1,\ldots,U_n$ be $(\alpha,a,\epsilon)$-block-encodings of $A_1,\ldots,A_n$, respectively, where $\alpha \ge \max_{i\in[n]} \|A_i\|$, $a\in\N_0$, and $\epsilon\ge0$.
    Then, we can implement a $(\|\vecy\|_1 \alpha, a + \lceil \lg n \rceil, \|\vecy\|_1 \epsilon)$-block-encoding of $A$, using the controlled versions of $U_1,\ldots,U_n$ once each, and $O(n)$ elementary gates.
    \label{th:BESum}
\end{theorem}

\begin{proof}
    $W$ in Lemma 29 in Ref.~\cite{Gilyen2019} can be implemented by sequentially using controlled $U_1,...,U_n$.
    Both $P_L$ and $P_R$ in the lemma are now the oracle to generate $\frac{1}{\|\vecy\|_1}\sum_{i=1}^n \sqrt{y_i}\ket{i}$, with $\beta=\|\vecy\|_1$ and $\epsilon_1=0$.
    This oracle can be constructed by $O(n)$ elementary gates~\cite{mottonen2005}.
    Combining these things, we obtain the claim\footnote{There is a typo in Lemma 29 in Ref.~\cite{Gilyen2019}: ``$(\alpha\beta,a+b,\alpha\epsilon_1+\alpha\beta\epsilon_2)$-block-encoding" in the last sentence should be ``$(\alpha\beta,a+b,\alpha\epsilon_1+\beta\epsilon_2)$-block-encoding".}.
\end{proof}

Second, we can adjust the normalization factor and the ancilla qubit number to larger values. 

\begin{lemma}
    If we have access to a $(\alpha,a,\epsilon)$-block-encoding $U_A$ of a matrix $A$, where $\alpha \ge \|A\|$, $a\in\N_0$, and $\epsilon\ge0$, we also have access to $(\alpha^\prime,a+1,\epsilon)$-block-encoding of $A$ for any $\alpha^\prime>\alpha$.
    \label{lem:AdjustBEScFac}
\end{lemma}

\begin{proof}
    We add one qubit to the system S on which $U_A$ acts.
    By operating the Y-rotation gate $R_Y\left(2\arccos\left(\frac{\alpha}{\alpha^\prime}\right)\right)=
    \begin{pmatrix}
        \frac{\alpha}{\alpha^\prime} & -\sqrt{1-\left(\frac{\alpha}{\alpha^\prime}\right)^2} \\
        \sqrt{1-\left(\frac{\alpha}{\alpha^\prime}\right)^2} & \frac{\alpha}{\alpha^\prime}
    \end{pmatrix}$
    on the added qubit and then $U_A$ on the original system, the following transformation is realized
    \begin{align}
        \ket{0}\ket{\psi} \rightarrow \ket{0}\frac{\alpha}{\alpha^\prime}U_A\ket{\psi} + \sqrt{1-\left(\frac{\alpha}{\alpha^\prime}\right)^2}\ket{1}U_A\ket{\psi},
    \end{align}
    where $\ket{\psi}$ is an arbitrary state on S.
    Since the upper-left block of $\frac{\alpha}{\alpha^\prime}U_A$ is $\frac{1}{\alpha^\prime}\tilde{A}$, with $\tilde{A}$ being a matrix such that $\|\tilde{A}-A\|\le\epsilon$, the entire operation is a $(\alpha^\prime,a+1,\epsilon)$-block-encoding of $A$.
\end{proof}

\begin{lemma}
    If we have access to a $(\alpha,a,\epsilon)$-block-encoding $U_A$ of a matrix $A$, where $\alpha \ge \|A\|$, $a\in\N_0$, and $\epsilon\ge0$, we also have access to a $(\alpha,a^\prime,\epsilon)$-block-encoding of $A$ for any integer $a^\prime>a$.
    \label{lem:AddBEAnc}
\end{lemma}

\begin{proof}
    Adding $a^\prime-a$ qubits to the system on which $U_A$ acts and leaving them unused, we can regard $U_A$ as a $(\alpha,a^\prime,\epsilon)$-block-encoding of $A$.
\end{proof}

Third, given block-encodings of matrices, we can construct the block-encoding of their direct sum. 

\begin{lemma}
    Let $n\in\N$, $A_1,\ldots,A_n$ be complex matrices, which may have different sizes with row and column numbers at most $2^l$.
    Let $U_1,\ldots,U_n$ be $(\alpha,a,\epsilon)$-block-encodings of $A_1,\ldots,A_n$, respectively, where $\alpha \ge \max_{i\in[n]} \|A_i\|$, $a\in\N_0$, and $\epsilon\ge0$.
    Then, we can implement a $(\alpha, O(l+a+\log n), \epsilon)$-block-encoding of $A=A_1 \oplus \ldots \oplus A_n = \begin{pmatrix} A_1 && \\ &\ddots& \\ &&A_n \end{pmatrix}$, using the controlled versions of $U_1,\ldots,U_n$ once each, and $O(n(l+a+\log n))$ elementary gates.
    \label{lem:BEDirectSum}
\end{lemma}

\begin{proof}
    Suppose that $U_1,\ldots,U_n$ act on $d_1,\ldots,d_n$ qubits, and let $d=\max_{i\in[n]}d_i \le l+a$.
    We consider the system $\mathrm{S}$ consisting of a $\lceil \lg n \rceil$-qubit register $\mathrm{R}_\mathrm{c}$ and a $d$-qubit register $\mathrm{R}_\mathrm{t}$.
    For $i\in[n]$, we denote by $\tilde{U}_i$ the unitary on $\mathrm{R}_\mathrm{t}$ that corresponds to operating $U_i$ on the last $d_i$ qubits in $\mathrm{R}_\mathrm{t}$, that is,
    \begin{align}
        \tilde{U}_i \coloneqq I_{2^{d-d_i}} \otimes U_i =
        \begin{tikzpicture}[baseline=(m.center)]
        \node (m) {
        $
        \begin{pmatrix}
            U_i & & \\ & \ddots & \\ & & U_i~~
        \end{pmatrix}.
        $
        };
        \draw[decorate,decoration={brace,amplitude=5pt}]
          (-0.7,0.75) -- (0.85,-0.5)
          node[midway,xshift=13pt, yshift=10pt] {$\text{$2^{d-d_i}$}$};
        \end{tikzpicture}
    \end{align}
    By sequentially operating
    \begin{align}
        \ket{i}\!\bra{i} \otimes \tilde{U}_i + \sum_{\substack{j\in\left[\lceil \lg n \rceil\right] \\ j \ne i}} \ket{j}\!\bra{j} \otimes I_{2^d}, ~ i=1,\ldots,n
    \end{align}
    on $\mathrm{S}$, we implement the unitary $U\coloneqq\tilde{U}_1 \oplus \ldots \oplus \tilde{U}_n \oplus I_{(2^{\lceil \lg n \rceil}-n)2^d}$ on $\mathrm{S}$.
    We note that $U$ has the following form
    \begin{align}
        U=
        \begin{pmatrix}
            \frac{1}{\alpha}
            \begin{pmatrix}
                A_1 & * \\
                * & *
            \end{pmatrix}
            &&& \\
            &\ddots&&\\
            &&
            \frac{1}{\alpha}
            \begin{pmatrix}
                A_n & * \\
                * & *
            \end{pmatrix}
            & \\
            &&& \multirow{3}{*}{$I_{(2^{\lceil \lg n \rceil}-n)2^d}$} \\
            &&& \\
            &&&
        \end{pmatrix}.
    \end{align}
    This can be transformed into
    \begin{align}
        \left(
        \begin{array}{
            >{\centering\arraybackslash$}p{2.5cm}<{$}
            >{\centering\arraybackslash$}p{2.5cm}<{$}
            }
            \frac{1}{\alpha}
            \begin{pmatrix}
            A_1 & & \\
            & \ddots & \\
            & & A_n
            \end{pmatrix}
            & * \\
            \multirow{4}{*}{*}& \multirow{4}{*}{*} \\
             & \\
             & \\
             & \\
        \end{array}
        \right),
    \end{align}
    which is nothing but the block-encoding of $A$, by interchanging rows and columns with $O(n)$ modular additions.
    Each modular addition costs $O(\log n + d)$ elementary gates, so interchanging rows and columns costs $O(n(\log n + d))=O(n(l+a+\log n))$ elementary gates in total.
    Combining the above things, we get the claim.
\end{proof}

\subsubsection{Quantum linear system solver}

There are many quantum algorithms for solving linear systems, from the pioneering Harrow-Hassidim-Lloyd algorithm \cite{HHL} to recent block-encoding-based ones with improved complexities.
In this paper, we invoke the one presented in Ref.~\cite{Martyn2021}.

\begin{theorem}[Theorem 2 in Ref.~\cite{miyamoto2026quantum}, a modification of Algorithm 5 in \cite{Martyn2021}]
    Let $\epsilon>0$.
    Let $P$ be an invertible matrix with singular values contained in $[\alpha_P/\kappa_P,\alpha_P]$ with $\alpha_P>0,\kappa_P \ge 1$.
    Suppose that we have access to an $(\alpha_P,a_P,0)$-block-encoding $V_P$ of $P$ and its inverse.
    Then, we have access to an $(1,a_P,\epsilon)$-block-encoding $V_{P^{-1}}$ of $\frac{\alpha_P}{2\kappa_P}P^{-1}$, which uses $V_P^\dagger$
    \begin{align}
        O\left(\kappa_P\log\left(\frac{1}{\epsilon}\right)\right)
        \label{eq:BECompMatInv}
    \end{align}
    times and
    \begin{align}
        O\left(a_P \kappa_P\log\left(\frac{1}{\epsilon}\right)\right)
        \label{eq:GateCompMatInv}
    \end{align}
    additional elementary gates.
    \label{eq:MatInvQSVT}
\end{theorem}

\subsubsection{Overlap estimation}

Most of the quantum solvers for numerical problems such as linear equation systems and differential equations output quantum states that encode solutions in amplitudes, and reading out information of interest as a number is a separate issue \cite{aaronson2015read}.
In this paper, as Ref.~\cite{miyamoto2026quantum}, we use the method proposed in Ref.~\cite{Knill2007} for estimating the overlap between two quantum states.

\begin{theorem}
    Let $\epsilon,\delta\in(0,1)$.
    Suppose that we have access to the oracles $U_\psi$ and $U_\phi$ that generate quantum states $\ket{\psi}$ and $\ket{\phi}$, respectively: $U_\psi\ket{0}=\ket{\psi}$ and $U_\phi\ket{0}=\ket{\phi}$.
    Then, there exists a quantum algorithm that outputs $\hat{a}\in(0,1)$ such that $|\hat{a}-\braket{\psi|\phi}|\le\epsilon$ with probability at least $1-\delta$, querying $U_\psi$ and $U_\phi$
    \begin{equation}
        O\left(\frac{1}{\epsilon}\log\left(\frac{1}{\delta}\right)\right)
    \end{equation}
    times.
    \label{th:OE}
\end{theorem}

\section{Quantum algorithm for nonlinear \acp{SDE} \label{sec:main}}

\subsection{Assumptions}

To construct the \ac{CL}-based quantum algorithm for the nonlinear \ac{SDE}~\eqref{eq:SDE}, we start by presenting some assumptions.
First, we assume the existence of the solution of Eq.~\eqref{eq:SDE} and the boundedness of its expected squared norm.
\begin{assum}
    There exists a unique adapted $\R^d$-valued stochastic process $\vecX$ that solves Eq.~\eqref{eq:SDEGen}.
    There exists $\sigma_\vecX>0$ such that
    \begin{align}
        \E[\|\vecX_t\|^2] \le \sigma_\vecX^2
    \end{align}
    holds for any $t\in[0,T]$.
    \label{ass:XVar}
\end{assum}

Next, we assume that the norms of $A$ and its derivative are bounded and that we have access to its time-dependent block-encoding. 
\begin{assum}
    There exists $\alpha_A,\alpha_{\drm A/\drm t}>0$ such that
    \begin{align}
        \max_{t\in[0,T]}\|A(t)\| \le \alpha_A
        \label{eq:alphaA}
    \end{align}
    and
    \begin{align}
        \max_{t\in[0,T]}\left\|\frac{\drm A}{\drm t}(t)\right\| \le \alpha_{\drm A/\drm t}
        \label{eq:alphadAdt}
    \end{align}
    hold.
    Besides, we have access to a $(\alpha_A,a_A,0)$-time-dependent-block-encoding $V_A$ of $A$, where $a_A\in\N$.
    \label{ass:alphaA}
\end{assum}

On $A$, we also assume that $A$'s logarithmic norm is negative.
\begin{assum}
    \begin{align}
        \max_{t\in[0,T]} \mu(A(t)) \le -\eta
    \end{align}
    holds for some $\eta>0$.
    \label{ass:muA}
\end{assum}

According to Eq.~\eqref{eq:PhiNormBnd}, this assumption implies that
\begin{align}
    \|\Phi_A(t,s)\| \le e^{-\eta (t-s)} \le 1
    \label{eq:e^AtNorm<1}
\end{align}
holds for any $0\le s \le t \le T$.

We also assume the boundedness of the norm of $F$ and access to its block-encoding.

\begin{assum}
    There exists $\alpha_F,\alpha_{\drm F/\drm t}>0$ such that
    \begin{align}
        \max_{t\in[0,T]}\|F(t)\| \le \alpha_F
        \label{eq:alphaF}
    \end{align}
    and
    \begin{align}
        \max_{t\in[0,T]}\left\|\frac{\drm F}{\drm t}(t)\right\| \le \alpha_{\drm F/\drm t}
        \label{eq:alphadFdt}
    \end{align}
    hold.
    Besides, we have access to a $(\alpha_F,a_F,0)$-time-dependent-block-encoding $V_F$ of $F(t)$, where $a_F\in\N$.
    \label{ass:alphaF}
\end{assum}

On $B$, we make the following assumption.

\begin{assum}
    There exists $\sigma,\alpha_{\drm B/\drm t}>0$ such that
    \begin{align}
        \max_{t\in[0,T]}\|B(t)\|_{S_1} \le \sigma
        \label{eq:BNorm}
    \end{align}
    and
    \begin{align}
        \max_{t\in[0,T]}\left\|\frac{\drm B}{\drm t}(t)B^T(t)\right\|_{\rm F} \le \alpha_{\drm B/\drm t}
        \label{eq:alphadBdt}
    \end{align}
    hold.
    Besides, we have access to an oracle $V_{BB^T}$ that acts as
    \begin{align}
        V_{BB^T}\ket{t} \ket{0} \ket{0}\ket{0} = \ket{t}\ket{0} \ket{B(t)B^T(t)} + \ket{0_\perp}
    \end{align}
    for any $t\in[0,T]$, where the unnormalized state $\ket{B(t)B^T(t)}$ is given as
    \begin{align}
        \ket{B(t)B^T(t)}=\frac{1}{\sigma^2}\sum_{i,j=1}^N (B(t)B^T(t))_{i,j}\ket{i}\ket{j},
        \label{eq:ketBBT}
    \end{align}
    and $\ket{0_\perp}$ is an unnormalized state such that $(I \otimes \ket{0}\!\bra{0}\otimes I\otimes I)\ket{0_\perp}=0$~\footnote{Hereafter, we repeatedly use the symbol $\ket{0_\perp}$ to denote some unnormalized state orthogonal to the states in the terms before it. Although the represented states may differ between different equations, we adopt this notation because the explicit form of the state is irrelevant, and we expect no confusion.}.
    \label{ass:BNorm}
\end{assum}

Note that because of Eq.~\eqref{eq:SchNormOrd}, Eq.~\eqref{eq:BNorm} leads to
\begin{align}
    \max_{t\in[0,T]}\|B(t)\| \le \sigma
\end{align}
and
\begin{align}
    \max_{t\in[0,T]}\|B(t)\|_{\rm F} \le \sigma.
\end{align}

Finally, we assume that we can generate a quantum state that encodes the initial value of $\vecX_t$.

\begin{assum}
    We have access to an oracle $V_{\vecX_{\rm ini}}$ that acts as
    \begin{align}
        V_{\vecX_{\rm ini}} \ket{0} = \ket{\vecX_{\rm ini}} \coloneqq \frac{1}{\|\vecX_{\rm ini}\|}\sum_{i=1}^N X_{\rm ini}^i\ket{i}.
    \end{align}
\label{ass:VXini}
\end{assum}

\subsection{Carleman linearization}

\subsubsection{Norm of the solution}

Now, let us consider \ac{CL} for the \ac{SDE} \eqref{eq:SDE}.
As a preparation, we bound the norm of $\vecX_t$.
This becomes an issue because the nonlinear term $F(t) \vecX_t^{\otimes s}$ is required to be smaller than the linear term $A(t) \vecX_t$ in some sense in order to guarantee the convergence of \ac{CL}, while $F(t) \vecX_t^{\otimes s}$ tends to be larger for larger $\|\vecX_t\|$.
More strictly, we now derive a bound on the tail probability of $\|\vecX_t\|$, that is, the probability that it exceeds a given value.

We start by presenting the following lemma, whose proof is deferred to Sec.~\ref{sec:ProofItoIntegTailPr}. 

\begin{lemma}
    Under Assumptions \ref{ass:alphaA} and \ref{ass:BNorm}, for any $\delta\in(0,1)$,
    \begin{align}
        \Prob\left\{\sup_{t\in[0,T]}\left\|\int_0^t \Phi_A(t,u) B(u) \drm\vecW_u\right\| \le \frac{U_{\delta}}{2}\right\} \ge 1-\delta
        \label{eq:OUsupTail}
    \end{align}
    holds, where
    \begin{align}
        U_{\delta} \coloneqq 2\left(\sqrt{3 \log\left(\frac{4\NumIntItoIntegUB }{\delta}\right)}+2\right)\sqrt{\sigma^2T}
        \label{eq:UForItoIntegUB}
    \end{align}
    and $\NumIntItoIntegUB  \coloneqq \max\{ \left\lceil 4\alpha_{A}^2T^2 \right\rceil,2000N\}$.
    \label{lem:ItoIntegTailPr}
\end{lemma}

Then, we obtain the following lemma on the tail probability of $\|\vecX_t\|$.

\begin{lemma}
Suppose that Assumptions \ref{ass:alphaA}, \ref{ass:muA}, \ref{ass:alphaF}, and \ref{ass:BNorm} hold.
Let $\delta\in(0,1)$.
Suppose that
\begin{align}
    \|\vecX_{\rm ini}\| \le \frac{U_{\delta}}{2}
    \label{eq:XiniNormBd}
\end{align}
and
\begin{align}
    2\alpha_F U_{\delta}^{s-1} \le \eta e^{-\eta T}
    \label{eq:SolNormDecCond}
\end{align}
holds.
Then,
\begin{align}
    \Prob\left\{\|\vecX_t\| \le U_{\delta},\forall t\in[0,T]\right\} \ge 1-\delta
    \label{eq:|X|pTail}
\end{align}
holds.
\label{lem:XtTail}
\end{lemma}

\begin{proof}
    From Eq.~\eqref{eq:SDESol}, we have
    \begin{align}
        \left\|\vecX_t\right\| & \le \|\Phi_A(t,0)\| \|\vecX_{\rm ini}\| + \int_0^t \|\Phi_A(t,u)\| \|F(u)\| \|\vecX_u\|^s \drm u + \left\|\int_0^t \Phi_A(t,u) B(u) \drm \vecW_u\right\| \nonumber \\
        & \le e^{-\eta t}\frac{U_{\delta}}{2} + \alpha_F\int_0^t \|\vecX_u\|^s \drm u + \left\|\int_0^t \Phi_A(t,u)B(u) \drm \vecW_u\right\|,
    \end{align}
    where we use Eqs.~\eqref{eq:e^AtNorm<1} and \eqref{eq:XiniNormBd}.
    Lemma \ref{lem:ItoIntegTailPr} implies that, with probability at least $1-\delta$ under $\Prob$,
    \begin{align}
        \left\|\vecX_t\right\| \le \frac{U_{\delta}}{2} + e^{-\eta t}\frac{U_{\delta}}{2} + \alpha_F\int_0^t \|\vecX_u\|^s \drm u
    \end{align}
    holds for any $t\in[0,T]$.
    When this holds, $\left\|\vecX_t\right\| \le y(t)$ holds with $y:[0,T]\rightarrow\R$, the solution of the integral equation
    \begin{align}
        y(t) = \frac{U_{\delta}}{2} + e^{-\eta t}\frac{U_{\delta}}{2} + \alpha_F\int_0^t (y(u))^s \drm u,
    \end{align}
    according to \cite[Theorem 2]{HEADLEY1974250}.
    $y$ also solves the differential equation
    \begin{align}
        y^\prime(t) = -\frac{1}{2}\eta U_{\delta}e^{-\eta t} +\alpha_F (y(t))^s, y(0)=U_{\delta},
    \end{align}
    which implies that $y(t) \le U_{\delta}$ holds for any $t\in[0,T]$ if Eq.~\eqref{eq:SolNormDecCond} holds.
    Summarizing these considerations, we see that under Eq. \eqref{eq:SolNormDecCond}, $\left\|\vecX_t\right\| \le U_{\delta}$ holds at least probability $1-\delta$ under $\Prob$, which is the claim.
\end{proof}

\subsubsection{Rescaling of the \ac{SDE}}

As another preparation for discussing \ac{CL}, we introduce rescaling of the solution of the \ac{SDE}.
That is, we define the rescaled random process
\begin{align}
    \hat{\vecX}_t \coloneqq \gamma\vecX_t
\end{align}
with $\gamma>0$.
Under this rescaling, the \ac{SDE} \eqref{eq:SDE} transforms into
\begin{align}
    \drm \hat{\vecX}_t = \left(A(t) \hat{\vecX}_t + \hat{F}(t)\hat{\vecX}_t^{\otimes s}\right) \drm t + \hat{B}(t) \drm \vecW_t,
    \label{eq:SDERes}
\end{align}
where
\begin{align}
    \hat{F} \coloneqq \gamma^{1-s}F, \hat{B} \coloneqq \gamma B.
    \label{eq:hatFhatB}
\end{align}

\subsubsection{Carleman linearization error}

Now, let us consider \ac{CL} of the \ac{SDE} \eqref{eq:SDE}.
Choosing an integer $\NC \ge s$, we define $\R^{N^j}$-valued random processes $\vecY^j=(Y^{j,1},\ldots,Y^{j,N^j})^T$ for $j\in[\NC]$ via the \ac{SDE}
\begin{align}
    &\drm \vecY^j_t = \nonumber \\
    & ~ 
    \left\{
    \begin{array}{>{\displaystyle}l >{\displaystyle}r}
        \calA^{1,1}(t) \vecY^1_t \drm t + \calA^{1,s}(t) \vecY^{s}_t \drm t + \hat{B}(t) \drm \vecW_t & ; ~ j=1 \\
        \sum_{l=1}^m\calA^{2,1,l}(t) \vecY^1_t \drm W^l_t +\calA^{2,2}(t) \vecY^2_t \drm t + \calA^{2,s+1}(t) \vecY^{s+1}_t \drm t + \hat{B}(t) \hat{B}^T(t) \drm t & ; ~ j=2 \\
        \calA^{j,j-2}(t) \vecY^{j-2}_t \drm t + \sum_{l=1}^m \calA^{j,j-1,l}(t) \vecY^{j-1}_t \drm W^l_t + \calA^{j,j}(t) \vecY^j_t \drm t + \calA^{j,j+s-1}(t) \vecY^{j+s-1}_t \drm t \\
        \multicolumn{2}{r}{; ~ j = 3,\ldots,\NC-s+1} \\
        \calA^{j,j-2}(t) \vecY^{j-2}_t \drm t + \calA^{j,j-1,l}(t) \vecY^{j-1}_t \drm W^l_t + \calA^{j,j}(t) \vecY^j_t \drm t  \\
        \multicolumn{2}{r}{; ~ j = \NC-s+2,\ldots,\NC}
    \end{array}
    \right.
    \label{eq:Y}
\end{align}
with the initial value $\vecY^j_0 = \hat{\vecX}_{\rm ini}^{\otimes j}$.
Here, $\calA^{j,j-2}:[0,T] \rightarrow \R^{N^j \times N^{j-2}}$ is given by $\calA^{j,j-2}(t)=\sum_{k=1}^{j-1} \sum_{l=k+1}^j \calA^{j,j-2,k,l}(t)$, where the entry of $\calA^{j,j-2,k,l}(t)$ indexed by $i_1,\ldots,i_j,i^\prime_1,\ldots,i^\prime_{j-2}\in[N]$ is
\begin{align}
    \calA^{j,j-2,k,l}_{i_1,\ldots,i_j,i^\prime_1,\ldots,i^\prime_{j-2}}(t) \coloneqq \delta_{i_1,i^\prime_1} \cdots \delta_{i_{k-1},i^\prime_{k-1}} \delta_{i_{k+1},i^\prime_k} \cdots \delta_{i_{l-1},i^\prime_{l-2}} \delta_{i_{l+1},i^\prime_{l-1}} \cdot \delta_{i_j,i^\prime_{j-2}} (\hat{B}(t)\hat{B}^T(t))_{i_k,i_l}.
    \label{eq:calAjj-2}
\end{align}
$\calA^{j,j-1,l}:[0,T]\rightarrow\R^{N^j \times N^{j-1}}$
is given by
\begin{align}
    \calA^{j,j-1,l}(t) \coloneqq \sum_{k=0}^{j-1} I^{\otimes k} \otimes \hat{B}_{:,l}(t) \otimes I^{\otimes (j-k-1)},
\end{align}
$\calA^{j,j}:[0,T]\rightarrow \R^{N^j \times N^j}$ is given by
\begin{align}
    \calA^{j,j}(t) \coloneqq \sum_{k=1}^j I^{\otimes (k-1)} \otimes A(t) \otimes I^{\otimes (j-k)},
    \label{eq:calAjj}
\end{align}
and $\calA^{j,j+s-1}:[0,T]\rightarrow \R^{N^j \times N^{j+s-1}}$ is given by
\begin{align}
    \calA^{j,j+s-1}(t) \coloneqq \sum_{k=1}^j I^{\otimes (k-1)} \otimes \hat{F}(t) \otimes I^{\otimes (j-k)}.
\end{align}
Collecting $\vecY^1,\ldots,\vecY^{\NC}$ in one $\R^{\calN}$-valued random process
\begin{align}
    \veccalY \coloneqq
    \begin{pmatrix}
    \vecY^1 \\
    \vdots \\
    \vecY^{\NC}
    \end{pmatrix},
\end{align}
where $\calN \coloneqq \sum_{j=1}^{\NC} N^j=\frac{N^{\NC+1}-1}{N-1}$, we have
\begin{align}
    \drm \veccalY_t = (\calC(t) \veccalY_t + \vecc(t)) \drm t + \sum_{l=1}^m (\calB^l(t) \veccalY_t + \vecb^l(t))\drm W^l_t,
    \label{eq:SDEcalY}
\end{align}
with the initial value
\begin{align}
    \veccalY_0 =
    \begin{pmatrix}
        \hat{\vecX}_{\rm ini}^{\otimes 1} \\
        \vdots \\
        \hat{\vecX}_{\rm ini}^{\otimes \NC}
    \end{pmatrix}
    .
    \label{eq:calY0}
\end{align}
Here, $\vecc:[0,T]\rightarrow\R^\calN$ is given by
\begin{align}
    \vecc(t) =
    \begin{pmatrix}
    c_1 \\ \vdots \\ c_\calN
    \end{pmatrix}
    \coloneqq
    \begin{pmatrix}
        \mathbf{0}_N \\
        {\rm vec}(\hat{B}(t)\hat{B}^T(t)) \\
        \mathbf{0}_{\calN - N - N^2}
    \end{pmatrix},
    \label{eq:vecc}
\end{align}
$\calC:[0,T]\rightarrow\R^{\calN \times \calN}$ is given by
\begin{align}
    \calC(t) \coloneqq
\left(
\begin{array}{ccccccc}
    \calA^{1,1}(t) & & & \calA^{1,s}(t) & & & \\
    & \calA^{2,2}(t) & & & \calA^{2,s+1}(t) & & \\
    \calA^{3,1}(t) & & \calA^{3,3}(t) & & & \ddots & \\
    & \ddots & & \ddots & & &  \calA^{\NC-s+1,\NC}(t)\\
    & & \ddots & & \ddots & &\\
    & & & \ddots & & \ddots &\\
    & & & & \calA^{\NC,\NC-2}(t) & & \calA^{\NC,\NC}(t)
\end{array}
\right),
\end{align}
$\vecb^l:[0,T]\rightarrow \R^{\calN}$ is given by
\begin{align}
    \vecb^l(t) \coloneqq
    \begin{pmatrix}
        \hat{B}_{:,l}(t) \\
        \mathbf{0}
    \end{pmatrix},
\end{align}
and $\calB^l:[0,T]\rightarrow\R^{\calN \times \calN}$ is given by
{\renewcommand{\arraystretch}{1.3}
\begin{align}
    \calB^l(t) \coloneqq
    \begin{pmatrix}
        O_{N,N} & & & \\
        \calA^{2,1,l}(t) & O_{N^2,N^2} & & \\
        & \ddots & \ddots & \\
        & & \calA^{\NC,\NC-1,l}(t) & O_{N^{\NC},N^{\NC}}
    \end{pmatrix}
    .
\end{align}
}
We note that $\hat{\vecX}_t$ satisfies
\begin{align}
    & \drm \hat{\vecX}_t^{\otimes j} = \nonumber \\ 
    ~&
    \begin{dcases}
        \calA^{1,1}(t) \hat{\vecX}_t \drm t + \calA^{1,s}(t) \hat{\vecX}^{\otimes s}_t \drm t + \hat{B}(t) \drm \vecW_t & ; ~ j=1 \\
        \sum_{l=1}^m \calA^{2,1,l}(t) \hat{\vecX}_t \drm W^l_t +\calA^{2,2}(t) \hat{\vecX}^{\otimes 2}_t \drm t + \calA^{2,s+1}(t) \hat{\vecX}^{\otimes(s+1)}_t \drm t + \hat{B}(t) \hat{B}^T(t) \drm t & ; ~ j=2 \\
        \calA^{j,j-2}(t) \hat{\vecX}^{\otimes(j-2)}_t \drm t + \sum_{l=1}^m \calA^{j,j-1,l}(t) \hat{\vecX}^{\otimes(j-1)}_t \drm W^l_t + \calA^{j,j}(t) \hat{\vecX}^{\otimes j}_t \drm t + \calA^{j,j+s-1}(t) \hat{\vecX}^{\otimes(j+s-1)}_t \drm t & ; ~ j \ge 3
    \end{dcases}
    \label{eq:X^j}
\end{align}
The \acp{SDE} of $\hat{\vecX}_t^{\otimes j}$ and $\vecY_t^j$ coincide up to the terms beyond $\NC$-th order in $\hat{\vecX}_t$.
We thus expect that $\vecY_t^j$ approximates $\hat{\vecX}_t^{\otimes j}$, and in fact, the error $\veceta^j_t \coloneqq \vecY_t^j - \hat{\vecX}_t^{\otimes j}$ is bounded with high probability as stated in the following lemma, whose proof is deferred to Sec.~\ref{sec:PrCarErrTail}.

\begin{lemma}
    Suppose that Assumptions \ref{ass:XVar} to \ref{ass:BNorm} hold.
    Let $\epsilon\in(0,\sqrt{\sigma^2T})$ and $c\in(0,1)$.
    Define 
    \begin{gather}
        \hat{\epsilon} \coloneqq \frac{c\epsilon}{24\sqrt{\sigma^2T}}\left(\log\left(\frac{12\sqrt{\sigma^2T}}{c\epsilon}\left(20\NumIntItoIntegUB  \max\left\{\frac{45c^2 \sigma_X^2}{64\sigma^2 T},90\right\}\right)^{\frac{1}{2}}\right)\right)^{-1}, \\
        \SampCarErrProof \coloneqq \left\lceil\max\left\{\frac{c^2 \sigma_X^2}{64\sigma^2 T},2\right\}\times\frac{45}{\hat{\epsilon}^2}\right\rceil, \\
        U \coloneqq 2\left(\sqrt{3 \log\left(20\NumIntItoIntegUB\SampCarErrProof\right)}+2\right)\sqrt{\sigma^2T}, \label{eq:U}\\
        \tilde{\epsilon} \coloneqq \hat{\epsilon}\min\left\{\frac{\log \left(\frac{1}{c}\right)}{9\sqrt{5(s-1)\SampCarErrProof}\eta T},1\right\}.
    \end{gather}
    Set 
    \begin{align}
        \NC = \NC_{\epsilon,c} \coloneqq \left\lceil\frac{1}{\log \left(\frac{1}{c}\right)}\left(\log\left(\frac{1}{\tilde{\epsilon}}\right)+\sqrt{2\left(\log\left(\frac{1}{\tilde{\epsilon}}\right)-1\right)}\right)\right\rceil.
        \label{eq:NCCond}
    \end{align}
    and
    \begin{align}
        \gamma = \gamma_{\epsilon,c} \coloneqq \frac{c}{U}.
        \label{eq:gamma}
    \end{align}
    Suppose that
    \begin{align}
        \left(\NC+\frac{1}{2}+\frac{1}{2\NC}\right)\gamma^2\sigma^2+\gamma^{1-s}\alpha_F < \eta,
        \label{eq:CarlConvCond1}
    \end{align}
    \begin{align}
        2\alpha_F U^{s-1} \le \eta e^{-\eta T},
        \label{eq:CarlConvCond2}
    \end{align}
    and
    \begin{align}
        \|\vecX_{\rm ini}\| \le \frac{U}{2}
        \label{eq:XiniBd}
    \end{align}
    hold.
    Then,
    \begin{align}
        \left\|\frac{1}{\gamma}\E[\vecY^1_t]-\E[\vecX_t]\right\| \le \epsilon
        \label{eq:etaNorm}
    \end{align}
    holds for any $t\in[0,T]$.
    \label{lem:CarErrTail}
\end{lemma}

\begin{remark}

    Now, let us comment on how the rescaling and the conditions that appear in Lemma~\ref{lem:CarErrTail} are related to those in Ref.~\cite{Liu2021}, which proposed the \ac{CL}-based quantum algorithm for \acp{ODE} in the form of $\frac{\drm}{\drm t}u=F_2 u^{\otimes 2} + F_1 u + F_0$.
    In Ref.~\cite{Liu2021}, the solution $u$ is rescaled by a factor $\gamma$ satisfying $\gamma < 1/\|u_{\rm in}\|$, where $u_{\rm in}$ is the initial value of $u$.
    Note that in the setting in Ref.~\cite{Liu2021}, $\|u(t)\|$ is non-increasing, with its maximum value being $\|u_{\rm in}\|$.
    On the other hand, in Lemma~\ref{lem:CarErrTail}, the factor $\gamma$ is set so that $\gamma<1/U$, where $U$ is a value such that under the conditions \eqref{eq:CarlConvCond2} and \eqref{eq:XiniBd}, $\|\vecX_t\|$ does not exceed $U$ with high probability (see the proof of Lemma~\ref{lem:CarErrTail} for the details).

    Adopting the above rescaling, as part of a sufficient condition for the convergence of \ac{CL}, the authors of Ref.~\cite{Liu2021} present the condition ${\rm R}<1$, which translates into\footnote{Strictly speaking, the inequality (2.3) in the full version of Ref.~\cite{Liu2021} has a form different from Eq.~\eqref{eq:LiuetalCond} in the following points. First, there are notational differences: in Ref.~\cite{Liu2021}, both the original $F_2$ (resp. $F_0$) and the rescaled one are denoted by $F_2$ (resp. $F_0$) without a hat, and the rescaling factor $\gamma$ is omitted in Eq.~(2.3). Second, in Eq.~(2.3), $|{\rm Re}(\lambda_1)|$, where $\lambda_1$ is the eigenvalue of $A$ (denoted by $F_1$ in Ref.~\cite{Liu2021}) with the largest real part, replaces $\eta$ in our Eq.~\eqref{eq:LiuetalCond}, although $|{\rm Re}(\lambda_1)|=\eta$ holds in some cases, such as when $A$ is normal.}
    \begin{align}
        \gamma\|F_0\| + \gamma^{-1}\|F_2\| < \eta.
        \label{eq:LiuetalCond}
    \end{align}
    We may find the similarity between Eqs.~\eqref{eq:CarlConvCond1} and \eqref{eq:LiuetalCond}.
    Both of them require the norm of the nonlinear term to be sufficiently small.
    Additionally, Eq.~\eqref{eq:LiuetalCond} requires the norm of the inhomogeneous term to be small as well, whereas Eq.~\eqref{eq:CarlConvCond1} requires the norm of the noise coefficient matrix $B$ to be small.
    We can understand these conditions as follows.
    If the \ac{ODE} in Ref.~\cite{Liu2021} does not have the quadratic and inhomogeneous terms but only the linear term, the evolution equations for the tensor products $u^{\otimes j}$ decouple, namely, $u^{\otimes 1}, u^{\otimes 2}, \ldots$ evolve in time without affecting each other.
    In this situation, the finite-order truncation in \ac{CL} does not introduce any error.
    On the other hand, $F_2$ and $F_0$ make the time-evolution of $u^{\otimes 1}, u^{\otimes 2}, \ldots$ couples, because of which the truncation in \ac{CL} yields an error. 
    A similar discussion also applies in \ac{CL} of the \ac{SDE}.
    As noted from Eq.~\eqref{eq:X^j}, $\calA^{j,j+s-1}$, which originates from the nonlinear term, and $\calA^{j,j-2}$, which the noise term yields via applying It\^{o}'s formula to $\vecX_t^{\otimes j}$, introduce the coupling among $\vecX_t^{\otimes j}$ in their time evolution, which makes the truncation in \ac{CL} erroneous.
    Conversely, the truncation error can be made small by suppressing the terms that lead to the coupling, which translates into the conditions \eqref{eq:CarlConvCond1} or \eqref{eq:LiuetalCond}.
    
\end{remark}

\begin{remark}
    Lemma~\ref{lem:CarErrTail} shows that in a specific setting where $\gamma$ and $J$ are taken as above and Eqs.~\eqref{eq:CarlConvCond1}, \eqref{eq:CarlConvCond2}, and \eqref{eq:XiniBd} hold, the convergence of \ac{CL} is guaranteed in the sense that Eq.~\eqref{eq:etaNorm} holds.
    Note that this is a sufficient condition and that being outside this setting does not necessarily mean that \ac{CL} does not work.
    In fact, Ref.~\cite{Liu2021} observed that even if their sufficient condition ${\rm R}<1$ is broken, \ac{CL} for the Burgers equation works, at least in their numerical demonstration.
    Keeping this fact in mind, in our numerical experiment conducted in Sec.~\ref{sec:exp}, we set $\gamma$ to a fixed value and change $J$ to see that the error between $\frac{1}{\gamma}\E[\vecY^1_t]$ and $\E[\vecX_t]$ actually converges to 0 for larger $J$, even though our sufficient condition does not hold.
\end{remark}

\subsection{Estimating the expectation}

Having the \ac{SDE} \eqref{eq:SDEcalY} that is linear with respect to $\veccalY_t$, we might expect that the quantum algorithm proposed in Ref.~\cite{miyamoto2026quantum} could be used to generate the quantum state encoding realizations of the \ac{SDE}'s solution and calculate expectations of polynomials of the solution.
However, the \ac{SDE} contains the multiplicative noise, which prevents the application of the method in Ref.~\cite{miyamoto2026quantum}, which assumes that the noise is additive.
Thus, in this paper, we aim at the more limited objective of calculating expectations of linear functions of $\vecX_t$, rather than polynomials.
In fact, even this task is nontrivial for nonlinear \acp{SDE}, since we do not have any closed-form expression for $\E[\vecX_t]$, unlike in the linear case, where $\E[\vecX_t]=\Phi_A(t,0)\vecX_0$.

With the approximation of $\E[\vecX_t]$ by $\frac{1}{\gamma}\E[\vecY^1_t]$ in mind, we consider $\vecm_{\veccalY}(t)\coloneqq\E[\veccalY_t]$, which embeds $\E[\vecY^1_t]$ as its first $N$ entries.
According to Eq. (4.8.9) in Ref.~\cite{KloedenPlaten}, $\vecm_{\veccalY}(t)$ follows the \ac{ODE} 
\begin{align}
    \frac{\drm}{\drm t}\vecm_{\veccalY}(t) = \calC(t) \vecm_{\veccalY}(t) + \vecc(t), \vecm_{\veccalY}(0)=\veccalY_0,
    \label{eq:meanODE}
\end{align}
whose solution can be written as
\begin{align}
    \vecm_{\veccalY}(t) = \Phi_{\calC}(t,0)\veccalY_0 + \int_0^t \Phi_{\calC}(t,u)\vecc(u) \drm u.
    \label{eq:meanSol}
\end{align}
Setting $r+1$ time grid points $t_n \coloneqq n\Delta t,n\in[r]_0$ with $\Delta t \coloneqq \frac{T}{r}$, $\vecm_{\veccalY}$ at $t_0,\ldots,t_r$ is obtained via solving the following linear equation:
\begin{align}
    \begin{pmatrix}
    I & &  \\
    -\Phi_0 & I & & \\
    & \ddots & \ddots & \\
    & & -\Phi_{r-1} & I
    \end{pmatrix}
    \begin{pmatrix}
        \vecm_{\veccalY}(t_0) \\ \vecm_{\veccalY}(t_1) \\ \vdots \\ \vecm_{\veccalY}(t_r)
    \end{pmatrix}
    =
    \begin{pmatrix}
        \veccalY_0 \\ \vecDelta_0 \\ \vdots \\ \vecDelta_{r-1}
    \end{pmatrix},    
\end{align}
where $\Phi_n \coloneqq \Phi_\calC(t_{n+1},t_n)$, and
\begin{align}
    \vecDelta_n \coloneqq \int_{t_n}^{t_{n+1}} \Phi_{\calC}(t_{n+1},u)\vecc(u) \drm u.
\end{align}
Then, we can generate the quantum state that approximately encodes $\vecm_{\veccalY}(t_0),\ldots,\vecm_{\veccalY}(t_r)$ in a way similar to Ref.~\cite{Berry2024quantumalgorithm,An2026fastforwarding}, as mathematically stated in the following lemma and theorem.

\begin{lemma}
    Let $\epsilon\in(0,1]$.
    Suppose that 
    \begin{gather}
        \max_{t \in [0,T]} \|\calC(t)\| \le \alpha_\calC, \max_{t \in [0,T]} \left\|\frac{\drm \calC}{\drm t}(t)\right\| \le \alpha_{\drm \calC/\drm t},
        \max_{t\in[0,T]}\|\vecc(t)\| \le \alpha_\vecc, \max_{t\in[0,T]}\left\|\frac{\drm \vecc(t)}{\drm t}\right\| \le \alpha_{\drm \vecc/\drm t}
        \label{eq:NormBndsCc}
    \end{gather}
    hold for $\alpha_\calC,\alpha_{\drm \calC/\drm t},\alpha_\vecc,\alpha_{\drm \vecc/\drm t}>0$ and that
    \begin{align}
        \max_{t \in [0,T]} \mu(\calC(t)) \le 0
    \end{align}
    holds.
    Suppose that we have access to the following oracles:
    \begin{itemize}
        \item $V_{\calC}$: $(\alpha_\calC,a_\calC,0)$-time-dependent-block-encoding of $\calC$, where $a_\calC\in\N$.

        \item $V_{\vecc}$ that acts as
        \begin{align}
            V_{\vecc}\ket{t} \ket{0} \ket{0} = \ket{t} \ket{0} \ket{\vecc(t)} + \ket{0_\perp}
        \end{align}
        for any $t\in[0,T]$, where the unnormalized state $\ket{\vecc(t)}$ is given as
        \begin{align}
            \ket{\vecc(t)}=\frac{1}{\alpha_\vecc}\sum_{i=1}^\calN c_i(t)\ket{i},
        \end{align}
        and $\ket{0_\perp}$ is an unnormalized state such that $(I \otimes \ket{0}\!\bra{0}\otimes I)\ket{0_\perp}=0$.

        \item $V_{\veccalY_0}$ that acts as
        \begin{align}
            V_{\veccalY_0}\ket{0} = \ket{\veccalY_0} \coloneqq \frac{1}{\|\veccalY_0\|} \sum_{i=1}^\calN \calY_0^i \ket{i}.
        \end{align}
        
    \end{itemize}
    Let $r$, $K$, and $M$ be integers satisfying
    \begin{align}
        r\ge \max\left\{\alpha_\calC T,7\right\}, K\ge\max\left\{\log\left(\frac{3}{\varepsilon}\right),7\right\}, M\ge \max\left\{\frac{4\alpha_{\drm \calC/\drm t}}{\alpha_\calC^2 \varepsilon}, \left(\frac{3\alpha_\calC \alpha_{\drm \calC/\drm t} T^3}{r^3}+\frac{2\alpha_{\drm \vecc/\drm t}T^2}{r^2}\right)\frac{1}{\varepsilon}\right\},
        \label{eq:KrM}
    \end{align}
    where
    \begin{gather}
        \varepsilon \coloneqq \sqrt{\frac{1}{(r+1)^3\left(4\left(\|\veccalY_0\|^2+1\right)+\frac{3}{5}\alpha_\vecc^2T^2\right)}}U_{\vecm_{\veccalY}}\epsilon, \nonumber \\
        U_{\vecm_{\veccalY}} \coloneqq \sqrt{\|\veccalY_0\|^2+rU_{\widetilde{\vecDelta}}^2},~
        U_{\widetilde{\vecDelta}} \coloneqq \frac{\alpha_\vecc}{\alpha_\calC}\sum_{k=1}^K \frac{1}{k!} \left(\frac{\alpha_\calC \Delta t}{M}\right)^k.
        \label{eq:vareps}
    \end{gather}
    Then, we can construct an oracle $V_{\vecm_{\veccalY},\epsilon}^{\rm hist}$ that acts as
    \begin{align}
        V_{\vecm_{\veccalY},\epsilon}^{\rm hist}\ket{0}\ket{0}\ket{0}=
        \ket{\Psi_{\vecm_{\veccalY},\epsilon}^{\rm hist}} \coloneqq 
        \ket{0}\Ket{\tilde{\tilde{\calM}}_{\veccalY,{\rm hist}}} + \ket{0_\perp}.
        \label{eq:VHisY}
    \end{align}
    Here, $\ket{0_\perp}$ is an unnormalized state such that $(\ket{0}\!\bra{0}\otimes I\otimes I)\ket{0_\perp}=0$, and $\Ket{\tilde{\tilde{\calM}}_{\veccalY,{\rm hist}}}$ is an unnormalized state given by
    \begin{align}
        \Ket{\tilde{\tilde{\calM}}_{\veccalY,{\rm hist}}} \coloneqq \frac{1}{4rU_{\vecm_{\veccalY}}}\sum_{n=0}^r\sum_{i=1}^\calN\tilde{\tilde{\calM}}_{\veccalY,{\rm hist}}^{n,i}\ket{n}\ket{i}
    \end{align}
    with $\tilde{\tilde{\calM}}_{\veccalY,{\rm hist}}=\left(\tilde{\tilde{\calM}}_{\veccalY,{\rm hist}}^{n,i}\right)_{n\in[r]_0,i\in[\calN]}\in\mathbb{R}^{(r+1) \times \calN}$
    satisfying
    \begin{align}
        \left\|\tilde{\tilde{\calM}}_{\veccalY,{\rm hist}} - 
        \calM_{\veccalY,{\rm hist}}\right\|_{\rm F} \le U_{\vecm_{\veccalY}} \epsilon
        \label{eq:calMtiltilErr}
    \end{align}
    for
    \begin{align}
        \calM_{\veccalY,{\rm hist}}=\left(\calM_{\veccalY,{\rm hist}}^{n,i}\right)_{n\in[r]_0,i\in[\calN]} \coloneqq
        \begin{pmatrix}
            \left(\vecm_{\veccalY}(t_0)\right)^T \\ \vdots \\ \left(\vecm_{\veccalY}(t_r)\right)^T
        \end{pmatrix}.
    \end{align}
    $V_{\vecm_{\veccalY},\epsilon}^{\rm hist}$ uses the controlled $U_\calC$
    \begin{align}
        O\left(rK\log\left(\frac{r}{\epsilon}\right)\right)
        \label{eq:QueryVcalCStateGen}
    \end{align}
    times, the controlled $V_\vecc$ once, the controlled $V_{\veccalY_0}$ once, and
    \begin{align}
        O\left(rK \left(\log K\log M+a_\calC\right) \log\left(\frac{r}{\epsilon}\right)\right)
        \label{eq:QueryElGateStateGen}
    \end{align}
    additional elementary gates.
    \label{lem:HistStatemY}
\end{lemma}

\begin{proof}
    
    As seen in Ref.~\cite{miyamoto2026quantum} (originally Refs.~\cite{Berry2024quantumalgorithm,An2026fastforwarding}), we can construct a $(\alpha_{\widetilde{\Phi}}, a_{\widetilde{\Phi}}, 0)$-block-encoding $V_{\widetilde{\Phi}}$ of
    \begin{align}
    \widetilde{\Phi} \coloneqq
    \begin{pmatrix}
    I & &  \\
    -\tilde{\Phi}^{K,r,M}_0 & I & & \\
    & \ddots & \ddots & \\
    & & -\tilde{\Phi}^{K,r,M}_{r-1} & I
    \end{pmatrix},
    \end{align}
    where
    \begin{align}
    \alpha_{\widetilde{\Phi}}=1+e,~a_{\widetilde{\Phi}}=O(K(a_\calC+\log K)),
    \label{eq:alphaaPhiTil}
    \end{align}
    by making $O(K)$ uses of the controlled $U_\calC$ and $O(K \log K \log M)$ uses of additional elementary gates.
    Here, the truncated Dyson series $\tilde{\Phi}^{K,r,M}_n$ is given by
    \begin{align}
        \tilde{\Phi}^{K,r,M}_n \coloneqq \sum_{k=0}^K \frac{1}{k!}\left(\frac{\Delta t}{M}\right)^k \sum_{l_1,\ldots,l_k=0}^{M-1} \mathcal{T} \calC\left(\tau_{n,l_1}\right) \cdots \calC\left(\tau_{n,l_k}\right)
        \label{eq:PhiTilde}
    \end{align}
    for $n\in[r-1]_0$, where $\tau_{n,l} \coloneqq t_n+l\frac{t_{n+1}-t_n}{M}$ for $l\in[M-1]_0$, and $\mathcal{T}$ is the time-ordering operator, and it satisfies
    \begin{align}
        \left\|\tilde{\Phi}^{K,r,M}_n - \Phi_\calC(t_{n+1},t_n)\right\| \le \varepsilon
        \label{eq:PhitilPhiErr}
    \end{align}
    for $r$, $K$, and $M$ satisfying Eq.~\eqref{eq:KrM}, as seen in Lemma 3 in Ref.~\cite{miyamoto2026quantum}.
    Besides, according to Ref.~\cite{Berry2024quantumalgorithm}, we can construct the oracle $V_{\widetilde{\vecDelta}}$ that acts for any $n\in[r-1]_0$ as
    \begin{align}
        V_{\widetilde{\vecDelta}} \ket{0}\ket{n}\ket{0} = \ket{0}\ket{n}\ket{\tilde{\vecDelta}_n} + \ket{0_\perp},
    \end{align}
    by making $O(K)$ uses of $V_\calC$, one use of $V_\vecc$, and $O(K \log K \log M)$ uses of additional elementary gates. 
    Here, $\ket{0_\perp}$ is an unnormalized state such that $(\ket{0}\!\bra{0}\otimes I \otimes I)\ket{0_\perp}=0$, and $\ket{\tilde{\vecDelta}_n}$ is an unnormalized state given by
    \begin{align}
        \ket{\tilde{\vecDelta}_n} \coloneqq \frac{1}{U_{\widetilde{\vecDelta}}} \sum_{i=1}^\calN \tilde{\Delta}^{K,r,M}_{n,i} \ket{i}
    \end{align}
    with
    \begin{align}
        \tilde{\vecDelta}^{K,r,M}_n = \begin{pmatrix}\tilde{\Delta}^{K,r,M}_{n,1} \\ \vdots \\ \tilde{\Delta}^{K,r,M}_{n,\calN}\end{pmatrix} \coloneqq \sum_{k=1}^K \frac{1}{k!}\left(\frac{\Delta t}{M}\right)^k \sum_{l_1,\ldots,l_k=0}^{M-1} \mathcal{T} \calC\left(\tau_{n,l_1}\right) \cdots \calC\left(\tau_{n,l_{k-1}}\right)\vecc(\tau_{n,l_k}),
        \label{eq:DeltaTil}
    \end{align}
    for which, by using Eqs. (14), (130), and (131) in Ref.~\cite{Berry2024quantumalgorithm} along with simple algebra, we have
    \begin{align}
        \left\|\tilde{\vecDelta}^{K,r,M}_n-\vecDelta_n\right\| \le \varepsilon
    \end{align}
    for $r$, $K$, and $M$ satisfying Eq.~\eqref{eq:KrM}.
    Then, by using the controlled versions of $V_{\widetilde{\vecDelta}}$ and $V_{\veccalY_0}$ once each along with arithmetic circuits, we can construct an oracle $V_{\calD}$ that acts as
    \begin{align}
        V_{\calD} \ket{0}\ket{0}\ket{0} = \ket{0}\ket{\calD} + \ket{0_\perp},
        \label{eq:VcalD}
    \end{align}
    where $\ket{0_\perp}$ is an unnormalized state such that $(\ket{0}\!\bra{0}\otimes I \otimes I)\ket{0_\perp}=0$, and $\ket{\calD}$ is an unnormalized state given by
    \begin{align}
        \ket{\calD} \coloneqq \frac{1}{U_{\vecm_{\veccalY}}} \sum_{n=0}^r \sum_{i=1}^\calN \calD_{n,i}\ket{n}\ket{i}, ~ \calD = (\calD_{n,i})_{n\in[r]_0,i\in[\calN]}\coloneqq
        \begin{pmatrix}
        \veccalY_0^T \\
        \left(\tilde{\vecDelta}^{K,r,M}_0\right)^T \\
        \vdots \\
        \left(\tilde{\vecDelta}^{K,r,M}_{r-1}\right)^T
        \end{pmatrix}
    \end{align}
    (see Eq.~(155) in Ref.~\cite{miyamoto2026quantum}).
    
    We define $\tilde{\mathbf{m}}_{\veccalY,{\rm hist}} \coloneqq \widetilde{\Phi}^{-1}{\rm vec}(\calD)$ and write it in block form as $\tilde{\mathbf{m}}_{\veccalY,{\rm hist}} = \begin{pmatrix}\tilde{\mathbf{m}}_{\veccalY,0} \\ \vdots \\ \tilde{\mathbf{m}}_{\veccalY,r}\end{pmatrix}$ with $\tilde{\mathbf{m}}_{\veccalY,0},\ldots,\tilde{\mathbf{m}}_{\veccalY,r}\in\R^\calN$.
    We then define $\tilde{\calM}_{\veccalY,{\rm hist}} \in \R^{(r+1) \times \calN}$ by
    \begin{align}
        \tilde{\calM}_{\veccalY,{\rm hist}} = 
        \begin{pmatrix}
            \tilde{\mathbf{m}}_{\veccalY,0}^T \\ \vdots \\ \tilde{\mathbf{m}}_{\veccalY,r}^T
        \end{pmatrix}.
    \end{align}
    By a discussion similar to that in the proof of Theorem 8 in Ref.~\cite{An2026fastforwarding}, we have
    \begin{align}
        \left\|\tilde{\calM}_{\veccalY,{\rm hist}} - \calM_{\veccalY,{\rm hist}}\right\|_{\rm F} \le \sqrt{3} \left(\frac{1}{3}(r+1)^3\varepsilon^2 \|\veccalY_0\|^2 + \frac{1}{20}(r+1)^3\varepsilon^2 \alpha_\vecc^2T^2 + \frac{1}{3}(r+1)^3\varepsilon^2\right)^{1/2},
    \end{align}
    which translates into
    \begin{align}
         \left\|\tilde{\calM}_{\veccalY,{\rm hist}} - \calM_{\veccalY,{\rm hist}}\right\|_{\rm F} \le \frac{U_{\vecm_{\veccalY}} \epsilon}{2}.
        \label{eq:mYhistErr}
    \end{align}

    As we will see at the end of this proof, Eq.~\eqref{eq:PhitilPhiErr} leads to
    \begin{align}
        \|\widetilde{\Phi}^{-1}\| \le 2r.
        \label{eq:TilPhiInvNorm}
    \end{align}
    We also have $\left\|\widetilde{\Phi}\right\| \le \alpha_{\widetilde{\Phi}}$.
    These mean that the singular values of $\widetilde{\Phi}$ lie in 
    $\left[\frac{1}{2r},\alpha_{\widetilde{\Phi}}\right]$, and thus, Theorem~\ref{eq:MatInvQSVT} implies that we can construct an $(1,O(K(a_\calC+\log K)),\epsilon_3)$-block-encoding $V_{\widetilde{\Phi}^{-1}}$ of $\frac{\alpha_{\widetilde{\Phi}}}{2\kappa_{\widetilde{\Phi}}}\widetilde{\Phi}^{-1}$ by making 
    \begin{align}
        O\left(\kappa_{\widetilde{\Phi}}\log\left(\frac{1}{\epsilon_3}\right)\right)
        \label{eq:PhiTilInvComp1}
    \end{align}
    uses of $V_{\widetilde{\Phi}}$ and
    \begin{align}
        O\left(a_{\widetilde{\Phi}}\kappa_{\widetilde{\Phi}}\log\left(\frac{1}{\epsilon_3}\right)\right)
        \label{eq:PhiTilInvComp2}
    \end{align}
    uses of additional elementary gates, where
    \begin{align}
        \kappa_{\widetilde{\Phi}} \coloneqq 2\alpha_{\widetilde{\Phi}}r,~\epsilon_3\coloneqq\frac{\epsilon}{8r}.
        \label{eq:kappaPhiTilEps3}
    \end{align}
    Then, on a system consisting of two registers, where the first one contains the ancillary qubits for $V_{\widetilde{\Phi}^{-1}}$ and those corresponding to the first register in Eq.~\eqref{eq:VcalD}, we can operate $V_{\calD}$ and $V_{\widetilde{\Phi}^{-1}}$ in this order to obtain the state written as
    \begin{align}
        \ket{0}\frac{\alpha_{\widetilde{\Phi}}}{2\kappa_{\widetilde{\Phi}}}\calL\ket{\calD} + \ket{0_\perp},
        \label{eq:ResState}
    \end{align}
    where the matrix $\calL$ satisfies
    \begin{align}
        \left\|\frac{\alpha_{\widetilde{\Phi}}}{2\kappa_{\widetilde{\Phi}}}\calL - \frac{\alpha_{\widetilde{\Phi}}}{2\kappa_{\widetilde{\Phi}}}\widetilde{\Phi}^{-1}\right\| \le \epsilon_3.
        \label{eq:MatInvErr}
    \end{align}
    We then regard the state in Eq.~\eqref{eq:ResState} as the state in Eq.~\eqref{eq:VHisY}.
    Combining Eqs.~\eqref{eq:mYhistErr} and \eqref{eq:MatInvErr} yields
    \begin{align}
         \left\|\tilde{\tilde{\calM}}_{\veccalY,{\rm hist}} - 
        \calM_{\veccalY,{\rm hist}}\right\|_{\rm F}=& \left\|4rU_{\vecm_{\veccalY}}\frac{\alpha_{\widetilde{\Phi}}}{2\kappa_{\widetilde{\Phi}}}\calL\ket{\calD} - {\rm vec}\left(\calM_{\veccalY,{\rm hist}}\right)\right\| \nonumber \\
        \le & 4rU_{\vecm_{\veccalY}} \times \left\|\frac{\alpha_{\widetilde{\Phi}}}{2\kappa_{\widetilde{\Phi}}}\calL - \frac{\alpha_{\widetilde{\Phi}}}{2\kappa_{\widetilde{\Phi}}}\widetilde{\Phi}^{-1}\right\| \times \left\|\ket{\calD}\right\| + \left\|\widetilde{\Phi}^{-1}\calD-{\rm vec}\left(\calM_{\veccalY,{\rm hist}}\right)\right\| \nonumber \\
        \le & 4rU_{\vecm_{\veccalY}}\times \epsilon_3 + \frac{U_{\vecm_{\veccalY}} \epsilon}{2} \nonumber \\
        \le & U_{\vecm_{\veccalY}} \epsilon,
        \label{eq:UBCErr}
    \end{align}
    which means that the state in Eq.~\eqref{eq:ResState} is actually in the form of Eq.~\eqref{eq:VHisY}, including the condition \eqref{eq:calMtiltilErr}.
    This means that the stated oracle $V_{\vecm_{\veccalY},\epsilon}^{\rm hist}$ is constructed by using $V_{\calD}$ and $V_{\widetilde{\Phi}^{-1}}$ once each.

    Now, let us prove the claimed complexity.
    We recall that $V_{\widetilde{\Phi}^{-1}}$ calls $V_{\widetilde{\Phi}}$ and additional elementary gates the numbers of times given in Eqs.~\eqref{eq:PhiTilInvComp1} and \eqref{eq:PhiTilInvComp2}, respectively, and that $V_{\widetilde{\Phi}}$ is constructed by $O(K)$ uses of the controlled $U_\calC$ and $O(K \log K \log M)$ uses of additional elementary gates.
    By combining these, we see that $V_{\widetilde{\Phi}^{-1}}$ uses the controlled $U_\calC$ and additional elementary gates the numbers of times given in Eqs.~\eqref{eq:QueryVcalCStateGen} and \eqref{eq:QueryElGateStateGen}, respectively.
    Adding the complexity of $V_{\calD}$, which is a combination of the controlled versions of $V_{\widetilde{\vecDelta}}$ and $V_{\veccalY_0}$, leaves these evaluations unchanged, except one call to the controlled $V_\vecc$ and one call to the controlled $V_{\veccalY_0}$.

    Lastly, let us prove Eq.~\eqref{eq:TilPhiInvNorm}.
    According to Lemma 2 in Ref.~\cite{miyamoto2026quantum}, we have
    \begin{align}
        \widetilde{\Phi}^{-1} =
        \begin{pmatrix}
            (\widetilde{\Phi}^{-1})_{0,0} & \cdots & (\widetilde{\Phi}^{-1})_{r,0} \\
            \vdots & \ddots & \vdots \\
            (\widetilde{\Phi}^{-1})_{r,0} & \cdots & (\widetilde{\Phi}^{-1})_{r,r}
        \end{pmatrix},
    \end{align}
    where each block $(\widetilde{\Phi}^{-1})_{n,n^\prime}\in \mathbb{R}^{\calN \times \calN},n,n^\prime\in[r]_0$ is given by
    \begin{align}
        (\widetilde{\Phi}^{-1})_{n,n^\prime} =
        \begin{cases}
            I &  ; ~ n = n^\prime \\
            \tilde{\Phi}^{K,r,M}_{n-1} \cdots \tilde{\Phi}^{K,r,M}_{n^\prime} & ; ~ n^\prime < n \\
            O & ; ~ n < n^\prime
        \end{cases}
        .
    \end{align}
    We also have
    \begin{align}
        \left\|\tilde{\Phi}^{K,r,M}_n\right\| \le \left\|\Phi(t_{n+1},t_n)\right\| + \left\|\tilde{\Phi}^{K,r,M}_n-\Phi(t_{n+1},t_n)\right\| \le 1+\varepsilon \le 1+\frac{1}{5r},
    \end{align}
    where we use $\varepsilon \le \frac{1}{5r}$, which follows from Eq.~\eqref{eq:vareps} by simple algebra.
    We thus have
    \begin{align}
        \|(\widetilde{\Phi}^{-1})_{n,n^\prime}\| \le \left(1+\frac{1}{5r}\right)^{n-n^\prime} \le \left(1+\frac{1}{5r}\right)^r \le e^{\frac{1}{5}}, 
    \end{align}
    where we use the inequality $\forall x>0, \left(1+\frac{1}{x}\right)^x \le e$.
    Then, Lemma 6 in Ref.~\cite{An2026fastforwarding} yields $\left\|\widetilde{\Phi}^{-1}\right\| \le e^{\frac{1}{5}}(r+1) \le 2r$.
\end{proof}

\begin{theorem}

    Let $\epsilon\in(0,1]$.
    Suppose that Assumptions \ref{ass:alphaA}, \ref{ass:muA}, \ref{ass:alphaF}, \ref{ass:BNorm}, and \ref{ass:VXini} hold.
    Suppose that Eq.~\eqref{eq:CarlConvCond1} holds for some $\NC\in\N$ and $\gamma>0$.
    Set $r$, $K$, and $M$ as
    \begin{align}
        r & = r(J) \coloneqq \max\left\{\left\lceil\tilde{\alpha}_\calC T\right\rceil,7\right\}, \nonumber \\
        K & = K(J,\gamma,\epsilon) \coloneqq \max\left\{\left\lceil\log\left(\frac{3}{\tilde{\varepsilon}}\right)\right\rceil,7\right\}, \nonumber \\
        M & = M(J,\gamma,\epsilon) \coloneqq \left\lceil\max\left\{\frac{4\tilde{\alpha}_{\drm \calC/\drm t}}{\tilde{\alpha}_\calC^2 \tilde{\varepsilon}}, \left(\frac{3\tilde{\alpha}_\calC \tilde{\alpha}_{\drm \calC/\drm t} T^3}{r^3(J)}+\frac{2\tilde{\alpha}_{\drm \vecc/\drm t}T^2}{r^2(J)}\right)\frac{1}{\tilde{\varepsilon}}\right\}\right\rceil,
        \label{eq:KrM2}
    \end{align}
    where
    \begin{align}
        \tilde{\alpha}_\calC \coloneqq 3\NC\alpha_A,
        \label{eq:tilalphaCalC}
    \end{align}
    \begin{align}
        \tilde{\alpha}_{\drm \calC/\drm t} \coloneqq \sqrt{3} \NC\left(\alpha_{\drm A/\drm t} + (\NC-1)\gamma^2 \alpha_{\drm B/\drm t} + \gamma^{1-s}\alpha_{\drm F/\drm t}\right),
    \end{align}
    \begin{align}
        \tilde{\alpha}_\vecc \coloneqq \gamma^2\sigma^2
    \end{align}
    \begin{align}
        \tilde{\alpha}_{\drm \vecc/\drm t} \coloneqq 2\gamma^2 \alpha_{\drm B/\drm t}
    \end{align}
    \begin{align}
        \tilde{\varepsilon} \coloneqq \sqrt{\frac{1}{(r(J)+1)^3\left(4\left(\|\veccalY_0\|^2+1\right)+\frac{3}{5}\tilde{\alpha}_\vecc^2T^2\right)}}\tilde{U}_{\vecm_{\veccalY}}\epsilon
    \end{align}
    \begin{align}
        \tilde{U}_{\vecm_{\veccalY}} \coloneqq \sqrt{\|\veccalY_0\|^2+r(J)\tilde{U}_{\widetilde{\vecDelta}}^2},
    \end{align}
    \begin{align}
        \tilde{U}_{\widetilde{\vecDelta}} \coloneqq \frac{\tilde{\alpha}_\vecc}{\tilde{\alpha}_\calC}\sum_{k=1}^{K(J,\gamma,\epsilon)} \frac{1}{k!} \left(\frac{\tilde{\alpha}_\calC \Delta t}{M(J,\gamma,\epsilon)}\right)^k.
    \end{align}
    Then, we can construct an oracle $V_{\vecm_{\veccalY},\epsilon}^{\rm hist}$, which acts as Eq.~\eqref{eq:VHisY}, with $U_{\vecm_{\veccalY}}$ replaced by $\tilde{U}_{\vecm_{\veccalY}}$.
    $V_{\vecm_{\veccalY},\epsilon}^{\rm hist}$ makes
    \begin{align}
        O\left(\NC^2 r(J) K(J,\gamma,\epsilon) \log\left(\frac{r(J)}{\epsilon}\right)\right)
    \end{align}
    uses of the controlled $V_A$ and the controlled $V_F$,
    \begin{align}
        O\left(\NC^3 r(J) K(J,\gamma,\epsilon) \log\left(\frac{r(J)}{\epsilon}\right)\right)
    \end{align}
    uses of the controlled $V_{BB^T}$,
    \begin{align}
        O\left(\NC\right)
    \end{align}
    uses of the controlled $V_{\vecX_{\rm ini}}$, and
    \begin{align}
        O\left(r(J) K(J,\gamma,\epsilon) \left(\NC^2(\NC+\log N)+\log K(J,\gamma,\epsilon) \log M(J,\gamma,\epsilon) + a_A + a_F\right) \log\left(\frac{r(J)}{\epsilon}\right)\right)
    \end{align}
    uses of elementary gates.
    \label{th:HistStatemY}
\end{theorem}

\begin{proof}
    To use Lemma~\ref{lem:HistStatemY}, we consider the following points.\\
    
    \noindent \underline{\textbf{Bounding norms}}
    
    Because of Lemma \ref{lem:dcalC/dtpNorm} in Sec.~\ref{sec:Lem4ThHistStatemY}, Eqs.~\eqref{eq:alphadAdt}, \eqref{eq:alphadFdt}, and \eqref{eq:alphadBdt} leads to
    \begin{align}
        \left\|\frac{\drm \calC(t)}{\drm t}\right\| \le \sqrt{3} \NC\left(\alpha_{\drm A/\drm t} + (\NC-1)\gamma^2 \alpha_{\drm B/\drm t} + \gamma^{1-s}\alpha_{\drm F/\drm t}\right)=\tilde{\alpha}_{\drm \calC/\drm t}.
    \end{align}
    Because of Lemma~\ref{lem:muc1<0} in Sec.~\ref{sec:PrCarErrTail}, Eq.~\eqref{eq:CarlConvCond1} leads to $\mu(\calC(t)) \le 0$.
    We also have
    \begin{align}
        \left\|\frac{\drm \vecc(t)}{\drm t}\right\| = \left\|{\rm vec}\left(\frac{\drm}{\drm t}\hat{B}(t)\hat{B}^T(t)+\hat{B}(t)\left(\frac{\drm}{\drm t}\hat{B}(t)\right)^T\right)\right\|  \le 2 \left\|\frac{\drm}{\drm t}\hat{B}(t)\hat{B}^T(t)\right\|_{\rm F} \le 2\gamma^2 \alpha_{\drm B/\drm t}=\tilde{\alpha}_{\drm \vecc/\drm t}.
    \end{align}

    \noindent \underline{\textbf{Constructing oracles}}
    
    Let us construct the oracles assumed in Lemma~\ref{lem:HistStatemY} from the currently given oracles.
    We first consider $V_\calC$, decomposing $\calC(t)$ as
    \begin{align}
        & \calC(t) = \calC_{\rm LL}(t) + \calC_{\rm D}(t) + \calC_{\rm UR}(t) \nonumber \\
        & \calC_{\rm LL}(t) =
        \begin{pmatrix}
        & & & & \\
        & & & & \\
        \calA^{3,1}(t) & & & & \\
        & \ddots & & & \\
        & & \calA^{\NC,\NC-2}(t) & & 
        \end{pmatrix},
        \calC_{\rm D}(t) = \begin{pmatrix} \calA^{1,1}(t) & & \\ & \ddots & \\ & & \calA^{\NC,\NC}(t) \end{pmatrix}, \\
        & \calC_{\rm UR}(t)=
        \begin{pmatrix}
            & & \calA^{1,s}(t) & & \\
            & & & \ddots & \\
            & & & & \calA^{\NC+s-1,\NC}(t)\\
            & & & &
        \end{pmatrix}.
    \end{align}
    For each block $\calA^{j,j-2}(t)$ in $\calC_{\rm LL}(t)$, each term $\calA^{j,j-2,k,l}(t)$ in it is block-encoded by the circuit $V_{\calA^{j,j-2,k,l}}$ shown in Fig.~\ref{fig:calAjj-2}.
    In fact, $V_{BB^T}$ in the circuit transforms $\ket{t}\ket{0}\ket{0}\ket{0}\ket{i_1}\cdots\ket{i_{j-2}}$ into $\ket{t}\ket{0}\ket{B(t)B^T(t)}\ket{i_1}\cdots\ket{i_{j-2}}+\ket{0_\perp}$, where $t\in[0,T]$ and $i_1,\ldots,i_{j-2}\in[N]$, and then the subsequent swap gates yields
    \begin{align}
        &\ket{t}\ket{0}\frac{1}{\sigma^2}\sum_{i_k,i_l=1}^N (B(t)B^T(t))_{i_k,i_l}\ket{i_1}\cdots\ket{i_k}\cdots\ket{i_l}\cdots\ket{i_{j-2}}+\ket{0_\perp} \nonumber \\
        &=\ket{t}\ket{0}\frac{1}{\gamma^2\sigma^2}\sum_{i_k,i_l=1}^N (\hat{B}(t)\hat{B}^T(t))_{i_k,i_l}\ket{i_1}\cdots\ket{i_k}\cdots\ket{i_l}\cdots\ket{i_{j-2}}+\ket{0_\perp}
    \end{align}
    which means that the circuit acts as the $(\gamma^2\sigma^2,1,0)$-time-dependent-block-encoding of $\calA^{j,j-2,k,l}(t)$.
    According to Theorem~\ref{th:BESum}, given these circuits for $\{\calA^{j,j-2,k,l}\}_{\substack{k=1,\ldots,j-1 \\ l=k+1,\ldots,j}}$, we can combine them to obtain a $\left(\frac{\NC(\NC-1)}{2}\gamma^2\sigma^2,1+\lceil\lg j \rceil,0\right)$-time-dependent-block-encoding $V_{\calA^{j,j-2}}$ of $\calA^{j,j-2}$, which can be deemed as a $\left(\frac{\NC(\NC-1)}{2}\gamma^2\sigma^2,1+\lceil\lg \NC \rceil,0\right)$-time-dependent-block-encoding (c.f. Lemma~\ref{lem:AddBEAnc}).
    Then, according to Lemma~\ref{lem:BEDirectSum}, combining the controlled versions of $V_{\calA^{3,1}},\ldots,V_{\calA^{\NC,\NC-2}}$ yields a $\left(\frac{\NC(\NC-1)}{2}\gamma^2\sigma^2,O(\NC\log N),0\right)$-time-dependent-block-encoding of $\begin{pmatrix}
    \calA^{3,1}(t) & &\\
    & \ddots & \\ 
    & & \calA^{\NC,\NC-2}(t) 
    \end{pmatrix}$,
    and applying row shifting to it yields a $\big(\frac{\NC(\NC-1)}{2}\gamma^2\sigma^2, \allowbreak O(\NC\log N),0\big)$-time-dependent-block-encoding $V_{\calC_{\rm LL}}$ of $\calC_{\rm LL}$, with $O(\NC \log N)$ elementary gates used.
    $V_{\calA^{j,j-2,k,l}}$ makes $O(1)$ uses of $V_{BB^T}$ and elementary gates, $V_{\calA^{j,j-2}}$ uses $\{V_{\calA^{j,j-2,k,l}}\}_{k,l}$ once each along with $O(\NC^2)$ elementary gates, and $V_{\calC_{\rm LL}}$ uses $V_{\calA^{3,1}},\ldots,V_{\calA^{\NC,\NC-2}}$ once each along with $O(\NC^2\log N)$ elementary gates.
    By piling up these, we see that $V_{\calC_{\rm LL}}$ makes $O(\NC^3)$ uses of $V_{BB^T}$ and $O(\NC^2(\NC + \log N))$ elementary gates.

    \begin{figure}[t]
    \begin{center}
    
    \begin{quantikz}[wire types={q,q,q,q,n,q,n,q,n,q}, transparent]
    \lstick{$\ket{t}$} & \qwbundle{} & \gate[4]{V_{BB^T}} & & \qw & \qw \\
    \lstick{ancilla} & \qw &  & \qw  & \qw & \qw\\
    \lstick{1st} & \qwbundle{} & & \swap{3} & \qw & \qw \\
    \lstick{2nd} & \qwbundle{} & & & \swap{4}  & \qw \\
    & \vdots & & \\
    \lstick{$k$-th} & \qwbundle{} & & \targX{} & \qw & \qw\\
    & \vdots & & \\
    \lstick{$l$-th} & \qwbundle{} & & & \targX{} & \qw \\
    & \vdots & & \\
    \lstick{$j$-th} & \qwbundle{} & & & & \\
    \end{quantikz}
    \caption{The quantum circuit for block-encoding of $\calA^{j,j-2,k,l}$.}
    \label{fig:calAjj-2}
    \end{center}
    \end{figure}
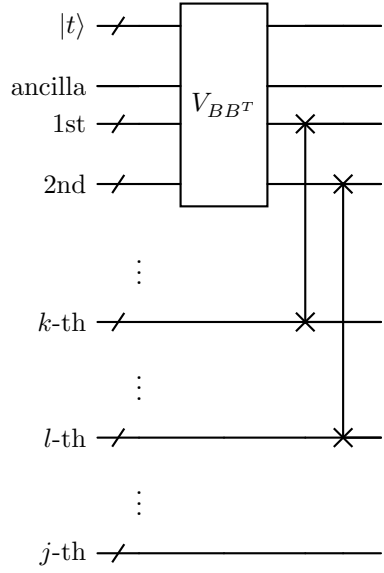

    $\calC_{\rm D}$ can be constructed as follows.
    Because of the form of $\calA^{j,j}$ shown in Eq.~\eqref{eq:calAjj}, we can construct the $(\NC\alpha_A,a_A+\lceil\lg j\rceil,0)$-time-dependent-block-encoding $V_{\calA^{j,j}}$ of $\calA^{j,j}$ by using $V_A$ $j$ times and $O(j)$ elementary gates, as implied by Theorem~\ref{th:BESum}.
    Then, as implied by Lemma~\ref{lem:BEDirectSum}, using $V_{\calA^{1,1}},\ldots,V_{\calA^{\NC,\NC}}$ once each along with $O(\NC^2\log N)$ elementary gates, we obtain the $(\NC\alpha_A,O(\NC \log N + a_A),0)$-time-dependent-block-encoding $V_{\calC_{\rm D}}$ of $\calC_{\rm D}$.
    In total, $V_{\calC_{\rm D}}$ makes $O(\NC^2)$ uses of $V_A$ and $O(\NC^2 \log N)$ uses of elementary gates.

    The construction of $\calC_{\rm UR}$ is similar: we obtain the $(\NC\gamma^{1-s}\alpha_F,O(\NC \log N + a_F),0)$-time-dependent-block-encoding $V_{\calC_{\rm UR}}$ of $\calC_{\rm UR}$, using $V_F$ $O(\NC^2)$ times and $O(\NC^2 \log N)$ elementary gates.

    Finally, according to Theorem~\ref{th:BESum}, by combining $V_{BB^T}$, $V_{\calC_{\rm D}}$, and $V_{\calC_{\rm UR}}$, whose normalization factor is adjusted to $\NC\alpha_A$\footnote{Note that $\max\left\{\frac{\NC(\NC-1)}{2}\gamma^2\alpha_{B,{\rm F}},\NC\alpha_A,\NC\gamma^{1-s}\alpha_F\right\}=\NC\alpha_A$ under Eq.~\eqref{eq:CarlConvCond1}.} as in Lemma~\ref{lem:AdjustBEScFac}, we obtain $(\tilde{\alpha}_\calC,O(\NC\log\NC+a_A+a_F),0)$-time-dependent-block-encoding $V_{\calC}$ of $\calC$.
    In it, $V_A$, $V_F$, $V_{BB^T}$, and elementary gates are used $O(\NC^2)$, $O(\NC^2)$, $O(\NC^3)$, and $O(\NC^2 (\NC+\log N))$ times, respectively.\\

    $V_\vecc$ can be constructed from $V_{BB^T}$ with slight modification.
    Note that the state $\ket{i}\ket{j}$ in Eq.~\eqref{eq:ketBBT} can be regarded as the state $\ket{N(i-1)+j}$, with the two registers deemed as one combined register.
    Then, $V_{BB^T}$ followed by addition of $N$ to the integer yields $\frac{1}{\gamma^2\sigma^2}\sum_{k=1}^\calN c_k \ket{t}\ket{0}\ket{k}+\ket{0_\perp}$ with $\vecc$ in Eq.~\eqref{eq:vecc}, which means that this operation is nothing but $V_\vecc$ with $\alpha_\vecc=\gamma^2\sigma^2=\tilde{\alpha}_\vecc$.
    In this, $V_{BB^T}$ is used once, and $O(\log N)$ elementary gates are used.

    Since $\calY_0$ is given as Eq.~\eqref{eq:calY0}, we can construct $V_{\calY_0}$ by using $V_{\vecX_{\rm ini}}$ $O(\NC)$ times along with $O(\log \NC)$ elementary gates, according to the proof of Lemma 4 in Ref.~\cite{Liu2021}.\\

    \noindent \underline{\textbf{Complexity}}
    
    We now have $\tilde{\alpha}_\calC,\tilde{\alpha}_{\drm \calC/\drm t},\tilde{\alpha}_\vecc,\tilde{\alpha}_{\drm \vecc/\drm t}$, which corresponds to $\alpha_\calC,\alpha_{\drm \calC/\drm t},\alpha_\vecc,\alpha_{\drm \vecc/\drm t}$ in Lemma~\ref{lem:HistStatemY}, the oracles needed in the lemma, and $\mu(\calC(t))\le0$, so we can apply the lemma.
    Setting $r$, $K$, and $M$ as stated, we can generate the claimed quantum state.
    The claimed query numbers are obtained by combining the query numbers of the controlled $V_\calC$, $V_\vecc$, and $V_{\veccalY_0}$, and elementary gates in Lemma~\ref{lem:HistStatemY} with the aforementioned complexities for constructing these oracles from the given ones.
\end{proof}

\begin{theorem}
    Suppose that Assumptions \ref{ass:XVar} to \ref{ass:VXini} hold.
    Let $\delta,c\in(0,1)$, $\bar{C}>0$, and $\epsilon \in (0,\sqrt{\sigma^2T}\bar{C}]$.
    Set $J=J_{\epsilon_{\rm C},c}$ and $\gamma=\gamma_{\epsilon_{\rm C},c}$, where
    \begin{align}
        \epsilon_{\rm C} \coloneqq \frac{7\epsilon \min\left\{\log \left(\frac{1}{c}\right),1\right\}}{648 \bar{C} \max\{\alpha_A T,1\}}\left(\log\left(\frac{7776 \bar{C} \sqrt{\sigma^2T} \max\{\alpha_A T,1\}}{7c\log \left(\frac{1}{c}\right)\epsilon\min\left\{\frac{\log \left(1/c\right)}{9\sqrt{5}\eta T},1\right\}} \right)\right)^{-1}.
    \end{align}
    Set $r=r(J_{\epsilon_{\rm C},c})$.
    Suppose that Eq.~\eqref{eq:CarlConvCond1} hold with this $\NC$ and $\gamma$.
    Let $C=(C_{n,i})$ be an $r \times N$ real matrix with $\|C\|_{\rm F} = \sqrt{r}\bar{C}$, and assume that we have access to the oracle $U_C$ that acts as
    \begin{align}
        U_C \ket{0}\ket{0} = \ket{C} \coloneqq \frac{1}{\|C\|_{\rm F}} \sum_{n=1}^r \sum_{i=1}^N C_{n,i} \ket{n}\ket{i}.
    \end{align}
    Then, there is a quantum algorithm that, with probability at least $1-\delta$, outputs an estimate $\hat{\mu}_f$ of $\mu_f=\mathbb{E}[f]$, where
    \begin{align}
        f \coloneqq \sum_{n=1}^r \sum_{i=1}^N C_{n,i} X^i_{t_n},
    \end{align}
    with error at most $\epsilon$.
    This algorithm uses the controlled $V_A$ and $V_F$
    \begin{align}
    O\left(\frac{(\alpha_A T)^{\frac{5}{2}}\bar{C}\sqrt{\sigma^2T}}{c(1-c)^{\frac{9}{2}}\epsilon}\left(\log\Lambda\right)^{\frac{11}{2}}\log\log\Lambda\log\left(\frac{1}{\delta}\right)\right)
    \label{eq:CompExaUA}
    \end{align}
    times each, the controlled $V_{BB^T}$ 
    \begin{align}
    O\left(\frac{(\alpha_A T)^{\frac{5}{2}}\bar{C}\sqrt{\sigma^2T}}{c(1-c)^{\frac{9}{2}}\epsilon}\left(\log\Lambda\right)^{\frac{13}{2}}\log\log\Lambda\log\left(\frac{1}{\delta}\right)\right)
    \label{eq:CompVB}
    \end{align}
    times, the controlled $V_{\vecX_{\rm ini}}$ 
    \begin{align}
    O\left(\frac{(\alpha_A T)^{\frac{3}{2}}\bar{C}\sqrt{\sigma^2T}}{c(1-c)^{\frac{3}{2}}\epsilon}\left(\log\Lambda\right)^{\frac{3}{2}}\log\left(\frac{1}{\delta}\right)\right)
    \label{eq:CompXini}
    \end{align}
    times, and
    \begin{align}
    O\left(\frac{(\alpha_A T)^{\frac{5}{2}}\bar{C}\sqrt{\sigma^2T}}{c(1-c)^{\frac{5}{2}}\epsilon}\left(\frac{\left(\log\Lambda\right)^2}{(1-c)^2}+\log(\Lambda\Pi)\log\log\Lambda+a_A+a_F\right)(\log\Lambda)^2\log\left(\frac{1}{\delta}\right)\right)
    \label{eq:CompElemGate}
    \end{align}
    additional elementary gates, where $\Lambda,\Pi>0$ are of order
    \begin{align}
        \Lambda = O\left(\frac{(\sqrt{\sigma^2 T}+\sigma_\vecX) (\alpha_A T + \sqrt{N}) \bar{C}}{\epsilon c (1-c)}\right),~
        \Pi = O\left(\frac{\alpha_{\drm A/\drm t} + \frac{\sigma^2T}{c^2} \alpha_{\drm B/\drm t} + \left(\frac{\sigma^2T}{c^2}\right)^{\frac{1-s}{2}}\alpha_{\drm F/\drm t}}{\alpha_A^2}\right).
    \end{align}
    
    \label{th:MeanEstim}
\end{theorem}

\begin{proof}
We here present the algorithm as Algorithm~\ref{alg:main} and defer the rest of the proof to Appendix~\ref{app:PrME}.

\begin{algorithm}[H]
\caption{Proposed quantum algorithm for estimating $\mu_f$}\label{alg:main}
\begin{algorithmic}[1]

\State Construct the oracle $V_{{\rm hist},\epsilon^\prime}$ as in Theorem~\ref{th:HistStatemY}, where
\begin{align}
    \epsilon^\prime \coloneqq \frac{\gamma\epsilon}{3\tilde{U}_{\vecm_{\veccalY}}\|C\|_{\rm F}}.
    \label{eq:epsPr}
\end{align}

\State Using the algorithm in Theorem \ref{th:OE} with $V_{{\rm hist},\epsilon^\prime}$ and $V_C$, get an estimate $\hat{\mu}^\prime_f$ of $\Braket{\widetilde{C} | \Psi_{\vecm_{\veccalY},\epsilon}^{\rm hist}}$ with accuracy
\begin{align}
    \epsilon_{\rm OE}=\frac{\gamma\epsilon}{12r\tilde{U}_{\vecm_{\veccalY}}\|C\|_F}
\end{align}
and success probability $1-\delta$, where $\Ket{\widetilde{C}}\coloneqq\ket{0}\ket{C}$.

\State Output
\begin{align}
    \hat{\mu}_f = 
    4r\tilde{U}_{\vecm_{\veccalY}}\|C\|_F \hat{\mu}^\prime_f.
\end{align}

\end{algorithmic}
\end{algorithm}
    
\end{proof}

\begin{remark}
    The query complexity bounds in Theorem~\ref{th:MeanEstim} scale linearly with respect to $\sigma$ except for logarithmic factors.
    Note that $\sigma$, an upper bound of $\|B(t)\|_{S_1}$, can scale polynomially with respect to $N$.
    Since $\|B(t)\|_{S_1}$ is the sum of all of $B(t)$'s singular values, $\|B(t)\|_{S_1} \sim N$ holds if, for example, $B(t)$ is full-rank and its singular values have the same order of magnitude.
    One situation in which $\|B(t)\|_{S_1}=\tilde{O}(1)$ holds is when the noise is introduced into only a limited number of entries of $\vecX_t$.
    Even in this case, the noise can affect all the entries in $\vecX_t$ through its propagation via $A$, which means that the noise on a limited number of degrees of freedom affects the entire system through coupling.
\end{remark}

\section{Numerical experiment \label{sec:exp}}

\begin{figure}[t]
    \centering

    \begin{subfigure}{0.45\textwidth}
        \centering
        \includegraphics[width=\linewidth]{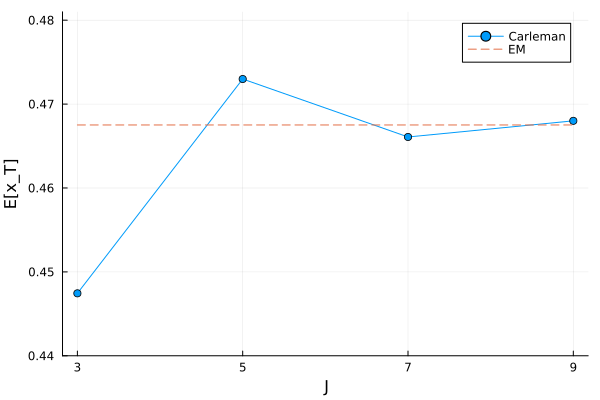}
        \caption{$\E[x_T]$}
        \label{fig:xT}
    \end{subfigure}
    \hfill
    \begin{subfigure}{0.45\textwidth}
        \centering
        \includegraphics[width=\linewidth]{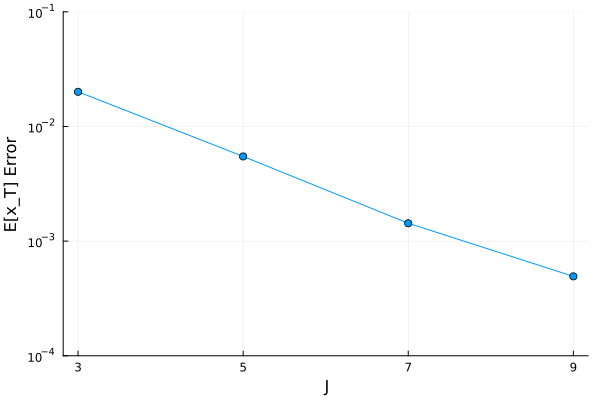}
        \caption{Error in $\E[x_T]$}
        \label{fig:xTErr}
    \end{subfigure}

    \vspace{1em}

    \begin{subfigure}{0.45\textwidth}
        \centering
        \includegraphics[width=\linewidth]{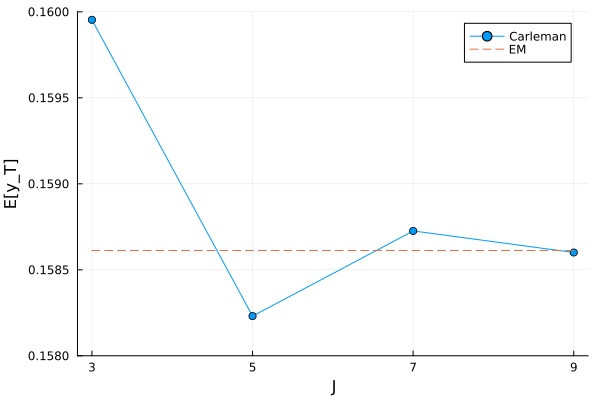}
        \caption{$\E[y_T]$}
        \label{fig:yT}
    \end{subfigure}
    \hfill
    \begin{subfigure}{0.45\textwidth}
        \centering
        \includegraphics[width=\linewidth]{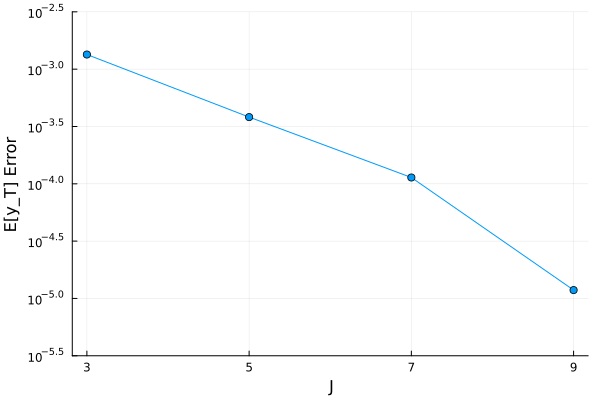}
        \caption{Error in $\E[y_T]$}
        \label{fig:yTErr}
    \end{subfigure}

    \caption{\ac{CL}-based approximation of the expectation $\E[\vecX_T]$ of $\vecX_T$ that follows Eq.~\eqref{eq:SDEEg}. In (a) (resp. (c)), the \ac{CL}-based approximation $\widehat{x_T}$ (resp. $\widehat{y_T}$) of $\E[x_T]$ (resp. $\E[y_T]$) for various $J$ is plotted in blue solid, along with the \ac{EM}-based Monte-Carlo approximation $\overline{x_T}$ (resp. $\overline{y_T}$) as a reference in red dashed. In (b) (resp. (d)), we plot $|\widehat{x_T}-\overline{x_T}|$ (resp. $|\widehat{y_T}-\overline{y_T}|$).}
    \label{fig:res}
\end{figure}

To demonstrate that \ac{CL} for \acp{SDE} works, we consider the following two-dimensional \ac{SDE}
\begin{align}
    \begin{dcases}
        \drm x_t & = -(m_x^2 x_t + m_{xy}^2 y_t + \lambda_x x_t^3) \drm t + \sigma_x \drm W^1_t \\
        \drm y_t & = -(m_y^2 x_t + m_{xy}^2 x_t + \lambda_y y_t^3) \drm t + \sigma_y \drm W^2_t
    \end{dcases}
    ,
    \label{eq:SDEEg}
\end{align}
which is written as Eq.~\eqref{eq:SDE} with
\begin{align}
    A=
    \begin{pmatrix}
    m_x^2  &  m_{xy}^2 \\
    m_{xy}^2 & m_y^2
    \end{pmatrix},~
    B=
    \begin{pmatrix}
    \sigma_x  &  0 \\
    0 & \sigma_y
    \end{pmatrix},~
    F=
    \begin{pmatrix}
    \lambda_x  &  0 & 0 & 0 & 0 & 0 & 0 & 0 \\
    0 & 0 & 0 & 0 & 0 & 0 & 0 & \lambda_y
    \end{pmatrix}
    ,
\end{align}
and numerically calculate the expectation of $\vecX_t=\begin{pmatrix} x_t \\ y_t \end{pmatrix}$ at $t=T=0.5$.
The constant parameters are set to $m_x^2=m_y^2=1$, $m_{xy}^2=0.5$, $\sigma_x=\sigma_y=0.5$, $\lambda_x=\lambda_y=0.5$, and the initial value are set to $x_0=1,y_0=0.5$
This \ac{SDE} corresponds to the overdamped Langevin equation $\drm\vecX_t=-\mathbf{\nabla}V(\vecX_t)+B\drm\vecW_t$ \cite{pavliotis2014stochastic} with the potential $V(\vecX_t)=\frac{1}{2}m_x^2 x_t^2+\frac{1}{2}m_y^2 y_t^2+m_{xy}^2 x_ty_t+\frac{1}{4}\lambda_x x_t^4+\frac{1}{4}\lambda_y y_t^4$.

First, as a reference, we integrate the \ac{SDE} \eqref{eq:SDEEg} in time up to $T$ by the \ac{EM} method~\cite{Maruyama1955} with time step width $\Delta t=0.0001$, using \texttt{DifferentialEquations.jl}~\cite{rackauckas2017differentialequations}.
The average of $\vecX_T$ over $10^7$ sample paths is $\overline{\vecX_T}=\begin{pmatrix} \overline{x_T} \\ \overline{y_T}\end{pmatrix}=\begin{pmatrix} 0.467513 \\ 0.158612 \end{pmatrix}$, which we use as a surrogate for the exact $\E[\vecX_T]$.

Then, we numerically solve Eq.~\eqref{eq:meanODE} to obtain $\widehat{\vecX_T}=\begin{pmatrix} \widehat{x_T} \\ \widehat{y_T}\end{pmatrix}\coloneqq\frac{1}{\gamma}\vecm_{\vecY^1}(T)$, where $\vecm_{\vecY^1}(T)$ is the first $N=2$ entries in $\vecm_{\veccalY}(T)$, as a \ac{CL}-based approximation of $\E[\vecX_T]$.
Here, we fix $\gamma=1/U$ with $U=5\sqrt{\sigma^2T}$ and change $J$, to see the convergence of \ac{CL}.
For this calculation, we use the 5/4 Runge-Kutta method \cite{TSITOURAS2011770} implemented as \texttt{Tsit5} in \texttt{DifferentialEquations.jl} with its default setting.

The result is as shown in Fig.~\ref{fig:res}, from which we see that the error in \ac{CL}-based approximation measured by $|\widehat{x_T}-\overline{x_T}|$ or $|\widehat{y_T}-\overline{y_T}|$, as expected, decreases exponentially as $J$ increases.
In fact, in this calculation, the sufficient condition for the convergence of \ac{CL} does not hold: the left-hand side of Eq.~\eqref{eq:CarlConvCond1} is $\left(\NC+\frac{1}{2}+\frac{1}{2\NC}\right)\gamma^2\sigma^2+\gamma^{1-s}\alpha_F \approx 9$ with $\sigma=\|B\|_{S_1}$ and $\alpha_F=\|F\|$ for the examined $J$, which is larger than $|\mu(A)|=0.5$.
Nevertheless, Fig.~\ref{fig:res} indicates that the \ac{CL}-based approach works at least in this case, which suggests its applicability beyond the currently obtained sufficient condition. 

\section{Summary}
In this paper, we have proposed a \ac{CL}-based quantum algorithm for high-dimensional nonlinear \acp{SDE}.
Starting from the \ac{SDE} in the form of Eq.~\eqref{eq:SDE}, we have derived the \ac{SDE}~\eqref{eq:SDEcalY} for the processes $\vecY^1_t,\ldots,\vecY^\NC_t$ approximating $\vecX_t,\ldots,\vecX_t^{\otimes \NC}$ and showed that $\E[\vecY^1_t]$, the expectation of $\vecY^1_t$, converges to $\E[\vecX_t]$ as the truncation order $J$ increases under a set of conditions \eqref{eq:CarlConvCond1} to \eqref{eq:XiniBd}, which means that the nonlinear and noise terms in the \ac{SDE} are sufficiently small.
Then, we have derived the \ac{ODE}~\eqref{eq:meanODE} that the expectations $\E[\vecY^j_t]$ follow, and applied the quantum \ac{ODE} solver in Ref.~\cite{Berry2024quantumalgorithm} to it, obtaining the history state, which encodes the values of $\E[\vecY^1_{t_n}]$ at time grid points $t_n$.
Finally, we have used overlap estimation to get an estimate of $\E\left[\sum_{n,i} C_{n,i}X^i_{t_n}\right]$, the expectation of a linear function of $\{\vecX_{t_n}\}$.
We have conducted a numerical demonstration on an \ac{SDE} related to overdamped Langevin dynamics, seeing that the \ac{CL} error actually converges exponentially with $J$ in this test case.

We have adopted a setting and a goal that serve as a first step in this work, and extending our method is an important direction for future work.
For example, it would be worthwhile to explore amplitude encoding of sample paths of $\vecX_t$, as in Ref.~\cite{miyamoto2026quantum}, since this would allow us to estimate expectations of more general test functions than linear ones.
As mentioned above, this requires extending the method in Ref.~\cite{miyamoto2026quantum} to \acp{SDE} with multiplicative noise, which would warrant a separate study.

\section*{Acknowledgements}

This work is supported by MEXT Quantum Leap Flagship Program (MEXT Q-LEAP) Grant No. JPMXS0120319794, JST COI-NEXT Program Grant No. JPMJPF2014, and JST [Moonshot R\&D] [Grant No. JPMJMS256J].

\section*{Code and data Availability}

The code and data used to generate the simulations and figures can be found at https://github.com/Koichi-Miyamoto/SDECarleman.

\section*{Authors' contributions}

Koichi Miyamoto, as the sole author of the manuscript, conceived, designed, and performed the analysis; he also wrote the paper.
ChatGPT was used for grammar checking and to provide assistance with proofs, all of which were independently verified by the author.

\bibliography{reference}
\bibliographystyle{unsrturl}

\appendix

\section{Proofs}

\subsection{Proof of Lemma \ref{lem:ItoIntegTailPr} \label{sec:ProofItoIntegTailPr}}

\begin{proof}
    We denote $\int_0^t \Phi_A(t,u) B(u) \drm\vecW_u$ by $\vecJ_t=(J^1_t,\ldots,J^N_t)^T$ for $t\in[0,T]$.
    This has mean $\mathbf{0}$ and, because of the It\^{o} isometry, covariance
    \begin{align}
        \E[\vecJ_t \vecJ^T_t] = \int_0^t \Phi_A(t,u)B(u)B^T(u)\Phi_A^T(t,u) \drm u.
        \label{eq:JCov}
    \end{align}
    We thus have
    \begin{align}
        \E[\|\vecJ_t\|^2] & = \int_0^t {\rm Tr}\left(\Phi_A(t,u)B(u)B^T(u)\Phi_A^T(t,u)\right) \drm u \nonumber \\
        & = \int_0^t {\rm Tr}\left(\Phi_A^T(t,u)\Phi_A(t,u)B(u)B^T(u)\right) \drm u \nonumber \\
        & \le \int_0^t \|\Phi_A^T(t,u)\Phi_A(t,u)\| \|B(u)B^T(u)\|_{S_1}\drm u \nonumber \\
        & \le \sigma^2t,
        \label{eq:JNormBd}
    \end{align}
    where we use Eqs.~\eqref{eq:TrBound}, \eqref{eq:e^AtNorm<1}, and \eqref{eq:BNorm}.
    
    For later use, let us bound $\E[\|\vecJ_{t^\prime}-\vecJ_t\|^2]$ for $0 \le t \le t^\prime \le T$ as well.
    We have
    \begin{align}
        \E[\|\vecJ_{t^\prime}-\vecJ_t\|^2] & = \E\left[\left\|\left(\Phi_A(t^\prime,t)-I\right)\vecJ_t+\int_t^{t^\prime} \Phi_A(t^\prime,\tau) B(\tau) d\vecW_\tau\right\|^2\right] \nonumber \\
        & = \E\left[\left\|\left(\Phi_A(t^\prime,t)-I\right)\vecJ_t\right\|^2\right] + \E\left[\left\|\int_t^{t^\prime} \Phi_A(t^\prime,\tau) B(\tau) d\vecW_\tau\right\|^2\right] \nonumber \\
        & = \int_0^t{\rm Tr}\left(\Phi_A(t,\tau)B(\tau)B^T(\tau)\Phi_A^T(t,\tau)\left(\Phi_A^T(t^\prime,t)-I\right)\left(\Phi_A(t^\prime,t)-I\right)\right)\drm \tau \nonumber \\
        & \quad + \int_t^{t^\prime}{\rm Tr}\left(\Phi_A(t^\prime,\tau)B(\tau)B^T(\tau)\Phi_A^T(t^\prime,\tau)\right)\drm \tau \nonumber \\
        & \le \int_0^t \|\Phi_A^T(t,\tau)\left(\Phi_A^T(t^\prime,t)-I\right)\left(\Phi_A(t^\prime,t)-I\right)\Phi_A(t,\tau)\| \|B(\tau)B^T(\tau)\|_{S_1} \drm\tau \nonumber \\
        & \quad + \int_t^{t^\prime} \|\Phi_A^T(t^\prime,\tau)\Phi_A(t^\prime,\tau)\|\|B(\tau)B^T(\tau)\|_{S_1}\drm\tau \nonumber \\
        & \le \left\|\Phi_A(t^\prime,t)-I\right\|^2 \sigma^2t + \sigma^2 (t^\prime-t),
        \label{eq:JtDiffNormBd}
    \end{align}
    where we use Eq.~\eqref{eq:TrBound} at the first inequality and Eqs.~\eqref{eq:e^AtNorm<1} and \eqref{eq:BNorm} at the second inequality.
    $\left\|\Phi_A(t^\prime,t)-I\right\|$ is bounded as $\left\|\Phi_A(t^\prime,t)-I\right\|\le\|\Phi_A(t^\prime,t)\|+\|I\|\le2$ for any $0 \le t \le t^\prime \le T$, and when $t^\prime-t\le\frac{1}{\alpha_A}$,
    \begin{align}
        \left\|\Phi_A(t^\prime,t)-I\right\| & = \left\|\sum_{n=1}^\infty \int_t^{t^\prime} \drm \tau_1 \int_t^{\tau_1} \drm \tau_2 \cdots \int_t^{\tau_{n-1}} \drm \tau_n A(\tau_1) \cdots A(\tau_n)\right\| \nonumber \\
        & \le \sum_{n=1}^\infty \int_t^{t^\prime} \drm \tau_1 \int_t^{\tau_1} \drm \tau_2 \cdots \int_t^{\tau_{n-1}} \drm \tau_n \|A(\tau_1)\| \cdots \|A(\tau_n)\| \nonumber \\
        & \le \sum_{n=1}^\infty \frac{1}{n!}\alpha_{A}^k(t^\prime-t)^k  \nonumber \\
        & = e^{\alpha_{A}(t^\prime-t)}-1 \nonumber \\
        & \le 2\alpha_{A}(t^\prime-t)
    \end{align}
    holds, where we use $e^x-1 \le 2x, \forall x\in[0,1]$ at the last inequality.
    Consequently, we have $\E[\|\vecJ_{t^\prime}-\vecJ_t\|^2] \le \frac{1}{2}(d_{[0,T]}(t,t^\prime))^2$, where
    \begin{align}
        d_{[0,T]}(t,t^\prime) \coloneqq
        \begin{dcases}
            \sqrt{2(1+4 \alpha_A^2 T(t^\prime-t)) \sigma^2(t^\prime-t)} & ; ~ t^\prime-t \le \frac{1}{\alpha_A} \\
            \sqrt{10\sigma^2T} & ; ~ {\rm otherwise}
        \end{dcases}
        .
    \end{align}

    We define
    \begin{align}
        U^\prime_{\delta} \coloneqq \sqrt{\frac{8}{3}\sigma^2T \log\left(\frac{4\NumIntItoIntegUB }{\delta}\right)}
        \label{eq:Uprdel}
    \end{align}
    and
    \begin{align}
        U^{\prime\prime}_{\delta} \coloneqq \left(2+\sqrt{\frac{4}{\NumIntItoIntegUB}\log \left(\frac{2\NumIntItoIntegUB }{\delta}\right)}\right)\sqrt{\sigma^2T},
        \label{eq:Uprprdel}
    \end{align}
    for which we have $U^\prime_{\delta} + U^{\prime\prime}_{\delta} < \frac{U_{\delta}}{2}$.
    Setting $\tau_n \coloneqq \frac{n}{\NumIntItoIntegUB}T$ for $n\in[\NumIntItoIntegUB ]_0$, we note that
    \begin{align}
        \left\{\sup_{t\in[0,T]} \|\vecJ_t\| \le \frac{U_{\delta}}{2}\right\} \supset  \bigcap_{n=0}^{\NumIntItoIntegUB -1} \left\{\|\vecJ_{\tau_n}\| \le U_\delta^\prime \right\} \cap \bigcap_{n=0}^{\NumIntItoIntegUB -1} \left\{\sup_{t\in\mathcal{I}_n} \|\Delta \vecJ^n_t\| \le U_{\delta}^{\prime\prime} \right\}
    \end{align}
    holds, where $\mathcal{I}_n \coloneqq [\tau_n,\tau_{n+1}]$ and $\Delta \vecJ^n_t \coloneqq \vecJ_t - \vecJ_{\tau_n}$.
    This implies that
    \begin{align}
        \Prob\left\{\sup_{t\in[0,T]} \|\vecJ_t\| \le \frac{U_{\delta}}{2}\right\}
        \ge &  \Prob\left(\bigcap_{n=0}^{\NumIntItoIntegUB -1} \left\{\|\vecJ_{\tau_n}\| \le U_\delta^\prime \right\} \cap \bigcap_{n=0}^{\NumIntItoIntegUB -1} \left\{\sup_{t\in\mathcal{I}_n} \|\Delta \vecJ^n_t\| \le U_\delta^{\prime\prime} \right\}\right) \nonumber \\
        = & 1 - \Prob\left(\bigcup_{n=0}^{\NumIntItoIntegUB -1} \left\{\|\vecJ_{\tau_n}\| > U_\delta^\prime \right\} \cup \bigcup_{n=0}^{\NumIntItoIntegUB -1} \left\{\sup_{t\in\mathcal{I}_n} \|\Delta \vecJ^n_t\|> U_\delta^{\prime\prime} \right\}\right) \nonumber \\
        \ge & 1 - \sum_{n=0}^{\NumIntItoIntegUB -1} \Prob\left\{\|\vecJ_{\tau_n}\| > U_\delta^\prime \right\} - \sum_{n=0}^{\NumIntItoIntegUB -1} \Prob\left\{\sup_{t\in\mathcal{I}_n} \|\Delta \vecJ^n_t\|> U_\delta^{\prime\prime} \right\}.
        \label{eq:supJtTail}
    \end{align}

    To bound the second term in Eq.~\eqref{eq:supJtTail}, we note that for each $n\in[\NumIntItoIntegUB -1]_0$, $\|\vecJ_{\tau_n}\|^2$ has the subexponential norm bounded as
    \begin{align}
        \big\|\|\vecJ_{\tau_n}\|^2\big\|_{\psi_1} = \left\|\sum_{i=1}^N (J^i_{\tau_n})^2\right\|_{\psi_1} \le \sum_{i=1}^N \left\|(J^i_{\tau_n})^2\right\|_{\psi_1} \le \sum_{i=1}^N \left\|J^i_{\tau_n}\right\|_{\psi_2}^2 \le \frac{8}{3} \sum_{i=1}^N \E[(J^i_{\tau_n})^2] \le \frac{8}{3} \sigma^2T,
    \end{align}
    where we use Lemma 2.8.5 and Exercise 2.24(d) in Ref.~\cite{vershynin2018high} at the second and third inequalities, respectively, and Eq.~\eqref{eq:JNormBd} at the last inequality.
    This means
    \begin{align}
        \E\left[\exp\left(\frac{\|\vecJ_{\tau_n}\|^2}{\frac{8}{3} \sigma^2T}\right)\right] \le \E\left[\exp\left(\frac{\|\vecJ_{\tau_n}\|^2}{\big\|\|\vecJ_{\tau_n}\|^2\big\|_{\psi_1}}\right)\right] \le 2.
        \label{eq:EexpJ^2}
    \end{align}
    By using Markov's inequality along with Eqs.~\eqref{eq:Uprdel} and \eqref{eq:EexpJ^2}, we have
    \begin{align}
        \Prob\left\{\|\vecJ_{\tau_n}\| \ge U_\delta^\prime \right\} & = \Prob\left\{\exp\left(\frac{\|\vecJ_{\tau_n}\|^2}{\frac{8}{3} \sigma^2T}\right) \ge \exp\left(\frac{(U_\delta^\prime)^2}{\frac{8}{3} \sigma^2T}\right) \right\} \nonumber \\
        & \le \exp\left(-\frac{(U_\delta^\prime)^2}{\frac{8}{3} \sigma^2T}\right)\E\left[\exp\left(\frac{\|\vecJ_{\tau_n}\|^2}{\frac{8}{3} \sigma^2T}\right)\right] \nonumber \\
        & \le \frac{\delta}{2\NumIntItoIntegUB }.
        \label{eq:PrJtaun}
    \end{align}

    Next, let us bound the third term in Eq.~\eqref{eq:supJtTail}.
    Noting that $\|\Delta\vecJ^n_t\|$ can be written as
    \begin{align}
        \|\Delta\vecJ^n_t\| = \sup_{\mathbf{u} \in \calS^N} J^{n,\mathbf{u}}_t,  J^{n,\mathbf{u}}_t \coloneqq \mathbf{u} \cdot \Delta\vecJ^n_t
    \end{align}
    and that $J^{n,\mathbf{u}}_t$ is a centered Gaussian with variance
    \begin{align}
      \E[(J^{n,\mathbf{u}}_t)^2] \le \E[\|\Delta\vecJ^n_t\|^2] \le \left(1+\frac{4 \alpha_A^2 T^2}{\NumIntItoIntegUB}\right)\frac{\sigma^2 T}{\NumIntItoIntegUB} \le \frac{2\sigma^2 T}{\NumIntItoIntegUB},
    \end{align}
    we have
    \begin{align}
        \Prob\left\{\sup_{t\in\mathcal{I}_n} \|\Delta\vecJ^n_t\| >  U_\delta^{\prime\prime}\right\} \le \exp\left(-\frac{\NumIntItoIntegUB \left(U_\delta^{\prime\prime}-\E\left[\sup_{t\in\mathcal{I}_n} \|\Delta\vecJ^n_t\|\right]\right)^2}{4\sigma^2 T}\right)
        \label{eq:BTISDelJ}
    \end{align}
    by the Borell–TIS inequality \cite[Theorem 2.1.1]{adler2007random} (originally from Ref.~\cite{borell1975brunn,cirel1976norms}).
    To bound $\E\left[\sup_{t\in\mathcal{I}_n} \|\Delta\vecJ^n_t\|\right]=\E\left[\sup_{(t,\mathbf{u})\in \mathcal{I}^\prime_n} J^{n,\mathbf{u}}_t\right]$, where $\mathcal{I}^\prime_n\coloneqq\mathcal{I}_n\times\calS_2^{N-1}$, we invoke Dudley's inequality 
    \cite[Theorem 10.1]{lifshits2012lectures}
    (originally from Ref.~\cite{dudley1967sizes}):
    \begin{align}
        \E\left[\sup_{(t,\mathbf{u})\in\mathcal{I}^\prime_n} J^{n,\mathbf{u}}_t\right] \le 4\sqrt{2} \int^\infty_0 \sqrt{\log \mathcal{N}(\mathcal{I}^\prime_n,d_{\mathcal{I}^\prime_n};\epsilon)} \drm\epsilon.
        \label{eq:EsupDelJ}
    \end{align}
    Here, $d_{\mathcal{I}^\prime_n}:\mathcal{I}^\prime_n \times \mathcal{I}^\prime_n \rightarrow \R$ is given by
    \begin{align}
        d_{\mathcal{I}^\prime_n}((t,\mathbf{u}),(t^\prime,\mathbf{u}^\prime)) = \sqrt{\E[(J^{n,\mathbf{u}}_t-J^{n,\mathbf{u}^\prime}_{t^\prime})^2]},    
    \end{align}
    and $\mathcal{N}(\mathcal{I}^\prime_n,d_{\mathcal{I}^\prime_n};\epsilon)$ denotes the smallest number of balls of radius $\epsilon$ in the metric $d_{\mathcal{I}^\prime_n}$ whose union covers $\mathcal{I}^\prime_n$.
    Then, using Eqs.~\eqref{eq:JNormBd} and \eqref{eq:JtDiffNormBd}, for $(t,\mathbf{u}),(t^\prime,\mathbf{u}^\prime) \in \mathcal{I}^\prime_n$ such that $t \le t^\prime$, we have
    \begin{align}
        \left(d_{\mathcal{I}^\prime_n}((t,\mathbf{u}),(t^\prime,\mathbf{u}^\prime))\right)^2
        = & \E[(\mathbf{u} \cdot \Delta\vecJ^n_t-\mathbf{u}^\prime \cdot \Delta\vecJ^n_{t^\prime})^2] \nonumber \\
        \le & 2\E[(\mathbf{u} \cdot (\Delta\vecJ^n_t-\Delta\vecJ^n_{t^\prime})^2]+2\E[((\mathbf{u}-\mathbf{u}^\prime)\cdot\Delta\vecJ^n_{t^\prime})^2] \nonumber \\
        \le & 2\E[\|\Delta\vecJ^n_t-\Delta\vecJ^n_{t^\prime}\|^2] + 2 \|\mathbf{u}-\mathbf{u}^\prime\|^2\E[\|\Delta\vecJ^n_{t^\prime}\|^2]\nonumber \\
        \le & (d_{[0,T]}(t,t^\prime))^2 + (d_{\calS^{N-1}}(\mathbf{u},\mathbf{u}^\prime))^2,
        \label{eq:metric}
    \end{align}
    where
    \begin{align}
        d_{\calS^{N-1}}(\mathbf{u},\mathbf{u}^\prime) \coloneqq 2\sqrt{\frac{\sigma^2 T}{\NumIntItoIntegUB}} \|\mathbf{u}-\mathbf{u}^\prime\|,
    \end{align}
    and this implies
    \begin{align}
        d_{\mathcal{I}_n^\prime}((t,\mathbf{u}),(t^\prime,\mathbf{u}^\prime)) \le d_{\rm UB} \coloneqq \sqrt{\frac{26 \sigma^2 T}{\NumIntItoIntegUB}}. 
    \end{align}
    Now, for $\epsilon\in(0,d_{\rm UB}]$, we take $\mathcal{P}_{\mathcal{I}_n}^\epsilon \coloneqq \{\tau^\epsilon_i\}_{i \in [N_\epsilon-1]_0}$, a set of $N_\epsilon$ points in $\mathcal{I}_n$, where 
    \begin{align}
        \tau^\epsilon_i \coloneqq ih_\epsilon,
        h_\epsilon \coloneqq \frac{\epsilon^2}{8\sigma^2},
        N_\epsilon\coloneqq \left\lceil \frac{T}{h_\epsilon \NumIntItoIntegUB } \right\rceil,
    \end{align}
    and $\mathcal{P}_{\calS_2^{N-1}}^\epsilon$, the smallest set of points on $\calS_2^{N-1}$ such that $N$-dimensional Euclidean balls of radius 
    \begin{align}
        \sqrt{\frac{\NumIntItoIntegUB}{8\sigma^2 T}}\epsilon 
    \end{align}
    centered at these points cover $\calS_2^{N-1}$.
    We can see that for any point in $\mathcal{I}_n^\prime$, there exists at least one element of $\mathcal{P}_{\mathcal{I}_n}^\epsilon\times\mathcal{P}_{\calS_2^{N-1}}^\epsilon$ such that the distance between these two points with respect to the metric $d_{\mathcal{I}^\prime_n}$ is at most $\epsilon$.
    This means that
    \begin{align}
        \mathcal{N}(\mathcal{I}^\prime_n,d_{\mathcal{I}^\prime_n};\epsilon) \le \left\lceil\frac{8\sigma^2T}{\epsilon^2\NumIntItoIntegUB } \right\rceil\left(\frac{2}{\epsilon}\sqrt{\frac{8 \sigma^2 T}{\NumIntItoIntegUB}}+1\right)^N
    \end{align}
    where we use $|\mathcal{P}_{\mathcal{I}_n}^\epsilon| = N_\epsilon$ and \cite[Corollary 4.2.11]{vershynin2018high}
    \begin{align}
        |\mathcal{P}_{\calS_2^{N-1}}^\epsilon|\le\left(\frac{2}{\epsilon}\sqrt{\frac{8 \sigma^2 T}{\NumIntItoIntegUB}}+1\right)^N.
    \end{align}
    For $\epsilon \le d_{\rm UB}$, some algebra yields
    \begin{align}
        \mathcal{N}(\mathcal{I}^\prime_n,d_{\mathcal{I}^\prime_n};\epsilon) \le \frac{26\sigma^2T}{\epsilon^2 \NumIntItoIntegUB }\left(\frac{4}{\epsilon}\sqrt{\frac{8 \sigma^2 T}{\NumIntItoIntegUB}}\right)^N.
    \end{align}
    We thus get
    \begin{align}
        \E\left[\sup_{(t,\mathbf{u})\in\mathcal{I}^\prime_n} J^{n,\mathbf{u}}_t\right]
        \le & 4\sqrt{2} \int^{d_{\rm UB}}_0 \sqrt{\log \left(\frac{26\sigma^2T}{\epsilon^2 \NumIntItoIntegUB }\right)+N\log\left(\frac{4}{\epsilon}\sqrt{\frac{8 \sigma^2 T}{\NumIntItoIntegUB}}\right)} \drm\epsilon \nonumber \\
        \le & 4\sqrt{2}\left[\int^{d_{\rm UB}}_0 \sqrt{\log \left(\frac{26\sigma^2T}{\epsilon^2 \NumIntItoIntegUB }\right)}\drm\epsilon + \int^{d_{\rm UB}}_0 \sqrt{N\log\left(\frac{4}{\epsilon}\sqrt{\frac{8 \sigma^2 T}{\NumIntItoIntegUB}}\right)}\drm\epsilon\right] \nonumber \\
        \le & 4\sqrt{2}\left[\sqrt{\frac{52\sigma^2T}{\NumIntItoIntegUB}}\int^1_0 \sqrt{\log \left(\frac{1}{x}\right)}\drm x + 4\sqrt{\frac{8N\sigma^2 T}{\NumIntItoIntegUB}}\int^{\frac{1}{4}\sqrt{\frac{5}{2}}}_0 \sqrt{\log\left(\frac{1}{x}\right)}\drm x\right] \nonumber \\
        \le & 8\sqrt{\frac{26\sigma^2T}{\NumIntItoIntegUB}}\frac{\sqrt{\pi}}{2}+64\sqrt{\frac{N\sigma^2 T}{\NumIntItoIntegUB}}\frac{1}{4}\sqrt{\frac{5}{2}}\left(\sqrt{\log\left(4\sqrt{\frac{2}{5}}\right)}+\frac{\sqrt{\pi}}{2}\right)
        \label{eq:supJnu}
    \end{align}
    where we use\footnote{This can be seen by noting
    \begin{align}
        \int^a_0 \sqrt{\log \left(\frac{1}{x}\right)}\drm x = \int^1_0 \sqrt{\log \left(\frac{1}{ay}\right)}a\drm y \le a\int^1_0 \left( \sqrt{\log\left(\frac{1}{a}\right)} + \sqrt{\log\left(\frac{1}{y}\right)}\right)\drm y = a\sqrt{\log\left(\frac{1}{a}\right)}+a\int^1_0 \sqrt{\log\left(\frac{1}{y}\right)}\drm y
         \nonumber
    \end{align}
    and
    \begin{align}
        \int^1_0 \sqrt{\log \left(\frac{1}{y}\right)}\drm y = \int_0^\infty \sqrt{z}e^{-z}\drm z=\Gamma\left(\frac{3}{2}\right)= \frac{\sqrt{\pi}}{2}, \nonumber
    \end{align}
    where we make the substitution $z=\log(1/y)$.
    }
    \begin{align}
        \int^a_0 \sqrt{\log \left(\frac{1}{x}\right)}\drm x \le a\left(\sqrt{\log\left(\frac{1}{a}\right)}+\frac{\sqrt{\pi}}{2}\right), ~ \forall a\in(0,1].
    \end{align}
    Due to the definition of $\NumIntItoIntegUB $, Eq.~\eqref{eq:supJnu} leads to
    \begin{align}
        \E\left[\sup_{(t,\mathbf{u})\in\mathcal{I}^\prime_n} J^{n,\mathbf{u}}_t\right] \le 2\sqrt{\sigma^2T}.
        \label{eq:EsupJnu}
    \end{align}
    We finally plug Eqs.~\eqref{eq:Uprprdel} and \eqref{eq:EsupJnu} into Eq.~\eqref{eq:BTISDelJ} to get
    \begin{align}
        \Prob\left\{\sup_{t\in\mathcal{I}_n} \|\Delta\vecJ^n_t\| >  U_\delta^{\prime\prime}\right\} \le \frac{\delta}{2\NumIntItoIntegUB }.
        \label{eq:PsupDelJ}
    \end{align}
    
    To complete the proof, we plug Eqs. \eqref{eq:PrJtaun} and \eqref{eq:PsupDelJ} into Eq. \eqref{eq:supJtTail}, which yields Eq.~\eqref{eq:OUsupTail}.

\end{proof}

\subsection{Proof of Lemma \ref{lem:CarErrTail} \label{sec:PrCarErrTail}}

We need some supporting lemmas.

\begin{lemma}
    If
    \begin{align}
        \mu(A(t))+\frac{\NC}{2}\|\hat{B}(t)\hat{B}^T(t)\|+\left(\frac{\NC}{2}+\frac{1}{2}+\frac{1}{2\NC}\right)\|\hat{B}(t)\hat{B}^T(t)\|_{\rm F}+\|\hat{F}(t)\| < 0
        \label{eq:muC1<0Cond}
    \end{align}
    holds for $t\in[0,T]$, then
    \begin{align}
        \mu(\tilde{\calC}(t)) \le 0
        \label{eq:muC1<0}
    \end{align}
    holds, where
    \begin{align}
        \tilde{\calC}(t) \coloneqq \calC(t) + \frac{1}{2}\sum_{l=1}^m \calB^l(t) (\calB^l(t))^T.
    \end{align}    
    \label{lem:mucalCtil<0}
\end{lemma}

\begin{proof}
    Note that $\tilde{\calC}(t)$ is equal to $\calC(t)$ except that $\tilde{\calC}(t)$'s $j$-th diagonal block is
    \begin{align}
        \tilde{\calA}^{j,j}(t)\coloneqq
        \begin{dcases}
            \calA^{1,1}(t) & ; ~ j=1 \\
            \calA^{j,j}(t)+\frac{1}{2}\sum_{l=1}^m \calA^{j,j-1,l}(t)(\calA^{j,j-1,l}(t))^T & ; ~ {\rm otherwise}
        \end{dcases}.
    \end{align}
    For each $j\in[\NC]$, a straightforward calculation gives
    \begin{align}
        \mu(\tilde{\calA}^{j,j}(t)) \le \mu\left(\calA^{j,j}(t)\right) + \frac{1}{2}\left\|\sum_{l=1}^m \calA^{j,j-1,l}(t) (\calA^{j,j-1,l}(t))^T\right\| \le j\mu\left(A(t)\right) + \frac{j^2}{2} \|\hat{B}(t)\hat{B}^T(t)\|
        \label{eq:mucalAjj}
    \end{align}
    \begin{align}
        \|\calA^{j,j-2}(t)\| \le \frac{j(j-1)}{2} \|{\rm vec}(\hat{B}(t)\hat{B}^T(t))\| = \frac{j(j-1)}{2} \|\hat{B}(t)\hat{B}^T(t)\|_{\rm F}, 
        \label{eq:calAjj-2Norm}
    \end{align}
    \begin{align}
        \|\calA^{j,j+s-1}(t)\| \le j\|\hat{F}(t)\|,
        \label{eq:calAjj+s-1}
    \end{align}
    and, by using these along with Eq.~\eqref{eq:muC1<0Cond}, we have
    \begin{align}
        \mu(\tilde{\calA}^{j,j}) + \frac{1}{2}\|\calA^{j,j-2}\| + \frac{1}{2}\|\calA^{j+2,j}\| + \frac{1}{2}\|\calA^{j,j+s-1}\| + \frac{1}{2}\|\calA^{j-s+1,j}\| < 0
        \label{eq:muC1<0CondTemp}
    \end{align}
    Here and hereafter, $\calA$ with superscripts out of range (e.g., $\calA^{1,-1}$) is regarded as $O$.

    Let
    \begin{align}
    \vecx=\begin{pmatrix} \vecx_1 \\ \vdots \\ \vecx_{\NC}  \end{pmatrix}\in\calS^{\calN}_2,
    \label{eq:xUnitBlock}
    \end{align}
    where for each $i\in[\NC]$, the $i$-th block $\vecx_i$ lies in $\R^{N^i}$.
    We then have an expression
    \begin{align}
        \|(I+h\tilde{\calC}) \vecx\| = \left(\sum_{j=1}^{\NC}\left\|(I+h\tilde{\calA}^{j,j})\vecx_j+h\calA^{j,j-2}\vecx_{j-2}+h\calA^{j,j+s-1}\vecx_{j+s-1}\right\|^2\right)^{1/2},
        \label{eq:I+hC1x}
    \end{align}
    interpreting $\vecx$ with a subscript out of range (e.g., $\vecx_{-1}$ and $\vecx_{\NC+1}$) as $\mathbf{0}$ here and hereafter.
    We consider the Taylor expansion of this with respect to sufficiently small $h>0$:
    \begin{align}
        & \left\|(I+h\tilde{\calA}^{j,j})\vecx_j+h\calA^{j,j-2}\vecx_{j-2}+h\calA^{j,j+s-1}\vecx_{j+s-1}\right\|^2 \nonumber \\
        \le & \left(\|I+h\tilde{\calA}^{j,j}\| \|\vecx_j\| + h\|\calA^{j,j-2}\|\|\vecx_{j-2}\|+h\|\calA^{j,j+s-1}\|\|\vecx_{j+s-1}\|\right)^2 \nonumber \\
        = & \left(1+h\mu(\tilde{\calA}^{j,j})+o(h) + h\|\calA^{j,j-2}\|\frac{\|\vecx_{j-2}\|}{ \|\vecx_j\|}+h\|\calA^{j,j+s-1}\|\frac{\|\vecx_{j+s-1}\|}{ \|\vecx_j\|}\right)^2 \|\vecx_j\|^2  \nonumber \\
        = & \left(1+2h\left(\mu(\tilde{\calA}^{j,j})+\|\calA^{j,j-2}\|\frac{\|\vecx_{j-2}\|}{ \|\vecx_j\|}+\|\calA^{j,j+s-1}\|\frac{\|\vecx_{j+s-1}\|}{ \|\vecx_j\|}\right)+o(h)\right)\|\vecx_j\|^2
    \end{align}
    if $\vecx_j \ne \mathbf{0}$, where we use Eq.~\eqref{eq:|I+hA|smallh}, and
    \begin{align}
        \left\|(I+h\tilde{\calA}^{j,j})\vecx_j+h\calA^{j,j-2}\vecx_{j-2}+h\calA^{j,j+s-1}\vecx_{j+s-1}\right\|^2 = o(h)
    \end{align}
    otherwise.
    These can be collectively written as
    \begin{align}
        & \left\|(I+h\tilde{\calA}^{j,j})\vecx_j+h\calA^{j,j-2}\vecx_{j-2}+h\calA^{j,j+s-1}\vecx_{j+s-1}\right\|^2 \le \nonumber \\
        & \quad \|\vecx_j\|^2 + 2h\left(\mu(\tilde{\calA}^{j,j})\|\vecx_j\|^2+\|\calA^{j,j-2}\|\|\vecx_{j-2}\|\|\vecx_j\|+\|\calA^{j,j+s-1}\|\|\vecx_{j+s-1}\|\|\vecx_j\|\right)+o(h).
        \label{eq:IneqFormuC}
    \end{align}
    By plugging this into Eq.~\eqref{eq:I+hC1x}, we get
    \begin{align}
        \|(I+h\tilde{\calC}) \vecx\|
        \le & 1 + \sum_{j=1}^{\NC} h \mu(\tilde{\calA}^{j,j})\|\vecx_j\|^2 +\sum_{j=1}^{\NC-s+1} h\|\calA^{j,j+s-1}\|\|\vecx_{j+s-1}\|\|\vecx_j\| \nonumber \\
        & ~ +\sum_{j=3}^{\NC} h\|\calA^{j,j-2}\|\|\vecx_{j-2}\|\|\vecx_j\| + o(h).
        \label{eq:|(I+hC1)x|Temp}
    \end{align}
    This can be transformed as
    \begin{align}
        & \frac{\|(I+h\tilde{\calC}) \vecx\|-1}{h} \nonumber \\
        \le & \sum_{j=1}^{\NC} \mu(\tilde{\calA}^{j,j})\|\vecx_j\|^2 +\sum_{j=1}^{\NC-s+1} \|\calA^{j,j+s-1}\|\|\vecx_{j+s-1}\|\|\vecx_j\|+\sum_{j=3}^{\NC} \|\calA^{j,j-2}\|\|\vecx_{j-2}\|\|\vecx_j\| + o(1)  \nonumber \\
        \le & \sum_{j=1}^{\NC-s+1} \|\calA^{j,j+s-1}\|\left(\|\vecx_{j+s-1}\|\|\vecx_j\|-\frac{1}{2}\|\vecx_j\|^2-\frac{1}{2}\|\vecx_{j+s-1}\|^2\right) \nonumber \\
        & + \sum_{j=3}^{\NC} \|\calA^{j,j-2}\|\left(\|\vecx_{j-2}\|\|\vecx_j\|-\frac{1}{2}\|\vecx_j\|^2-\frac{1}{2}\|\vecx_{j-2}\|^2\right) + o(1) \nonumber \\
        \le & o(1),
    \end{align}
    where we use Eq.~\eqref{eq:muC1<0CondTemp} at the second inequality, and the last inequality follows from the AM-GM inequality.
    Taking the supremum with respect to $\vecx$ and the limit $h \rightarrow 0^+$ in this inequality yields Eq.~\eqref{eq:muC1<0}.
\end{proof}

\begin{lemma}
    Under the same assumption as Lemma~\ref{lem:mucalCtil<0}, $\mu(\calC(t)) \le 0$ holds.
    \label{lem:muc1<0}
\end{lemma}

\begin{proof}
    Because of Eqs.~\eqref{eq:muAdd} and \eqref{eq:muRNProd}, we have
    \begin{align}
    \mu(\calC(t)) \le \mu(\tilde{\calC}(t)) + \frac{1}{2}\sum_{l=1}^m\mu\left(-\calB^l(t) (\calB^l(t))^T\right).
    \end{align}
    Eq.~\eqref{eq:mu2Formula} implies that $\mu\left(-\calB^l(t) (\calB^l(t))^T\right)\le0$.
    Thus, when $\mu(\tilde{\calC}(t)) \le 0$ holds, $\mu(\calC(t)) \le 0$ also holds.
\end{proof}

\begin{lemma}
    In the setting of Lemma~\ref{lem:CarErrTail},
    \begin{align}
        \E[\|\vecY^1_t\|^2] \le 2
    \end{align}
    holds.
    \label{lem:Y2ndMom}
\end{lemma}

\begin{proof}

    From Eq.~\eqref{eq:SDEcalY}, we have
    \begin{align}
        \drm \|\veccalY_t\|^2 & = 2\veccalY_t \cdot (\calC(t) \veccalY_t + \vecc(t)) \drm t + \sum_{l=1}^m \|\calB^l(t) \veccalY_t + \vecb^l(t)\|^2 \drm t + 2\veccalY_t \cdot \sum_{l=1}^m (\calB^l(t) \veccalY_t + \vecb^l(t))\drm W^l_t \nonumber \\
        & = 2\veccalY_t \cdot (\tilde{\calC}(t) \veccalY_t + \vecc(t)) \drm t + \sum_{l=1}^m \|\vecb^l(t)\|^2 \drm t + 2\veccalY_t \cdot \sum_{l=1}^m (\calB^l(t) \veccalY_t + \vecb^l(t))\drm W^l_t.
    \end{align}
    This leads to
    \begin{align}
        \drm \|\veccalY_t\|^2 \le & 2\mu(\tilde{\calC}(t)) \|\veccalY_t\|^2 \drm t + 2\left\|\vecc(t)\right\| \|\veccalY_t\| \drm t + \sum_{l=1}^m \|\vecb^l(t)\|^2 \drm t + 2\veccalY_t \cdot \sum_{l=1}^m (\calB^l(t) \veccalY_t + \vecb^l(t))\drm W^l_t,
    \end{align}
    where we use Eq.~\eqref{eq:mu2Formula}.
    We note that under Eq.~\eqref{eq:CarlConvCond1}, Eq.~\eqref{eq:muC1<0Cond} holds for every $t\in[0,T]$, and so does $\mu(\tilde{\calC}(t)) \le 0$.
    We also note
    \begin{align}
        \left\|\vecc(t)\right\| \le \|\hat{B}(t)\hat{B}^T(t)\|_{\rm F} \le \left(\frac{c\sigma}{U}\right)^2
    \end{align}
    and
    \begin{align}
        \sum_{l=1}^m \|\vecb^l(t)\|^2 \le \|\hat{B}(t)\|_{\rm F}^2 \le \left(\frac{c\sigma}{U}\right)^2.
    \end{align}
    Using these, we get
    \begin{align}
        \drm \|\veccalY_t\|^2 \le \left(\frac{c\sigma}{U}\right)^2(2\|\veccalY_t\|+1) \drm t + 2\veccalY_t \cdot \sum_{l=1}^m (\calB^l(t) \veccalY_t + \vecb^l(t))\drm W^l_t
    \end{align}
    and thus the following bound on the right derivative
    \begin{align}
        \partial^+_t \E[\|\veccalY_t\|^2]  \le \left(\frac{c\sigma}{U}\right)^2 (2\E[\|\veccalY_t\|]+1). 
    \end{align}
    Using H\"{o}lder's inequality and the inequality $a+b \le \sqrt{2(a^2+b^2)},~\forall a,b>0$, we obtain
    \begin{align}
        \partial^+_t \E[\|\veccalY_t\|^2]  \le \left(\frac{c\sigma}{U}\right)^2 \sqrt{8\E[\|\veccalY_t\|^2]+2}, 
    \end{align}
    and this yields
    \begin{align}
        \E[\|\veccalY_t\|^2]  \le \frac{1}{8}\left(4\left(\frac{c\sigma}{U}\right)^2T+\sqrt{8\|\veccalY_0\|^2+2}\right)^2-\frac{1}{4} \le 2, 
    \end{align}
    where we use Eq.~\eqref{eq:U}, $c<1$, and
    \begin{align}
        \left\|\veccalY_0\right\|=\sum_{j=1}^J\left\|\hat{\vecX}_0\right\|^j \le \sum_{j=1}^J\left(\frac{c}{2}\right)^j \le c < 1.
    \end{align}
    Since $\|\vecY^1_t\|^2 \le \|\veccalY_t\|^2$, the claim follows.

\end{proof}

\begin{lemma}
    For positive numbers $a$, $b$, and $x$ satisfying $x\le\frac{b}{2\log(a/b)}$ and $\frac{b}{a} \le 1$, 
    $x\log\left(\frac{a}{x}\right) \le b$ holds.
    \label{lem:loga/x}
\end{lemma}

\begin{proof}
    This can be seen by simple calculus.
\end{proof}

\begin{proof}[Proof of Lemma \ref{lem:CarErrTail}]
    Letting $\hat{\vecX}^{(k)}_t=(\hat{X}^{(k),1}_t,\ldots,\hat{X}^{(k),N}_t)^T,k\in[\SampCarErrProof]$ be $\SampCarErrProof$ i.i.d. copies of $\hat{\vecX}_t$, $\frac{1}{\SampCarErrProof}\sum_{k=1}^\SampCarErrProof\hat{\vecX}^{(k)}_t$ has the mean $\E[\hat{\vecX}_t]$, and 
    \begin{align}
        \E\left[\left\|\frac{1}{\SampCarErrProof}\sum_{k=1}^\SampCarErrProof\hat{\vecX}^{(k)}_t-\E[\hat{\vecX}_t]\right\|^2\right] & = \sum_{i=1}^N \E\left[\left(\frac{1}{\SampCarErrProof}\sum_{k=1}^\SampCarErrProof\hat{X}^{(k),i}_t-\E[\hat{X}^i_t]\right)^2\right] \nonumber \\
        & = \sum_{i=1}^N \frac{1}{\SampCarErrProof} {\rm Var}(\hat{X}^i_t)  \nonumber \\
        & \le \sum_{i=1}^N \frac{1}{\SampCarErrProof} \E[(\hat{X}^i_t)^2]  \nonumber \\
        & \le \frac{\E[\|\hat{\vecX}_t\|^2]}{\SampCarErrProof} \nonumber \\
        & \le \frac{c^2\sigma_X^2}{\SampCarErrProof U^2}  \nonumber \\
        & \le c^2\sigma_X^2 \cdot \frac{64\sigma^2T\hat{\epsilon}^2}{45c^2\sigma_X^2} \cdot \frac{1}{\left(8\sqrt{\sigma^2T}\right)^2} \nonumber \\
        & \le \frac{\hat{\epsilon}^2}{45}
    \end{align}
    holds.
    Thus, because of Chebyshev's inequality,
    \begin{align}
        \Prob\left\{\left\|\frac{1}{\SampCarErrProof}\sum_{k=1}^\SampCarErrProof\hat{\vecX}^{(k)}_t - \E[\hat{\vecX}_t]\right\|\ge\frac{\hat{\epsilon}}{3}\right\}\le\frac{1}{5}
        \label{eq:XAvErr}
    \end{align}
    holds.
    Besides, letting $\vecY^{1,(k)}_t=(Y^{1,(k),1}_t,\ldots,Y^{1,(k),N}_t)^T,k\in[\SampCarErrProof]$ be $\SampCarErrProof$ i.i.d. copies of $\vecY^1_t$, $\frac{1}{\SampCarErrProof}\sum_{k=1}^\SampCarErrProof\vecY^{1,(k)}_t$ has the mean $\E[\vecY^1_t]$ and satisfies
    \begin{align}
        \E\left[\left\|\frac{1}{\SampCarErrProof}\sum_{k=1}^\SampCarErrProof\vecY^{1,(k)}_t-\E[\vecY^1_t]\right\|^2\right] \le \frac{\E[\|\vecY^1_t\|^2]}{\SampCarErrProof} \le \frac{2}{\SampCarErrProof} \le \frac{\hat{\epsilon}^2}{45}
    \end{align}
    where we use Lemma~\ref{lem:Y2ndMom}, and thus Chebyshev's inequality gives
    \begin{align}
        \Prob\left\{\left\|\frac{1}{\SampCarErrProof}\sum_{k=1}^\SampCarErrProof\vecY^{1,(k)}_t-\E[\vecY^1_t]\right\|\ge\frac{\hat{\epsilon}}{3}\right\}\le\frac{1}{5}.
        \label{eq:YAvErr}
    \end{align}

    Let us consider an $\R^\calN$-valued random process $\widetilde{\veccalX}=(\tilde{\mathcal{X}}^1,\ldots,\tilde{\mathcal{X}}^\calN)^T$ that follows the \ac{SDE}
    \begin{align}
    \drm \widetilde{\veccalX}_t = (\calC(t) \widetilde{\veccalX}_t + \vecc(t)) \drm t + \sum_{l=1}^m \left(\calB^l(t) \widetilde{\veccalX}_t + \vecb^l(t)\right) \drm W^l_t + \veccalE_t \drm t
    \end{align}
    and has the initial value $\widetilde{\veccalX}_0 = \veccalY_0$.
    Here, an $\R^\calN$-valued random process $\boldsymbol{\mathcal{E}}$ is defined by
    \begin{align}
        \veccalE_t =
        \begin{pmatrix}
            \calE^1_t \\
            \vdots \\
            \calE^\calN_t
        \end{pmatrix}
        \coloneqq
        \begin{pmatrix}
            \mathbf{0}_N \\
            \vdots \\
            \mathbf{0}_{N^{\NC-s+1}} \\
            \calA^{\NC-s+2,\NC+1}(t) \vecXtrunc_t^{\otimes(\NC+1)} \\
            \vdots \\
            \calA^{\NC,\NC+s-1}(t) \vecXtrunc_t^{\otimes(\NC+s-1)}
        \end{pmatrix}
    \end{align}
    with
    \begin{align}
        \vecXtrunc_t \coloneqq \frac{\hat{\vecX}_t}{\max\left\{1,\frac{\|\hat{\vecX}_t\|}{c}\right\}},
        \label{eq:hatvecX}
    \end{align}
    which obviously satisfies
    \begin{align}
        \left\|\vecXtrunc_t\right\| \le c.
        \label{eq:XtilNormUB}
    \end{align}
    Note that, if $\vecXtrunc$ were replaced with $\vecX$, $\widetilde{\veccalX}$ would coincide with $\veccalX\coloneqq\begin{pmatrix} \hat{\vecX}^{\otimes 1} \\ \vdots \\ \hat{\vecX}^{\otimes \NC} \end{pmatrix}$.
    
    The norm of $\veccalE_t$ is bounded as
    \begin{align}
        \|\veccalE_t\| & \le
        \sqrt{\sum_{j=\NC-s+2}^\NC \|\calA^{j,j+s-1}(t)\|^2\cdot\|\vecXtrunc_t\|^{2(j+s-1)}}
        \nonumber \\
        & \le \sqrt{s-1}\NC\|\hat{F}(t)\| c^{\NC+1} \nonumber \\
        & \le \sqrt{s-1}\NC\alpha_F\left(\frac{U}{c}\right)^{s-1} c^{\NC+1} \nonumber \\
        & \le \sqrt{s-1}\eta \NC c^{\NC+1} \eqqcolon \calU,
        \label{eq:calENorm}
    \end{align}
    where we use Eqs.~\eqref{eq:calAjj+s-1} and \eqref{eq:XtilNormUB} at the second inequality and Eq.~\eqref{eq:CarlConvCond1} at the last inequality.
    Denoting $\veccalY_t - \widetilde{\veccalX}_t$ by $\widetilde{\veceta}_t=(\tilde{\eta}^1_t,\ldots,\tilde{\eta}^{\calN}_t)^T$, it follows
    \begin{align}
        \drm \widetilde{\veceta}_t = \calC(t) \widetilde{\veceta}_t \drm t + \sum_{l=1}^m \calB^l(t) \widetilde{\veceta}_t \drm W^l_t - \boldsymbol{\mathcal{E}}_t \drm t,
    \end{align}
    and this yields
    \begin{align}
        \drm \|\widetilde{\veceta}_t\|^2 & = 2 \widetilde{\veceta}_t \cdot (\tilde{\calC}(t) \widetilde{\veceta}_t  - \boldsymbol{\mathcal{E}}_t) \drm t + \widetilde{\veceta}_t \cdot \sum_{l=1}^m \calB^l(t) \widetilde{\veceta}_t \drm W^l_t \nonumber \\
        & \le 2 \|\widetilde{\veceta}_t\| \|\boldsymbol{\mathcal{E}}_t\| \drm t + \widetilde{\veceta}_t \cdot \sum_{l=1}^m \calB^l(t) \widetilde{\veceta}_t \drm W^l_t,
    \end{align}    
    where we use $\mu(\tilde{\calC}(t))\le0$.
    We thus have
    \begin{align}
        \partial^+_t \E[\|\widetilde{\veceta}_t\|^2] \le 2\E[ \|\widetilde{\veceta}_t\| \|\boldsymbol{\mathcal{E}}_t\|] \le 2 \calU \sqrt{\E[\|\widetilde{\veceta}_t\|^2]},
    \end{align}
    where we use Eq.~\eqref{eq:calENorm} and H\"{o}lder's inequality.
    This leads to
    \begin{align}
        \E[\|\widetilde{\veceta}_t\|^2] \le (\calU t)^2,
        \label{eq:etai2ndMom}
    \end{align}
    from which Markov's inequality yields 
    \begin{align}
        \Prob\left\{\|\widetilde{\veceta}_t\| \ge \sqrt{5\SampCarErrProof} \calU t\right\} \le \frac{1}{5\SampCarErrProof}.
    \end{align}
    By some algebra, we see that taking
    \begin{align}
        \NC \ge \frac{-W_{-1}(-\tilde{\epsilon})}{\log \frac{1}{c}},
        \label{eq:NCCondTemp}
    \end{align}
    where $W_{-1}$ is the secondary branch of the Lambert $W$ function, yields $\sqrt{5\SampCarErrProof} \calU t \le \frac{1}{3}\hat{\epsilon}$ and thus
    \begin{align}
        \Prob\left\{\|\vecY_t-\tilde{\vecX}_t\| \ge \frac{\hat{\epsilon}}{3}\right\} \le \frac{1}{5\SampCarErrProof},
        \label{eq:YXtilDiff}
    \end{align}
    where $\tilde{\vecX}_t\in\R^N$ consists of the first $N$ elements of $\widetilde{\veccalX}_t$.
    Eq.~\eqref{eq:YXtilDiff} implies
    \begin{align}
        \Prob\left\{\left\|\vecY^{1,(k)}_t-\tilde{\vecX}^{(1)}_t\right\| \le \frac{\hat{\epsilon}}{3},\ldots,\left\|\vecY^{1,(\SampCarErrProof)}_t-\tilde{\vecX}^{(\SampCarErrProof)}_t\right\| \le \frac{\hat{\epsilon}}{3}\right\} \ge \frac{4}{5},
    \end{align}
    where $\tilde{\vecX}^{(1)},\ldots,\tilde{\vecX}^{(\SampCarErrProof)}$ are $\SampCarErrProof$ i.i.d. copies of $\tilde{\vecX}$, and thus
    \begin{align}
        \Prob\left\{\left\|\frac{1}{\SampCarErrProof}\sum_{k=1}^\SampCarErrProof\vecY^{1,(k)}_t-\frac{1}{\SampCarErrProof}\sum_{k=1}^\SampCarErrProof\tilde{\vecX}^{(k)}_t\right\|\le\frac{\hat{\epsilon}}{3}\right\} \ge \frac{4}{5}.
        \label{eq:YAvXTilAvErr}
    \end{align}
    Using the inequality $0<-W_{-1}(-e^{-u-1})<1+\sqrt{2u}+u, \forall u>0$ \cite[Theorem 1]{Chatzigeorgiou2013}, we see that $\NC$ in Eq.~\eqref{eq:NCCond} satisfies Eq.~\eqref{eq:NCCondTemp}, leading to Eq.~\eqref{eq:YAvXTilAvErr}.

    Lemma~\ref{lem:XtTail} implies that under Eq.~\eqref{eq:CarlConvCond2},
    \begin{align}
        \Prob\left\{\|\hat{\vecX}_t\| \le c,\forall t\in[0,T]\right\} \ge 1-\frac{1}{5\SampCarErrProof}
    \end{align}
    holds.
    Since $\vecXtrunc_t=\hat{\vecX}_t$ if $\|\hat{\vecX}_t\| \le c$, we have
    \begin{align}
        \Prob\left\{\vecXtrunc_t=\hat{\vecX}_t,\forall t\in[0,T]\right\} \ge 1-\frac{1}{5\SampCarErrProof},
    \end{align}
    which leads to
    \begin{align}
        \Prob\left\{\widetilde{\veccalX}_t=\veccalX_t,\forall t\in[0,T]\right\} \ge 1-\frac{1}{5\SampCarErrProof},
        \label{eq:ProbXtil=X}
    \end{align}
    especially
    \begin{align}
        \Prob\left\{\tilde{\vecX}_t=\hat{\vecX}_t,\forall t\in[0,T]\right\} \ge 1-\frac{1}{5\SampCarErrProof}.
    \end{align}
    This implies
    \begin{align}
        \Prob\left\{\tilde{\vecX}^{(1)}_t=\hat{\vecX}^{(1)}_t,\ldots,\tilde{\vecX}^{(\SampCarErrProof)}_t=\hat{\vecX}^{(\SampCarErrProof)}_t\right\} \ge \frac{4}{5}
    \end{align}
    for any $t\in[0,T]$, and thus
    \begin{align}
        \Prob\left\{\frac{1}{\SampCarErrProof}\sum_{k=1}^\SampCarErrProof\tilde{\vecX}^{(k)}_t=\frac{1}{\SampCarErrProof}\sum_{k=1}^\SampCarErrProof\hat{\vecX}^{(k)}_t\right\} \ge \frac{4}{5}.
        \label{eq:XtilAvXAvErr}
    \end{align}

    By combining Eqs. \eqref{eq:XAvErr}, \eqref{eq:YAvErr}, \eqref{eq:YAvXTilAvErr}, and \eqref{eq:XtilAvXAvErr}, we see that for any $t\in[0,T]$,
    \begin{align}
        \left\|\E[\vecY^1_t]-\E[\hat{\vecX}_t]\right\|\le \hat{\epsilon}
        \label{eq:YhatXErr}
    \end{align}
    holds with a nonzero probability (at least $\frac{1}{5}$) under $\Prob$, which means that it holds with certainty, since both $\E[\vecY^1_t]$ and $\E[\hat{\vecX}_t]$ are nonrandom.

    Finally,
    let us see that 
    \begin{align}
        \frac{\hat{\epsilon}}{\gamma_{\epsilon,c}}\le\epsilon,    
    \end{align}
    by which Eq.~\eqref{eq:YhatXErr} translates into Eq.~\eqref{eq:etaNorm}.
    In fact, this can be seen as
    \begin{align}
        \frac{\hat{\epsilon}}{\gamma_{\epsilon,c}} & = \frac{2\sqrt{\sigma^2T}\hat{\epsilon}}{c}\left(\sqrt{3 \log\left(20\NumIntItoIntegUB \left\lceil\max\left\{\frac{c^2 \sigma_X^2}{64\sigma^2 T},2\right\}\times\frac{45}{\hat{\epsilon}^2}\right\rceil\right)}+2\right) \nonumber \\
        & \le \frac{6\sqrt{\sigma^2T}\hat{\epsilon}}{c}\log\left(20\NumIntItoIntegUB  \max\left\{\frac{45c^2 \sigma_X^2}{64\sigma^2 T},90\right\} \frac{1}{\hat{\epsilon}^2}\right) \nonumber \\
        & \le \frac{12\sqrt{\sigma^2T}\hat{\epsilon}}{c}\log\left(\left(20\NumIntItoIntegUB  \max\left\{\frac{45c^2 \sigma_X^2}{64\sigma^2 T},90\right\}\right)^{\frac{1}{2}}\frac{1}{\hat{\epsilon}}\right) \nonumber \\
        & \le \frac{12\sqrt{\sigma^2T}}{c} \cdot \frac{c\epsilon}{12\sqrt{\sigma^2T}} \nonumber \\
        & \le \epsilon
    \end{align}
    Here, the second-to-last inequality follows from the definition of $\hat{\epsilon}$ and Lemma~\ref{lem:loga/x}.
    
\end{proof}

\subsection{Lemmas for Theorem \ref{th:HistStatemY} \label{sec:Lem4ThHistStatemY}}

\begin{lemma}
    \begin{align}
        \|\calC(t)\| \le \sqrt{3} \left(\NC\|A(t)\| + \frac{\NC(\NC-1)\gamma^2}{2} \|B(t)B^T(t)\|_{\rm F} + \frac{\NC}{\gamma^{s-1}}\|F(t)\|\right)
    \end{align}
    holds for any $t\in[0,T]$.
    \label{lem:calCpNorm}
\end{lemma}

\begin{proof}
    Letting $\vecx$ be as in Eq.~\eqref{eq:xUnitBlock}, we have
    \begin{align}
        & \left\|\calA^{j,j}(t)\vecx_j+\calA^{j,j-2}(t)\vecx_{j-2}+\calA^{j,j+s-1}(t)\vecx_{j+s-1}\right\|^2\nonumber \\
        \le & \left(\|\calA^{j,j}(t)\| \|\vecx_j\| + \|\calA^{j,j-2}(t)\|\|\vecx_{j-2}\|+\|\calA^{j,j+s-1}(t)\|\|\vecx_{j+s-1}\|\right)^2 \nonumber \\
        \le & 3\left(\|\calA^{j,j}(t)\|^2\|\vecx_j\|^2+\|\calA^{j,j-2}(t)\|^2\|\vecx_{j-2}\|^2+\|\calA^{j,j+s-1}(t)\|^2\|\vecx_{j+s-1}\|^2\right) \nonumber \\
        \le & 3\left((\NC\left\|A(t)\right\|)^2\|\vecx_j\|^2+\left(\frac{\NC(\NC-1)}{2} \|\hat{B}(t)\hat{B}^T(t)\|_{\rm F}\right)^2\|\vecx_{j-2}\|^2+(\NC\|\hat{F}(t)\|)^2\|\vecx_{j+s-1}\|^2\right)
    \end{align}
    for any $j\in[\NC]$, where we use Jensen's inequality at the second inequality, and the last inequality follows from Eqs.~\eqref{eq:calAjj-2Norm} and \eqref{eq:calAjj+s-1} and
    \begin{align}
        \|\calA^{j,j}(t)\| \le \sum_{k=1}^j \left\|I^{\otimes (k-1)} \otimes A(t) \otimes I^{\otimes (j-k)}\right\| \le \sum_{k=1}^j \left\|A(t)\right\| \le j\left\|A(t)\right\|.
    \end{align}
    We thus have
    \begin{align}
        & \sum_{j=1}^{\NC}\left\|\calA^{j,j}(t)\vecx_j+\calA^{j,j-2}(t)\vecx_{j-2}+\calA^{j,j+s-1}(t)\vecx_{j+s-1}\right\|^2 \nonumber \\
        \le & 3 \sum_{j=1}^{\NC} \left((\NC\left\|A(t)\right\|)^2\|\vecx_j\|^2+\left(\frac{\NC(\NC-1)}{2} \|\hat{B}(t)\hat{B}^T(t)\|_{\rm F}\right)^2\|\vecx_{j-2}\|^2+(\NC\|\hat{F}(t)\|)^2\|\vecx_{j+s-1}\|^2\right) \nonumber \\
        \le & 3 \left((\NC\left\|A(t)\right\|)^2+\left(\frac{\NC(\NC-1)}{2} \|\hat{B}(t)\hat{B}^T(t)\|_{\rm F}\right)^2+(\NC\|\hat{F}(t)\|)^2\right),
    \end{align}
    where we use $\sum_{j=1}^{\NC} \|\vecx_j\|^2=1$, and then
    \begin{align}
        \|\calC \vecx\|
        &=\left(\sum_{j=1}^{\NC}\left\|\calA^{j,j}(t)\vecx_j+\calA^{j,j-2}(t)\vecx_{j-2}+\calA^{j,j+s-1}(t)\vecx_{j+s-1}\right\|^2\right)^{1/2} \nonumber \\
        &\le \sqrt{3} \left((\NC\left\|A(t)\right\|)^2+\left(\frac{\NC(\NC-1)}{2} \|\hat{B}(t)\hat{B}^T(t)\|_{\rm F}\right)^2+(\NC\|\hat{F}(t)\|)^2\right)^{1/2} \nonumber \\
        & \le \sqrt{3} \left(\NC\left\|A(t)\right\| + \frac{\NC(\NC-1)}{2} \|\hat{B}(t)\hat{B}^T(t)\|_{\rm F} + \NC\|\hat{F}(t)\|\right) \nonumber \\
        & \le \sqrt{3} \left(\NC\left\|A(t)\right\| + \frac{\NC(\NC-1)\gamma^2}{2} \|B(t)B^T(t)\|_{\rm F} + \frac{\NC}{\gamma^{s-1}}\|F(t)\|\right).
    \end{align}
    This implies that the claim holds.
\end{proof}

\begin{lemma}
    \begin{align}
        \left\|\frac{\drm \calC(t)}{\drm t}\right\| \le \sqrt{3} \left(\NC\left\|\frac{\drm A}{\drm t}(t)\right\| + \NC(\NC-1)\gamma^2\left\|\frac{\drm B}{\drm t}B(t) B^T(t)\right\|_{\rm F} + \frac{\NC}{\gamma^{s-1}}\left\|\frac{\drm F}{\drm t}(t)\right\|\right)
    \end{align}
    holds for any $t\in[0,T]$.
    \label{lem:dcalC/dtpNorm}
\end{lemma}

\begin{proof}
    Noting
    \begin{align}
        \left\|\frac{\drm}{\drm t}\calA^{j,j}(t)\right\|
        \le j\left\|\frac{\drm A}{\drm t}(t)\right\|,
    \end{align}
    \begin{align}
        \left\|\frac{\drm}{\drm t}\calA^{j,j-2}(t)\right\| \le j(j-1) \left\|\left(\frac{\drm \hat{B}}{\drm t}(t)\hat{B}^T(t)\right)\right\|_{\rm F}, 
    \end{align}
    and
    \begin{align}
        \left\|\frac{\drm}{\drm t}\calA^{j,j+s-1}(t)\right\| \le j\left\|\frac{\drm \hat{F}}{\drm t}(t)\right\|,
    \end{align}
    we obtain the claim by a discussion similar to Lemma~\ref{lem:calCpNorm}.
\end{proof}

\subsection{Rest of the proof of Theorem~\ref{th:MeanEstim} \label{app:PrME}}

\begin{proof}[Rest of the proof of Theorem~\ref{th:MeanEstim}]
~ \\
\noindent \underline{\textbf{Accuracy}}

We note
\begin{align}
    \epsilon_{\rm C} \sqrt{r} \|C\|_{\rm F} & = \epsilon_{\rm C} r \bar{C} \nonumber \\
    & \le \epsilon_{\rm C} \bar{C} \max\left\{\lceil 3 J_{\epsilon_{\rm C},c} \alpha_A T \rceil, 7\right\} \nonumber \\
    & \le \epsilon_{\rm C} \bar{C} \max\left\{\frac{24}{7} J_{\epsilon_{\rm C},c} \alpha_A T, 7\right\} \nonumber \\
    & \le \max\left\{\frac{24}{7} \alpha_A T \bar{C} \epsilon_{\rm C} J_{\epsilon_{\rm C},c}, \frac{\epsilon}{3}\right\}
    \label{eq:epsCSqrtrC}
\end{align}
and
\begin{align}
    \epsilon_{\rm C} J_{\epsilon_{\rm C},c} = \epsilon_{\rm C} \left\lceil\frac{1}{\log \left(\frac{1}{c}\right)}\left(\log\left(\frac{1}{\tilde{\epsilon}_{\rm C}}\right)+\sqrt{2\left(\log\left(\frac{1}{\tilde{\epsilon}_{\rm C}}\right)-1\right)}\right)\right\rceil \le \frac{2}{\log \left(\frac{1}{c}\right)} \epsilon_{\rm C}\log\left(\frac{1}{\tilde{\epsilon}_{\rm C}}\right)+1,
    \label{eq:epsCJ1/2}
\end{align}
where
\begin{align}
    \tilde{\epsilon}_{\rm C} \coloneqq \frac{\alpha \epsilon_{\rm C}}{\log\left(\frac{\beta}{\epsilon_{\rm C}}\right)},~
    \alpha \coloneqq \frac{c}{24\sqrt{\sigma^2T}}\min\left\{\frac{\log \left(1/c\right)}{9\sqrt{5}\eta T},1\right\},~
    \beta \coloneqq \frac{12\sqrt{\sigma^2T}}{c}\left(20\NumIntItoIntegUB  \max\left\{\frac{45c^2 \sigma_X^2}{64\sigma^2 T},90\right\}\right)^{\frac{1}{2}}.
\end{align}
Simple algebra yields
\begin{align}
    \log\left(\frac{1}{\tilde{\epsilon}_{\rm C}}\right) \le 2 \log\left(\frac{1}{\alpha \epsilon_{\rm C}}\right),
\end{align}
and combining this with Eq.~\eqref{eq:epsCJ1/2} yields
\begin{align}
    \epsilon_{\rm C} J_{\epsilon_{\rm C},c} \le \frac{4}{\log \left(\frac{1}{c}\right)} \epsilon_{\rm C}\log\left(\frac{1}{\alpha \epsilon_{\rm C}}\right)+\epsilon_{\rm C}.
\end{align}
This further translates into
\begin{align}
    \epsilon_{\rm C} J_{\epsilon_{\rm C},c} \le \frac{4}{\log \left(\frac{1}{c}\right)} \cdot \frac{7\epsilon \log \left(\frac{1}{c}\right)}{324 \bar{C} \max\{\alpha_A T,1\}} + \frac{7\epsilon}{648 \bar{C} \max\{\alpha_A T,1\}} \le \frac{7\epsilon}{72 \bar{C} \max\{\alpha_A T,1\}},
\end{align}
where Lemma~\ref{lem:loga/x} and the definition of $\epsilon_{\rm C}$ are used at the first inequality.
Combining this with Eq.~\eqref{eq:epsCSqrtrC} yields
\begin{align}
    \epsilon_{\rm C} \sqrt{r} \|C\|_{\rm F} \le \frac{\epsilon}{3}.
    \label{eq:epsCSqrtrtilCF}
\end{align}   
Noting
\begin{align}
    \vecm_{\veccalY}(t)=
    \begin{pmatrix}
        \E[\vecY^1_t] \\ \vdots \\ \E[\vecY^{\NC}_t]
    \end{pmatrix},
\end{align}
we have
\begin{align}
    \E[\tilde{f}] = \frac{1}{\gamma}\sum_{n=1}^r \sum_{i=1}^N C_{n,i} \calM_{\veccalY,{\rm hist}}^{n,i},~\tilde{f} \coloneqq \frac{1}{\gamma}\sum_{n=1}^r \sum_{i=1}^N C_{n,i} Y^{1,i}_{t_n}.
\end{align}
We thus have
\begin{align}
    |\E[\tilde{f}]-\mu_f| & = \left|\sum_{n=1}^r \sum_{i=1}^N C_{n,i} \E[\gamma^{-1}Y^{1,i}_{t_n}-X^i_{t_n}]\right| \nonumber \\
    & \le \|C\|_{\rm F} \sqrt{\sum_{n=1}^r \left\|\gamma^{-1}\E[\vecY^1_{t_n}]-\E[\vecX_{t_n}]\right\|^2} \nonumber \\
    & \le \|C\|_{\rm F} \sqrt{r} \epsilon_{\rm C} \nonumber \\
    & = \frac{\epsilon}{3},
    \label{eq:EXiErr}
\end{align}
where we use the Cauchy-Schwarz inequality at the first inequality, Lemma~\ref{lem:CarErrTail} at the second inequality, and Eq.~\eqref{eq:epsCSqrtrtilCF} at the last inequality.

The outputs $\hat{\mu}^\prime_f$ of the overlap estimation in STEP 3 satisfies
\begin{align}
    \left|\hat{\mu}^\prime_f-\frac{1}{4r\tilde{U}_{\vecm_{\veccalY}}\|C\|_{\rm F}}\sum_{n=1}^r\sum_{i=1}^N C_{n,i}\tilde{\tilde{\calM}}_{\veccalY,{\rm hist}}^{n,i}\right| \le \epsilon_{\rm OE}
\end{align}
with probability at least $1-\delta$, and if this holds, we have
\begin{align}
    &\left|\hat{\mu}_f-\E[\tilde{f}]\right| \nonumber \\
    &\le \left|\frac{4r\tilde{U}_{\vecm_{\veccalY}}\|C\|_{\rm F}}{\gamma}\hat{\mu}^\prime_f-\frac{1}{\gamma}\sum_{n=1}^r\sum_{i=1}^N C_{n,i}\tilde{\tilde{\calM}}_{\veccalY,{\rm hist}}^{n,i}\right| + \frac{1}{\gamma}\left|\sum_{n=1}^r\sum_{i=1}^N C_{n,i}\tilde{\tilde{\calM}}_{\veccalY,{\rm hist}}^{n,i}-\sum_{n=1}^r\sum_{i=1}^N C_{n,i}\calM_{\veccalY,{\rm hist}}^{n,i}\right| \nonumber \\
    &\le \frac{4r\tilde{U}_{\vecm_{\veccalY}}\|C\|_{\rm F}\epsilon_{\rm OE}}{\gamma} + \frac{1}{\gamma}\|C\|_{\rm F}\left\|\tilde{\tilde{\calM}}_{\veccalY,{\rm hist}} - \calM_{\veccalY,{\rm hist}}\right\|_{\rm F} \nonumber \\
    &\le \frac{4r\tilde{U}_{\vecm_{\veccalY}}\|C\|_{\rm F}\epsilon_{\rm OE}}{\gamma} + \frac{1}{\gamma}\|C\|_{\rm F} \tilde{U}_{\vecm_{\veccalY}} \epsilon^\prime \nonumber \\
    &\le \frac{2}{3}\epsilon,
    \label{eq:muXiEXiTil}
\end{align}
where we use Theorem~\ref{th:HistStatemY} at the third inequality.
Combining Eqs.~\eqref{eq:EXiErr} and \eqref{eq:muXiEXiTil} yields
\begin{align}
    \left|\hat{\mu}_f-\mu_f\right| \le \epsilon.
\end{align}

\noindent \underline{\textbf{Complexity}}

By noting that overlap estimation in Step 2 in Algorithm~\ref{alg:main} makes $O\left(\frac{1}{\epsilon_{\rm OE}}\log\left(\frac{1}{\delta}\right)\right)$ uses of $V_{\vecm_{\veccalY},\epsilon^\prime}^{\rm hist}$ and $V_C$ and that the complexity of $V_{\vecm_{\veccalY},\epsilon^\prime}^{\rm hist}$ is as stated in Theorem~\ref{th:HistStatemY} with $\epsilon$ replaced by $\epsilon^\prime$, we see that Algorithm~\ref{alg:main} makes
\begin{align}
    O\left(\frac{\NC^2 r^{\frac{5}{2}} K \tilde{U}_{\vecm_{\veccalY}} \bar{C}}{\gamma \epsilon} \log\left(\frac{r \tilde{U}_{\vecm_{\veccalY}} \bar{C}}{\gamma \epsilon}\right)\log\left(\frac{1}{\delta}\right)\right)
\end{align}
uses of the controlled $V_A$ and the controlled $V_F$,
\begin{align}
    O\left(\frac{\NC^3 r^{\frac{5}{2}} K \tilde{U}_{\vecm_{\veccalY}} \bar{C}}{\gamma \epsilon} \log\left(\frac{r \tilde{U}_{\vecm_{\veccalY}} \bar{C}}{\gamma \epsilon}\right)\log\left(\frac{1}{\delta}\right)\right)
\end{align}
uses of the controlled $V_{BB^T}$,
\begin{align}
    O\left(\frac{\NC r^{\frac{3}{2}} \tilde{U}_{\vecm_{\veccalY}} \bar{C}}{\gamma \epsilon}\log\left(\frac{1}{\delta}\right)\right)
\end{align}
uses of the controlled $V_{\vecX_{\rm ini}}$, and
\begin{align}
    O\left(\frac{r^{\frac{5}{2}} K \tilde{U}_{\vecm_{\veccalY}} \bar{C} \left(\NC^2+\log K \log M+a_A+a_F\right)}{\gamma \epsilon} \log\left(\frac{r \tilde{U}_{\vecm_{\veccalY}} \bar{C}}{\gamma \epsilon}\right)\log\left(\frac{1}{\delta}\right)\right)
\end{align}
uses of additional elementary gates, where we replace $\epsilon_{\rm OE}$ and $\epsilon^\prime$ with their definitions.
By some algebra, we see that in the current setting, the following holds:
\begin{align}
    J = O\left(\frac{1}{1-c}\log\left(\frac{(\sqrt{\sigma^2 T}+\sigma_\vecX) (\alpha_A T + \sqrt{N}) \bar{C}}{\epsilon c (1-c)}\right)\right),
\end{align}
\begin{align}
    r = O\left(\frac{\alpha_A T}{1-c} \log\left(\frac{(\sqrt{\sigma^2 T}+\sigma_\vecX) (\alpha_A T + \sqrt{N}) \bar{C}}{\epsilon c (1-c)}\right)\right),
\end{align}
\begin{align}
    \frac{1}{\gamma} = O\left(\frac{\sqrt{\sigma^2 T}}{c}\log\left(\frac{(\sqrt{\sigma^2 T}+\sigma_\vecX) (\alpha_A T + \sqrt{N}) \bar{C}}{\epsilon c (1-c)}\right)\right),
\end{align}
\begin{align}
    \tilde{U}_{\vecm_{\veccalY}} = O(1),
\end{align}
\begin{align}
    K=O\left(\log\left(\frac{(\sqrt{\sigma^2 T}+\sigma_\vecX) (\alpha_A T + \sqrt{N}) \bar{C}}{\epsilon c (1-c)}\right)\right),
\end{align}
\begin{align}
    \log M=O\left(\log\left(\frac{\alpha_{\drm A/\drm t} + \frac{\sigma^2T}{c^2} \alpha_{\drm B/\drm t} + \left(\frac{\sigma^2T}{c^2}\right)^{\frac{1-s}{2}}\alpha_{\drm F/\drm t}}{\alpha_A^2}\cdot\frac{(\sqrt{\sigma^2 T}+\sigma_\vecX) (\alpha_A T + \sqrt{N}) \bar{C}}{\epsilon c (1-c)}\right)\right).
\end{align}
Combining the above things, we obtain the claimed complexity.

\end{proof}

\end{document}